\documentclass[a4paper,11pt]{article}
\pdfoutput=1
\usepackage{amsmath}
\usepackage{amssymb}
\usepackage{color}
\usepackage{cite}
\usepackage{hyperref}
\usepackage{graphicx}
\usepackage{leftidx}
\usepackage{subfig}

\numberwithin{equation}{section}   

\def \be {\begin{equation}}
\def \ee {\end{equation}}
\def \ba {\begin{array}}
\def \ea {\end{array}}
\def \bea{\begin{eqnarray}}
\def \eea{\end{eqnarray}}

\begin{document}

\title{\textbf{An Observable in Classical Pure AdS$_3$ Gravity: \\ the Twist along a Geodesic}}
\author{
Xiao-Shuai Wang$^{1,2}$\footnote{wangxiaoshuai@itp.ac.cn}\,
Jie-qiang Wu$^{1,2}$\footnote{jieqiangwu@itp.ac.cn}
}
\date{}

\maketitle

\begin{center}
{\it
$^{1}$CAS Key Laboratory of Theoretical Physics, Institute of Theoretical Physics, Chinese Academy of Sciences, Beijing 100190, China \\
$^{2}$School of Physical Sciences, University of Chinese Academy of Sciences, \\ Beijing 100049, China
}
\vspace{10mm}
\end{center}

\begin{abstract}

In this paper, we consider a little-studied observable in classical pure AdS$_3$ gravity: the twist along a geodesic.
The motivation is that the twist only supports on the geodesic so may be a candidate element in the center of the algebra in either entanglement wedge associated to the geodesic.
We study the properties of the twist and get the following results.
First, we get the system's evolution generated by the twist, which exhibits a relative shift along the geodesic.
Second, we show that the twist commutes with the length of the same geodesic, which supports the proposal that the twist is a candidate element in the center.

\end{abstract}

\baselineskip 18pt
\thispagestyle{empty}

\newpage

\tableofcontents

\section{Introduction}

The construction of the Hilbert space of gravity is a long standing problem \cite{Arnowitt:1962hi,DeWitt:1967yk}, but is far from fully solved even in perturbative level. One significant challenge is to deal with diffeomorphism symmetries and gravitational constraints.

Constraints appear generally in theories with gauge redundancies, including gauge theories and gravity \cite{Dirac,Henneaux:1992ig}. They are some equations involving the initial data on a Cauchy surface, that are usually in terms of infinitesimally closed spacelike separated points. Because of the constraints, the degrees of freedom at spacelike separated points are not completely independent. More specifically, for a bipartite system, the Hilbert space of the whole system cannot be written as a tensor product of the Hilbert spaces of the individual subsystems \cite{Casini:2013rba,Donnelly:2014fua,Donnelly:2015hxa,Harlow:2016vwg,Akers:2018fow,Dong:2018seb}.\footnote{This statement is also true for multipartite system. However, for simplicity, we only focus on bipartite system in this paper.}

The Hilbert space for a bipartite system of a theory with constraints has the following structure
\cite{Casini:2013rba,Donnelly:2014fua,Donnelly:2015hxa,Harlow:2016vwg,Akers:2018fow,Dong:2018seb}.
We first need to introduce the notion of the center, which is formed by a set of commutative operators supporting on the interface between the two subsystems. Based on the center, the Hilbert space can be decomposed into a set of sub-Hilbert space with each of them being tensor factorizable.
Clearly, the center plays an important role in such Hilbert space decomposition. However, in gravity, the knowledge of the center is very limited, and the only known element in the center is the HRT-area.

In the framework of AdS/CFT correspondence, the HRT-area appears in the Ryu-Takayanagi (RT) formula \cite{Ryu:2006bv,Ryu:2006ef,Hubeny:2007xt} and the Jafferis-Lewkowycz-Maldacena-Suh (JLMS) formula \cite{Jafferis:2014lza,Jafferis:2015del,Dong:2016eik}.
The RT formula suggests that the entanglement entropy in field theory corresponds to the area of an extremal surface in gravity at classical order, where, in this paper, we refer to the extremal surface as HRT-surface and its area as HRT-area following the authors' name in \cite{Hubeny:2007xt}. Based on the RT formula, the JLMS formula furthermore suggests that the modular Hamiltonian in field theory also corresponds to the HRT-area in gravity at classical order.
It is in the JLMS formula that the HRT-area is treated as an operator in gravity and furthermore suggested as an element in the center.

Further studies have also been performed on the HRT-area from the perspective of treating it as an operator in gravity
\cite{Bousso:2019dxk,Bousso:2020yxi,Kaplan:2022orm}.
Motivated by the JLMS formula and the studies of modular flow \cite{Lewkowycz:2018sgn,Ceyhan:2018zfg,Faulkner:2018faa,Chen:2018rgz}, it is suggested and finally proved in \cite{Kaplan:2022orm} that the system's evolution generated by the HRT-area exhibits a kink transformation along the HRT-surface, which may be useful to study the structure of the Hilbert space of gravity in future.

Besides the HRT-area, we expect there should be some other operators in the center. We therefore raise up the problem to construct and study these operators in the center.
The operators in the center are diffeomorphism invariant operators \footnote{To be more precise, the diffeomorphism invariant operators/observables only need to be invariant under the non-physical diffeomorphisms whose parameters go to zero at the asymptotic boundary.} that support only on the interface which is the HRT-surface in this setup.
We may construct these operators based on some covariant approaches
\cite{Page:1983uc,Kuchar:1991qf,Marolf:1994nz,Giddings:2005id,Donnelly:2015hta,Giddings:2018umg,Giddings:2019wmj,Harlow:2018tng,Harlow:2021dfp}.
Moreover, assuming that we can indeed construct some of the operators in the center, we are also interested in studying their properties in the following two aspects. First, inspired by the story of the HRT-area, we are interested in studying the system's evolution generated by these operators. Second, in order to verify that these operators can indeed be elements in the center, we would like to compute the commutators both between these operators themselves and between these operators and the HRT-area.

As a preliminary attempt to the problem that we raise up, we focus on a simple model in this paper, that is the classical pure AdS$_3$ gravity which has the following action
\be\label{action} S=\frac{1}{16\pi G} \int d^3x \sqrt{-g}(R+2)+\mbox{boundary terms}, \ee
and the following equations of motion
\be R_{\mu\nu}-\frac{1}{2}Rg_{\mu\nu}-g_{\mu\nu}=0. \ee
In this model, the HRT-surface is a geodesic, and the operators' commutators correspond to observables' brackets.
For algebraic simplicity, we further restrict the discussion to the system with a planar asymptotic boundary.

We only consider one observable in this paper: the twist along the geodesic \cite{Castro:2014tta}.
The twist is defined as the rapidity of the relative boost between two normal frames to the geodesic.
Here, each of the normal frame is constructed by the following two steps: first, we prepare a normal frame at one of the geodesic endpoint by inheriting the frame from the asymptotic boundary, and second, by parallel transport we extend the normal frame to the whole geodesic.

The twist defined in this way is indeed a diffeomorphism invariant observable that only supports on the geodesic.
This is based on the following three facts.
First, the twist is a functional of the configuration, that means given a configuration of the theory we can always read out a number by evaluating the twist in this configuration.
Second, the functional is diffeomorphism invariant, which is because its construction only use the intrinsic property of the metric without referring to the coordinates.
Third, the functional only supports on the geodesic, in the sense that the variation of the metric away from the geodesic doesn't change the value of the twist.

The goal of this paper is to study the properties of the twist, including the system's evolution generated by the twist and the brackets with the twist.
We use the two approaches developed in \cite{Kaplan:2022orm} to study these properties.
Here, the first approach is based on the canonical formalism \cite{Arnowitt:1962hi,DeWitt:1967yk}, and the second approach is actually based on the covariant phase space formalism
\cite{Lee:1990nz,Iyer:1994ys,Wald:1999wa,Harlow:2019yfa}.
With these two approaches, we get the following two results:

\begin{itemize}
\item First, we get the system's evolution generated by the twist. Under a proper gauge choice, the system's evolution exhibits a relative shift along the geodesic.
\item Second, we show that the twist commutes with the length of the same geodesic. This supports the proposal that the twist is a candidate element in the center.
\end{itemize}

The plan for the rest of the paper is as follows.
In section \ref{definition}, we provide the definition of the twist along a geodesic.
In section \ref{evocan}, we study the system's evolution generated by the twist with the canonical formalism.
In section \ref{covariant}, we revisit the same problem with the covariant phase space formalism.
In section \ref{bracket}, we compute the bracket between the length and the twist of a same geodesic.
In section \ref{discussion}, we finish with conclusions and discussions.
Various technical details are left in appendices.

We emphasize that this paper is strongly inspired from the paper \cite{Kaplan:2023oaj} of Molly Kaplan, and the paper \cite{Held:2024bgs,ongoing2} of Jesse Held, Molly Kaplan, Donald Marolf, and one author of the current paper.

\section{The definition of the twist along a geodesic}\label{definition}

In this section, we provide the definition of the twist along a geodesic.

\begin{figure}
  \centering
  \includegraphics[width=5cm]{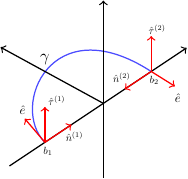}\\
  \caption{The geodesic and the normal frames}\label{twistfig}
\end{figure}

As in Fig.\ref{twistfig}, we consider a spacelike geodesic $\gamma$ with two endpoints $b_1$, $b_2$ at the asymptotic boundary. We parameterize the geodesic as
\be x^{\mu}=x^{\mu}(s), \ee
with $s$ being the proper length up to a shift. We also denote the tangent vector as $e^{\mu}$ which equals to
\be e^{\mu}= \frac{d x^{\mu}}{ds}. \ee

To define the twist, we first construct two normal frames to the geodesic $\gamma$ denoted as $(\hat{\tau}^{(1)},\hat{n}^{(1)})$, $(\hat{\tau}^{(2)},\hat{n}^{(2)})$ \footnote{In this paper, we always add a hat on a vector when we ignore its index to indicate its being a vector. For example, the $\hat{e}$, $\hat{\tau}^{(i)}$, $\hat{n}^{(i)}$ in (\ref{orthonormal}) are the same quantities as the $e^{\mu}$, $\tau^{(i)\mu}$, $n^{(i)\mu}$ in (\ref{transport}) respectively.} that satisfy the parallel transport conditions
\be\label{transport} \left. e^{\mu}\nabla_{\mu}\tau^{(i)\nu} \right|_{\gamma}
=\left. e^{\mu}\nabla_{\mu}n^{(i)\nu} \right|_{\gamma}=0, \ee
and the orthonormal conditions
\begin{align}\label{orthonormal}
&\left. \hat{\tau}^{(i)}\cdot \hat{e} \right|_{\gamma}=
\left. \hat{n}^{(i)}\cdot \hat{e} \right|_{\gamma}=
\left. \hat{\tau}^{(i)}\cdot \hat{n}^{(i)} \right|_{\gamma}=0 \notag \\
&\left. -\hat{\tau}^{(i)}\cdot \hat{\tau}^{(i)} \right|_{\gamma}=\left. \hat{n}^{(i)} \cdot \hat{n}^{(i)} \right|_{\gamma}=1,
\end{align}
 for $i=1,2$.
Here $\hat{\tau}^{(1)}$, $\hat{\tau}^{(2)}$ are future-pointing, and $\hat{n}^{(1)}$, $\hat{n}^{(2)}$ are right-pointing from the viewpoint of an observer facing along the geodesic $\gamma$.

The two normal frames $(\hat{\tau}^{(1)},\hat{n}^{(1)})$, $(\hat{\tau}^{(2)},\hat{n}^{(2)})$ are constructed by the following two steps.
First, we adopt a proper value for $(\hat{\tau}^{(1)},\hat{n}^{(1)})$/$(\hat{\tau}^{(2)},\hat{n}^{(2)})$ at a specific point close to the initial/final geodesic endpoint as in Fig.\ref{twistfig}.
Here, the adopting is by inheriting the frame from the asymptotic boundary together with an ignorable modification.
The adopted $(\hat{\tau}^{(1)},\hat{n}^{(1)})$, $(\hat{\tau}^{(2)},\hat{n}^{(2)})$, at the corresponding specific points, satisfy the orthonormal conditions (\ref{orthonormal}) and have the directions consistent with the requirement mentioned in the previous paragraph.\footnote{See subsection \ref{kinematic} for an explicit expression of the adopted value for $(\hat{\tau}^{(1)},\hat{n}^{(1)})$, $(\hat{\tau}^{(2)},\hat{n}^{(2)})$ at the corresponding specific points. There, we will also point out that the specific points are actually the intersections of the geodesic with the cutoff surface.}
Second, by taking a parallel transport, we extend $(\hat{\tau}^{(1)},\hat{n}^{(1)})$, $(\hat{\tau}^{(2)},\hat{n}^{(2)})$ to the whole geodesic $\gamma$.
Here, the constructed $(\hat{\tau}^{(1)},\hat{n}^{(1)})$, $(\hat{\tau}^{(2)},\hat{n}^{(2)})$ satisfy both the parallel transport conditions (\ref{transport}) and the orthonormal conditions (\ref{orthonormal}) on the whole geodesic $\gamma$.

With the two normal frames $(\hat{\tau}^{(1)},\hat{n}^{(1)})$, $(\hat{\tau}^{(2)},\hat{n}^{(2)})$, we now define the twist along the geodesic $\gamma$.
The two normal frames are related by a relative boost
\begin{align}\label{boost}
&\hat{\tau}^{(2)}-\hat{n}^{(2)} \big|_{\gamma}=e^{\zeta}(\hat{\tau}^{(1)}-\hat{n}^{(1)}) \big|_{\gamma}
\notag \\
&\hat{\tau}^{(2)}+\hat{n}^{(2)} \big|_{\gamma} =
e^{-\zeta} (\hat{\tau}^{(1)}+\hat{n}^{(1)}) \big|_{\gamma}.
\end{align}
Here, the rapidity $\zeta$ is constant along the geodesic $\gamma$, which can be shown by acting a tangent direction derivative $e^{\mu}\nabla_{\mu}$ on (\ref{boost}) and combining with the parallel transport condition (\ref{transport}).
We then define the twist along the geodesic as this constant rapidity $\zeta$.

\section{The system's evolution generated by the twist with the canonical formalism}\label{evocan}

Having provided the definition of the twist $\zeta$, we are now ready to study its properties in this and the following sections. In this section, we study the system's evolution generated by the twist $\zeta$ with the canonical formalism.

\subsection{A representation of the system's evolution generated by an observable}

We first explain what we mean by the system's evolution generated by an observable. This can be explained clearly by referring to quantum mechanics.

In quantum mechanics, for a given Hermitian operator $O$, we can view it as the Hamiltonian and use it to evolve the system.
In Heisenberg picture, the evolution is acted on the set of operators as
\be\label{evolutionquan} W(\lambda)=e^{i\lambda O} W e^{-i\lambda O}, \ee
where $W$ is an arbitrary operator and $\lambda$ is the evolution parameter.
We can also rewrite the evolution equation (\ref{evolutionquan}) as
\be\label{devolutionquan} \frac{d}{d\lambda} W(\lambda)=-i[W(\lambda),O], \ee
which has a direct correspondence in classical limit.

In classical limit, the operators correspond to observables, the operators' commutators correspond to observables' brackets, so the evolution equation (\ref{devolutionquan}) corresponds to
\be\label{devolutioncla} \frac{d}{d\lambda} W(\lambda)=\{W(\lambda),O\}. \ee
Simpler than the full quantum mechanics, the classical system can be completely described by the set of initial data. Therefore, we only need to apply the set of initial data to the evolution equation (\ref{devolutioncla}), which is sufficient to capture the evolution of the system.

In this paper, we only focus on the evolution (\ref{devolutioncla}) to the linear order with respect to $\lambda$,
and we try to compute the following quantity
\be\label{deltaW} \Delta W\equiv W(\lambda)-W(0)=\lambda \{W,O\}+o(\lambda), \ee
with $W$ taken through the set of initial data.
For pure AdS$_3$ gravity, the set of initial data is consist of the induced metric and the extrinsic curvature, denoted by $(\sigma_{ab},K_{ab})$, on a given Cauchy surface, denoted by $\Sigma$.
We therefore provide the following representation of the system's evolution generated by a given diffeomorphism observable $O$ as
\begin{align}\label{initialevo}
&\Delta \sigma_{ab}(t_0,x)=\lambda \{\sigma_{ab}(t_0,x),O\}+o(\lambda) \notag \\
&\Delta K_{ab}(t_0,x)=\lambda \{K_{ab}(t_0,x),O\}+o(\lambda).
\end{align}
Here, we have already introduced a coordinate system $(t,x^a)$ in which the Cauchy surface $\Sigma$ is at $t=t_0$, and $a$, $b$ run through the indices along the Cauchy surface $\Sigma$.
The brackets in the expressions (\ref{initialevo}) can be computed from (\ref{Poisson2})\footnote{One potential question here is which bracket we should use, the Poisson bracket or the Dirac bracket. The answer is that, if the observable $O$ in (\ref{initialevo}) is diffeomorphism invariant, there is no physical difference between these two, where the only difference is a gauge transformation. For algebraic simplicity, we use the Poisson bracket through out this paper.} with the chain rule as
\be\label{bracketO} \{\cdot,O\}=
\int d^2y \{\cdot,\sigma_{mn}(t_0,y)\} \frac{\delta O}{\delta \sigma_{mn}(t_0,y)}
+\int d^2y \{\cdot,K_{mn}(t_0,y)\} \frac{\delta O}{\delta K_{mn}(t_0,y)}. \ee
We also emphasize that, when applying the chain rule (\ref{bracketO}), we need to first represent the observable $O$ as a functional of the set of initial data on the Cauchy surface $\Sigma$.

\subsection{The kink transformation generated by the geodesic's length}\label{geodesiclength}

As an illustration of the representation (\ref{initialevo}) of the system's evolution generated by a given observable, we apply it to the geodesic's length in this subsection. We admit that this is only a review in our setup of the well known result \cite{Bousso:2019dxk,Bousso:2020yxi,Kaplan:2022orm} that the system's evolution generated by the HRT-area exhibits a kink transformation.\footnote{More precisely, by kink transformation, we refer to the boundary-condition-preserving kink transformation introduced in \cite{Kaplan:2022orm}.}
Nevertheless, the prescription and the conceptual clarification also apply in the study of the twist below.

The application is straightforward: we only need to apply the geodesic's length, denoted by $A$, to (\ref{initialevo}) and (\ref{bracketO}) in the position of $O$.
To simplify the computation, we further require that the geodesic $\gamma$ is contained in the Cauchy surface $\Sigma$ for the configuration that we consider.
This requirement can be achieved by a proper choice of the boundary of the Cauchy surface $\partial\Sigma$, together with a non-physical diffeomorphism acted on the configuration.\footnote{In this paper, we view the gravity as an effective field theory defined in a background coordinate system, and we locate the Cauchy surface only in terms of the background coordinates. In this setup, the geodesic with given end points on the asymptotic boundary is a functional of the configuration. By choosing the location of the Cauchy surface, we can at most make it containing the geodesic's endpoint for all configuration. However, after choosing such a Cauchy surface, no matter which configuration that we consider, we can always take a non-physical diffeomorphism acting on the configuration and mapping it to another configuration where the geodesic is contained in the Cauchy surface.}

To apply the geodesic's length $A$ in (\ref{initialevo}) and (\ref{bracketO}), we first compute the variation of the geodesic's length $A$ with respect to the variation of the set of initial data on the Cauchy surface $\Sigma$.
Under the requirement that the Cauchy surface $\Sigma$ contains the geodesic $\gamma$, the variation of the geodesic's length $A$ has the following simple expression\footnote{Note that we only make requirement for the metric $g_{\mu\nu}$ such that the Cauchy surface $\Sigma$ contains the geodesic $\gamma$, but we make no requirement for the metric $g_{\mu\nu}+\delta g_{\mu\nu}$.}
\begin{align}\label{deltaA}
\delta A=&\int_{\gamma} ds \frac{1}{2} \delta g_{\mu\nu} e^{\mu}e^{\nu} \notag \\
=& \int_{\gamma} ds \frac{1}{2} \delta \sigma_{ab} e^a e^b \notag \\
=&\int_{\Sigma} d^2 x \sqrt{\sigma} \frac{1}{2} \delta \sigma_{ab} e^a e^b \delta(\rho).
\end{align}
Here, we have used the first equation of (\ref{deltasigmaK}). $e^a$ is extended to a vector field of the Cauchy surface $\Sigma$ in an arbitrary way, which coincides with the tangent vector of the geodesic $\gamma$ when restricted there. $\rho$ is defined as a scalar field of the Cauchy surface $\Sigma$, which is positive in one side of the geodesic $\gamma$ and negative in the other side of the geodesic $\gamma$, and whose absolute value, for the region close enough to the geodesic $\gamma$, equals to the distance to the geodesic $\gamma$ through the Cauchy surface $\Sigma$.

Applying (\ref{deltaA}) to (\ref{initialevo}) and (\ref{bracketO}) and also taking use of (\ref{Poisson2}), we get the system's evolution represented as
\begin{align}\label{evolutionA}
&\Delta \sigma_{ab}(t_0,x)=o(\lambda)
\notag \\
&\Delta K_{ab}(t_0,x)=8\pi G \lambda \delta(\rho)n_an_b+o(\lambda).
\end{align}
Here $\rho$ is defined below (\ref{deltaA}). $n_a$ is defined as a one-form field of the Cauchy surface $\Sigma$, whose restriction on the geodesic $\gamma$ satisfies
\be\label{nadrho} n_a|_{\gamma}=\partial_a \rho |_{\gamma}. \ee
Or an equivalent expression is that the corresponding raised up vector
\be n^a=\sigma^{ab}n_b, \ee
when restricted on the geodesic $\gamma$, is the unit normal vector.

\begin{figure}[htbp]
\centering
\subfloat[The unevolved system]{\label{kink1}\includegraphics[width=6cm]{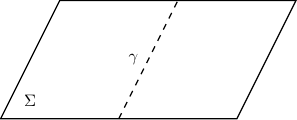}}
     \quad\quad
\subfloat[The evolved system]{\label{kink2}\includegraphics[width=6cm]{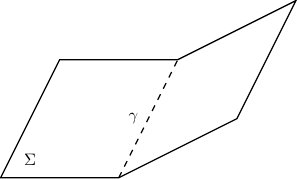}}
\caption{The kink transformation generated by the geodesic's length. Fig.\ref{kink1} and Fig.\ref{kink2} represent the systems before and after the evolution.
These figures should be interpreted in the way that the three dimensional metric is invariant and the Cauchy surface $\Sigma$ takes a kink transformation along the geodesic $\gamma$.
Moreover, in our mind, we should take an extra coordinate transformation on the evolved system in Fig.\ref{kink2}, which acts on both the metric and the Cauchy surface $\Sigma$, which maps the Cauchy surface $\Sigma$ back to the same location as the one in the original system in Fig.\ref{kink1} in terms of the coordinates, and which keeps the set of initial data on the Cauchy surface $\Sigma$ invariant.
The reason to use this illustration for system's evolution is because it is really hard to directly illustrate the evolution of the solution or the evolution of the set of initial data on a fixed Cauchy surface.
Moreover, we also point out that this illustration only works when the geodesic $\gamma$ is at the fixed points of a Killing field, which is indeed satisfied by the solutions of the pure AdS$_3$ gravity.
}\label{kink}
\end{figure}

The system's evolution in (\ref{evolutionA}) is the well known kink transformation illustrated in Fig.\ref{kink}.
It is the set of initial data on the Cauchy $\Sigma$ that evolves as (\ref{evolutionA}).

\subsection{The variation of the twist $\zeta$ with respect to the variation of the metric}\label{cdeltazeta}

We now return to our main topic: the system's evolution generated by the twist $\zeta$.
The study of the twist is parallel to the study of the geodesic's length $A$ reviewed in the previous subsection.
However, to provide more detials, we divide the study into several subsections.

In this subsection, we compute the variation of the twist $\zeta$ with respect to the variation of the metric.
Based on the definition of the twist $\zeta$ in section \ref{definition}, we transform this question to the computation of the variation of the geodesic and the normal frames.

We will compute the variation of the geodesic and the normal frames by solving some differential equations below, and we now derive these differential equations.
The geodesic $x^{\mu}(s)$ and the normal frames  $(\hat{\tau}^{(1)},\hat{n}^{(1)})$, $(\hat{\tau}^{(2)},\hat{n}^{(2)})$ satisfy the geodesic equation
\be\label{geo} \frac{d^2 x^{\mu}}{ds^2}+\Gamma^{\mu}_{\nu\rho}(x(s)) \frac{dx^{\nu}}{ds} \frac{dx^{\rho}}{ds}=0, \ee
and the parallel transport equation
\be\label{transportV} \left. \frac{dx^{\nu}}{ds}\nabla_{\nu} V^{\mu} \right|_{\gamma}=0,
\ee
where $V^{\mu}$ represents $\tau^{(1)\mu}$, $n^{(1)\mu}$, $\tau^{(2)\mu}$, $n^{(2)\mu}$.
By taking a variation of (\ref{geo}) and (\ref{transportV}), we get the differential equations for the variation of the geodesic and the normal frames as
\be\label{vargeo} e^{\alpha}\nabla_{\alpha}(e^{\beta}\nabla_{\beta}\delta x^{\mu})
+R_{\nu\alpha\phantom{\mu}\beta}^{\phantom{\nu\alpha}\mu} \delta x^{\nu}e^{\alpha}e^{\beta}
+\delta_g \Gamma^{\mu}_{\alpha\beta} e^{\alpha}e^{\beta}=0,
\ee
and
\be\label{varframe} e^{\alpha}\nabla_{\alpha} \delta^{(c)}V^{\mu}
+R_{\nu\alpha\phantom{\mu}\beta}^{\phantom{\nu\alpha}\mu} \delta x^{\nu} e^{\alpha} V^{\beta}
+\delta_g \Gamma^{\mu}_{\alpha\beta} e^{\alpha}V^{\beta}=0, \ee
where
\be \delta_g \Gamma^{\mu}_{\nu\rho}=
\frac{1}{2} g^{\mu\sigma}(\nabla_{\nu}\delta g_{\rho\sigma}
+\nabla_{\rho} \delta g_{\nu\sigma}-\nabla_{\sigma} \delta g_{\nu\rho}), \ee
and
\be \delta^{(c)} V^{\mu}(s)\equiv
\delta V^{\mu}(s)+\Gamma^{\mu}_{\nu\rho}(x(s)) \delta x^{\nu}(s) V^{\rho}(s). \ee
We have chosen the parameter $s$ to be the proper length up to a shift for the geodesic $x^{\mu}(s)$ in the unvaried metric $g_{\mu\nu}$; but note that, for the geodesic $x^{\mu}(s)+\delta x^{\mu}(s)$ in the varied metric $g_{\mu\nu}+\delta g_{\mu\nu}$, we only choose the parameter $s$ to be an affine parameter.\footnote{See appendix \ref{deriving} for a detailed derivation for (\ref{vargeo}) and (\ref{varframe}).}
In practice, we are only interested in the case that the unvaried metric $g_{\mu\nu}$ is the solution of the equations of motion so the Riemann tensor has the following simple behavior
\be\label{Riemann} R_{\mu\nu\rho\sigma}=g_{\mu\sigma}g_{\nu\rho}-g_{\mu\rho}g_{\nu\sigma}.
\ee
From now on, we will make this requirement and replace the Riemann tensor in (\ref{vargeo}) and (\ref{varframe}) by (\ref{Riemann}).

The solutions of the differential equations (\ref{vargeo}), (\ref{varframe}) for $\delta x^{\mu}$, $\delta^{(c)} \tau^{(1)\mu}$, $\delta^{(c)} n^{(1)\mu}$, $\delta^{(c)} \tau^{(2)\mu}$, $\delta^{(c)} n^{(2)\mu}$ would include some integrals along the geodesic of the variation of metric $\delta g_{\mu\nu}$ and its derivative. However, to fully solve these differential equations (\ref{vargeo}), (\ref{varframe}), we still need to adopt some boundary conditions on $\delta x^{\mu}$, $\delta^{(c)} \tau^{(1)\mu}$, $\delta^{(c)} n^{(1)\mu}$, $\delta^{(c)} \tau^{(2)\mu}$, $\delta^{(c)} n^{(2)\mu}$.

The boundary conditions should be read out from the definitions of the geodesic and the normal frames, especially on their aspects of near boundary behavior.
We can directly point out the locations of the adopted boundary conditions:
for $\delta x^{\mu}$, we adopt one boundary condition close to the initial endpoint and one boundary condition close to the final endpoint;
for $\delta^{(c)}\tau^{(1)\mu}$ or $\delta^{(c)}n^{(1)\mu}$, we adopt one boundary condition close to the initial endpoint;
for $\delta^{(c)}\tau^{(2)\mu}$ or $\delta^{(c)}n^{(2)\mu}$, we adopt one boundary condition close to the final endpoint.\footnote{The locations of the boundary conditions adopted on $\delta^{(c)}\tau^{(1)\mu}$, $\delta^{(c)}n^{(1)\mu}$, $\delta^{(c)}\tau^{(2)\mu}$, $\delta^{(c)}n^{(2)\mu}$ are obtained from the following arguments. We take $\delta^{(c)}\tau^{(1)\mu}$, $\delta^{(c)}n^{(1)\mu}$ as an example. In section \ref{definition}, we have defined $(\hat{\tau}^{(1)},\hat{n}^{(1)})$ as a parallel transport of the frame inherited from the asymptotic boundary at a specified point close to the initial endpoint of the geodesic. This inheriting should be interpreted as the boundary conditions adopted on $\tau^{(1)\mu}$, $n^{(1)\mu}$ close to the initial point. Under a variation, we would get the boundary conditions adopted on $\delta^{(c)} \tau^{(1)\mu}$, $\delta^{(c)} n^{(1)\mu}$ close to the initial endpoint. A similar argument also applies for $\delta^{(c)} \tau^{(2)\mu}$, $\delta^{(c)} n^{(2)\mu}$.}
We expect that the boundary conditions will in some sense set the corresponding quantities to zero in the corresponding locations.
But due to our lack of knowledge of the near boundary behavior, we cannot write down the explicit form of these boundary conditions.

\begin{figure}[htbp]
  \centering
  \includegraphics[width=7cm]{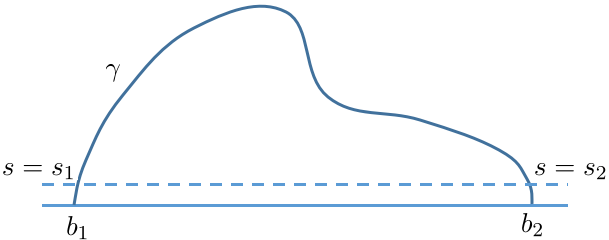}\\
  \caption{The cutoff surface and its intersections with the geodesic}\label{cutofffig}
\end{figure}

For the purpose of this paper, we can take a strategy that avoids taking use of the explicit form of the boundary conditions: we separate the contributions of the boundary conditions out of the relevant results, and finally ignore them by some assumptions.
We now explain this strategy in more detail.
We first introduce a cutoff surface as in Fig.\ref{cutofffig}. The cutoff surface intersects with the geodesic $\gamma$ at two intersections, denoted by their affine parameters $s_1$, $s_2$, close to the two geodesic's endpoints respectively.
We then solve the differential equations (\ref{vargeo}), (\ref{varframe}) in the region $s_1<s<s_2$ between the two intersections.
Instead of adopting some boundary conditions at $s_1$ or $s_2$, we allow the solutions depending on the values of the following quantities
$\delta x^{\mu}(s_1)$, $\delta x^{\mu}(s_2)$, $\delta^{(c)} \tau^{(1)\mu}(s_1)$, $\delta^{(c)} n^{(1)\mu}(s_1)$, $\delta^{(c)} \tau^{(2)\mu}(s_2)$, $\delta^{(c)} n^{(2)\mu}(s_2)$\footnote{Because of the orthonormal conditions, not all components of these quantities are independent.} supported at $s_1$ and $s_2$.\footnote{Here, this list of quantities is chosen in terms of the boundary conditions mentioned in the previous paragraph, where if we need to adopt a boundary condition for one quantity close to one geodesic's endpoint in the previous context, we instead, in the current context, allow the solutions depend on the value of the quantity at the corresponding intersection.}
Even though these quantities are not directly set to zero by the boundary conditions, they are still small in some sense because the intersections are close to the locations where the boundary conditions are adopted.
We will keep the dependence on these values until the step of computing the system's evolution, and then make the assumption that the total contributions supported at $s_1$ and $s_2$ can be ignored when removing the cutoff.

We are now ready to solve the differential equations (\ref{vargeo}), (\ref{varframe}), for $\delta x^{\mu}$, $\delta^{(c)} \tau^{(1)\mu}$, $\delta^{(c)} n^{(1)\mu}$, $\delta^{(c)} \tau^{(2)\mu}$, $\delta^{(c)} n^{(2)\mu}$. Here, we only sketch the solving process; see appendix \ref{solequations} for more details.
For convenience, we take a decomposition for $\delta x^{\mu}$, $\delta^{(c)} \tau^{(1)\mu}$, $\delta^{(c)} n^{(1)\mu}$, $\delta^{(c)} \tau^{(2)\mu}$, $\delta^{(c)} n^{(2)\mu}$ as
\begin{align}\label{decom}
\delta x^\mu(s)=&C^e(s) e^\mu(s)+C^{\tau^{(1)}}(s) \tau ^{(1)\mu}(s)+C^{n^{(1)}}(s) n^{(1)\mu}(s) \notag \\
=&C^e(s) e^\mu(s) +C^{\tau^{(2)}}(s) \tau ^{(2)\mu}(s)+ C^{n^{(2)}}(s) n^{(2)\mu}(s),
\end{align}
and
\begin{align}\label{deframe}
\delta ^{(c)} \tau ^{(1)\mu}(s)=&C_{(\tau^{(1)} )}^e(s) e^\mu(s)+
C_{(\tau^{(1)} )}^{\tau^{(1)}}(s) \tau ^{(1)\mu}(s) + C_{(\tau^{(1)} )}^{n^{(1)}}(s) n^{(1)\mu}(s) \notag \\
\delta ^{(c)} n^{(1)\mu}(s)=& C_{(n^{(1)})}^e(s) e^{\mu}(s)
+C_{(n^{(1)})}^{\tau^{(1)}}(s) {\tau^{(1)\mu}}(s)+C_{(n^{(1)})}^{n^{(1)}}(s) n^{(1)\mu}(s) \notag \\
\delta ^{(c)} \tau ^{(2)\mu}(s)=&C_{(\tau^{(2)} )}^e(s) e^\mu(s)+
C_{(\tau^{(2)} )}^{\tau^{(2)}}(s) \tau ^{(2)\mu}(s)+
C_{(\tau^{(2)} )}^{n^{(2)}}(s) n^{(2)\mu}(s) \notag \\
\delta ^{(c)} n^{(2)\mu}(s)=& C_{(n^{(2)})}^e(s) e^{\mu}(s)
+C_{(n^{(2)})}^{\tau^{(2)}}(s) \tau^{(2)\mu}(s)
+C_{(n^{(2)})}^{n^{(2)}}(s) n^{(2)\mu}(s).
\end{align}
Here, the sets $(\hat{e},\hat{\tau}^{(1)},\hat{n}^{(1)})$, $(\hat{e},\hat{\tau}^{(2)},\hat{n}^{(2)})$ are viewed as frames on the geodesic $\gamma$ that satisfy the orthonormal conditions
\begin{align}
&\hat{e}^2=-\hat{\tau}^{(i)2}=\hat{n}^{(i)2}=1 \notag \\
&\hat{e}\cdot \hat{\tau}^{(i)}=\hat{e}\cdot \hat{n}^{(i)}=\hat{\tau}^{(i)}\cdot \hat{n}^{(i)}=0,
\end{align}
and the parallel transport conditions
\be e^{\nu}\nabla_{\nu} e^{\mu}=e^{\nu}\nabla_{\nu} \tau^{(i)\mu}=e^{\nu}\nabla_{\nu} n^{(i)\mu}=0, \ee
for $i=1,2$.
Because of the relation (\ref{boost}), the $C^{\tau^{(1)}}$, $C^{n^{(1)}}$, $C^{\tau^{(2)}}$, $C^{n^{(2)}}$ in (\ref{decom}) satisfy
\begin{align}
&C^{\tau^{(2)}}(s)=\cosh \zeta \cdot C^{\tau^{(1)}}(s)+\sinh \zeta \cdot C^{n^{(1)}}(s) \notag \\
&C^{n^{(2)}}(s)=\sinh \zeta \cdot C^{\tau^{(1)}}(s)+\cosh \zeta \cdot C^{n^{(1)}}(s).
\end{align}
By applying (\ref{decom}), (\ref{deframe}) to the differential equations (\ref{vargeo}), (\ref{varframe}) and also taking a decomposition with respect to $(\hat{e},\hat{\tau}^{(i)},\hat{n}^{(i)})$, for $i=1,2$, we can decompose the differential equation (\ref{vargeo}) to (\ref{dxdecom}) and the differential equations (\ref{varframe}) to (\ref{dframe}). By solving the differential equations (\ref{dxdecom}), (\ref{dframe}) and also taking use of the orthonormal conditions, we get the components of $\delta x^{\mu}$ as in (\ref{soldeltax}) and the components of $\delta^{(c)}\tau^{(1)\mu}$, $\delta^{(c)} n^{(1)\mu}$, $\delta^{(c)}\tau^{(2)\mu}$, $\delta^{(c)} n^{(2)\mu}$ as in (\ref{Cftau1}), (\ref{Cfn1}), (\ref{Cftau2}), (\ref{Cfn2}) respectively.

Having got the variation of the geodesic and the normal frames, we are now ready to compute the variation of the twist with respect to the metric.
We first take a variation of (\ref{boost}) and get
\begin{align}\label{varboost}
&\delta^{(c)}\tau^{(2)\mu}-\delta^{(c)}n^{(2)\mu}= e^{\zeta} (\tau^{(1)\mu}-n^{(1)\mu}) \delta \zeta
+e^{\zeta}(\delta^{(c)}\tau^{(1)\mu} -\delta^{(c)} n^{(1)\mu}) \notag \\
&\delta^{(c)}\tau^{(2)\mu}+\delta^{(c)}n^{(2)\mu}= -e^{-\zeta} (\tau^{(1)\mu}+n^{(1)\mu}) \delta \zeta
+e^{-\zeta}(\delta^{(c)}\tau^{(1)\mu}+\delta^{(c)} n^{(1)\mu}).
\end{align}
By applying (\ref{Cftau1}), (\ref{Cfn1}), (\ref{Cftau2}), (\ref{Cfn2}) to (\ref{varboost}) through (\ref{deframe}), we get the variation of the twist $\zeta$ with respect to the metric as
\begin{align}\label{deltatwist}
\delta \zeta  =&\int_{{s_1}}^{{s_2}} ds
\frac{1}{2} n^{\alpha} \tau^{\beta} e^{\gamma}
\big( {\nabla _\alpha }\delta {g_{\beta \gamma }}(x(s))
-\nabla_{\beta} \delta g_{\alpha \gamma} (x(s)) \big) \notag \\
&+\big( \frac{1}{2}\delta {g_{\alpha \beta }}(x({s_1})) \tau^{(1)\alpha} n^{(1)\beta}
+C_{({\tau ^{(1)}})}^{{n^{(1)}}}({s_1}) \big)
-\big( \frac{1}{2}\delta {g_{\alpha \beta }}(x({s_2})) \tau^{(2)\alpha} n^{(2)\beta}
+C_{({\tau ^{(2)}})}^{{n^{(2)}}}({s_2}) \big).
    \end{align}
Here, $(\hat{\tau},\hat{n})$ is an arbitrary normal frame to the geodesic $\gamma$ with the same orientation as the frames $(\hat{\tau}^{(1)},\hat{n}^{(1)})$, $(\hat{\tau}^{(2)},\hat{n}^{(2)})$; the normal frame $(\hat{\tau},\hat{n})$ still satisfies the orthonormal conditions
\begin{align}\label{deltazeta}
&\hat{e}\cdot \hat{\tau}=\hat{e}\cdot \hat{n}=\hat{\tau} \cdot \hat{n}=0 \notag \\
&-\hat{\tau}^2=\hat{n}^2=1, \end{align}
but not necessarily satisfies the parallel transport conditions.
$C_{(\tau^{(1)})}^{n^{(1)}}(s_1)$, $C_{(\tau^{(2)})}^{n^{(2)}}(s_2)$ are given in (\ref{C1s1C2s2}); they are just some linear combinations of the components of $\delta^{(c)} \tau^{(1)\mu}(s_1)$, $\delta^{(c)} \tau^{(2)\mu}(s_2)$, which, as mentioned previously, are allowed to appear in the final expression.
In deriving (\ref{deltazeta}), we have also used (\ref{boost}), (\ref{cobrel}) and the following relation
\begin{align}
&n^{(1)\alpha} \tau^{(1)\beta} e^{\gamma}
\big( {\nabla _\alpha }\delta {g_{\beta \gamma }}(x(s)) - {\nabla _\beta }\delta {g_{\alpha \gamma }}(x(s)) \big) \notag \\
=&n^{(2)\alpha} \tau^{(2)\beta} e^{\gamma}
\big( {\nabla _\alpha }\delta {g_{\beta \gamma }}(x(s)) - {\nabla _\beta }\delta {g_{\alpha \gamma }}(x(s)) \big) \notag \\
=&n^{\alpha} \tau^{\beta} e^{\gamma}
\big( {\nabla _\alpha }\delta {g_{\beta \gamma }}(x(s)) - {\nabla _\beta }\delta {g_{\alpha \gamma }}(x(s)) \big),
\end{align}
which can be read out from the antisymmetry of
\be e^{\gamma} \big( {\nabla _\alpha }\delta {g_{\beta \gamma }}(x(s))
-\nabla_{\beta} \delta g_{\alpha \gamma} (x(s)) \big), \ee
in terms of the $\alpha$, $\beta$ indices.\footnote{For example, for a given anti-symmetric tensor $V_{\alpha\beta}$, we can directly check
\begin{align}
\tau^{(2)\alpha}n^{(2)\beta}V_{\alpha\beta}=&
(\cosh\zeta \tau^{(1)\alpha}-\sinh\zeta n^{(1)\alpha})(-\sinh \zeta \tau^{(1)\beta}+\cosh \zeta n^{(1)\beta})V_{\alpha\beta} \notag \\
=&\tau^{(1)\alpha}n^{(1)\beta}V_{\alpha\beta}.
\end{align}
}

\subsection{The system's evolution generated by the twist $\zeta$}

Having computed the variation of the twist $\zeta$ with respect to the variation of the metric (\ref{deltatwist}), we now study the system's evolution generated by the twist $\zeta$.

The study is parallel to the one of the geodesic's length $A$ in subsection \ref{geodesiclength}, and consists of the following steps. First, we make a choice of the boundary of the Cauchy surface $\partial \Sigma$ and take a non-physical diffeomorphism, such that the geodesic $\gamma$ is contained in the Cauchy surface $\Sigma$.
Second, we rewrite the variation of the twist (\ref{deltatwist}) in terms of the variation of the initial data. Third, by applying the variation of the twist to (\ref{initialevo}), (\ref{bracketO}), we get the system's evolution generated by the twist $\zeta$.

The first step is straightforward, so we directly go to the second step. We will rewrite the variation of the twist (\ref{initialevo}) in terms of the variation of the initial data in the following two paragraphs.

Here, the main issue is to deal with the integral in (\ref{deltatwist}).
We extract the integrand out as
\be\label{terms} \frac{1}{2} n^{\alpha} \tau^{\beta} e^{\gamma}
\big( \nabla_{\alpha} \delta g_{\beta\gamma}-\nabla_{\beta} \delta g_{\alpha\beta}), \ee
and deal with it now.
First, we restrict the expression (\ref{terms}) on the Cauchy surface $\Sigma$ and rewrite it as
\begin{align}\label{onCauchy}
&\frac{1}{2} n^{\alpha} \tau^{\beta} e^{\gamma}
\big( \nabla_{\alpha} \delta g_{\beta\gamma}-\nabla_{\beta} \delta g_{\alpha\beta})\bigg|_{\Sigma}  \notag \\
=&
\bigg[
-\frac{1}{2} e^{\alpha} n^{\beta} \tau^{\gamma} \nabla_{\gamma}
(\sigma_{\alpha}^{\phantom{\alpha}\mu} \sigma_{\beta}^{\phantom{\beta}\nu} \delta g_{\mu\nu})
-\frac{1}{2} n^{\alpha} K_{\alpha}^{\phantom{\alpha} \mu} e^{\nu} \delta g_{\mu\nu}
+\frac{1}{2} n^{\alpha} e^{\beta} D_{\alpha} (\tau^{\mu} \sigma_{\beta}^{\phantom{\beta}\nu} \delta g_{\mu\nu}) \notag \\
&+\frac{1}{2N} e^{\alpha}D_{\alpha} N n^{\mu}\tau^{\nu} \delta g_{\mu\nu}
+\frac{1}{2N} n^{\alpha} D_{\alpha} N e^{\mu} \tau^{\nu} \delta g_{\mu\nu}
-\frac{1}{2} e^{\alpha} n^{\beta}K_{\alpha\beta} \tau^{\mu} \tau^{\nu} \delta g_{\mu\nu} \bigg] \bigg|_{\Sigma} \notag \\
=& \bigg[ \frac{1}{2} e^{\mu} K_{\mu}^{\phantom{\mu}\alpha} n^{\beta} \delta \sigma_{\alpha\beta}
-e^{\alpha} n^{\beta} \delta K_{\alpha\beta}
-\frac{1}{2} e^{\alpha} n^{\beta} D_{\alpha} (\sigma_{\beta}^{\phantom{\beta}\mu} \tau^{\nu} \delta g_{\mu\nu}) \bigg] \bigg|_{\Sigma}
\end{align}
Here, we have extended the frame $(\hat{e},\hat{\tau},\hat{n})$ to the whole spacetime, such that, when restricted on the Cauchy surface $\Sigma$, the $\hat{\tau}$ coincides with the normal vector of the Cauchy surface $\Sigma$. In deriving these equations, we have taken an ADM decomposition and used (\ref{defKa}) in the first equation, and used (\ref{deltasigmaK}) in the second equation. Second, we further restrict (\ref{onCauchy}) on the geodesic $\gamma$ and rewrite it as
\begin{align}\label{ongeodesic}
&\frac{1}{2} n^{\alpha} \tau^{\beta} e^{\gamma}
\big( \nabla_{\alpha} \delta g_{\beta\gamma}-\nabla_{\beta} \delta g_{\alpha\beta})\bigg|_{\gamma}  \notag \\
=& \bigg[ \frac{1}{2} e^{\mu} K_{\mu}^{\phantom{\mu}\alpha} n^{\beta} \delta \sigma_{\alpha\beta}
-e^{\alpha} n^{\beta} \delta K_{\alpha\beta}
-\frac{1}{2} e^{\alpha} n^{\beta} D_{\alpha} (\sigma_{\beta}^{\phantom{\beta}\mu} \tau^{\nu} \delta g_{\mu\nu}) \bigg] \bigg|_{\gamma}
\notag \\
=&\bigg[
\frac{1}{2} e^{m} n^{n} K_{mn} n^{a}n^{b} \delta \sigma_{ab}-e^{a}n^{b} \delta K_{ab}
-\frac{1}{2} \frac{d}{ds}\big( \tau^{\mu}n^{\nu} \delta g_{\mu\nu}(x(s)) \big) \bigg] \bigg|_{\gamma}.
\end{align}
Here, we have used
\begin{align}\label{relationongeo} &e^{\nu} D_{\nu} e^{\mu} \big|_{\gamma}=0 \notag \\
&e^{\nu} D_{\nu} n^{\mu} \big|_{\gamma}=0 \notag \\
& K_{\mu\nu}e^{\mu} e^{\nu} \big|_{\gamma}=0, \end{align}
where the first and third equations of (\ref{relationongeo}) can be derived from an ADM decomposition of the geodesic equation
\be e^{\nu} \nabla_{\nu} e^{\mu}=0, \ee
and the second equation of (\ref{relationongeo}) can be checked component by component by taking use of the first equation of (\ref{relationongeo}) and the orthonormal conditions as
\begin{align}
&e^{\nu}D_{\nu} n^{\mu} n_{\mu}=\frac{1}{2} e^{\nu}D_{\nu}(n^{\mu} n_{\mu})=0 \notag \\
&e^{\nu}D_{\nu} n^{\mu} e_{\mu}=e^{\nu}D_{\nu}(n^{\mu} e_{\mu})-e^{\nu}D_{\nu} e^{\mu} n_{\mu}=0.
\end{align}

We rewrite the variation of the twist in terms of the variation of the initial data.
By applying (\ref{ongeodesic}) to (\ref{deltatwist}), we get
\begin{align}\label{deltainsertion}
\delta \zeta=&\int_{{s_1}}^{{s_2}} ds
\left( \frac{1}{2}e^{m} n^{n} K_{mn} n^a n^b \delta \sigma _{ab}
 - {e^a}{n^b}\delta {K_{ab}} \right) \notag \\
&+\left( \frac{1}{2} \delta g_{\alpha\beta}(x(s_1)) \tau^{(1)\alpha} n^{(1)\beta}
+C_{(\tau^{(1)})}^{n^{(1)}}(s_1)+\frac{1}{2} \delta g_{\alpha\beta}(x(s_1)) \tau^{\alpha} n^{\beta} \right) \notag \\
&-\left(
\frac{1}{2} \delta g_{\alpha\beta}(x(s_2)) \tau^{(2)\alpha} n^{(2)\beta}+C_{(\tau^{(2)})}^{n^{(2)}}(s_2)
+\frac{1}{2} \delta g_{\alpha\beta}(x(s_2)) \tau^{\alpha} n^{\beta} \right),
\end{align}
where the integral is already in terms of the variation of the initial data.
We now take the cutoff surface to the asymptotic boundary, where we also make an assumption that, under this limit, the contributions from the terms supporting at $s_1$, $s_2$ are ignorable.\footnote{We leave the proof of this assumption to the future study. However, in section \ref{covariant}, we will reproduce the system's evolution (\ref{evotwist}) with a different approach, which may be viewed as a support of the current assumption.}
Under this limit and based on this assumption, we then get the final expression of the variation of the twist with respect to the variation of the initial data
\begin{align}\label{deltazetafinal}
\delta \zeta=&\int_{\gamma}ds \left( \frac{1}{2} e^{m} n^{n} K_{mn} n^{a}n^{b} \delta\sigma_{ab}
- e^{a}n^{b} \delta K_{ab} \right)  \notag \\
=& \int_{\Sigma} d^2 x \sqrt{\sigma}
\left( \frac{1}{2} e^{m} n^{n} K_{mn} n^{a}n^{b} \delta\sigma_{ab} \delta(\rho)
- e^{a}n^{b} \delta K_{ab} \delta(\rho) \right).
\end{align}

Finally, by applying (\ref{deltazetafinal}) to (\ref{initialevo}), (\ref{bracketO}) and also by taking use of (\ref{Poisson2}), we get the system's evolution generated by the twist $\zeta$ as\footnote{Generally speaking, the system's evolution generated by a diffeomorphism invariant observable would preserve the constraints (\ref{constraints}). As a cross check of our computed system's evolution (\ref{evotwist}), we verify that it indeed preserves the constraints (\ref{constraintsequ}) in appendix \ref{evoconstraints}.}
\begin{align}\label{evotwist}
&\Delta \sigma_{ab}(t_0,x)=-8\pi G \lambda (e_an_b+n_ae_b) \delta(\rho) +o(\lambda) \notag \\
&\Delta K_{ab}(t_0,x)=-8\pi G \lambda e^m n^n K_{mn} n_an_b \delta(\rho) +o(\lambda).
\end{align}

\subsection{The geometric interpretation of the system's evolution generated by the twist $\zeta$: a relative shift along the geodesic}\label{twistill}

\begin{figure}[htbp]
  \centering
  \subfloat[The unevolved system]{\label{shiftm1}\includegraphics[width=6cm]{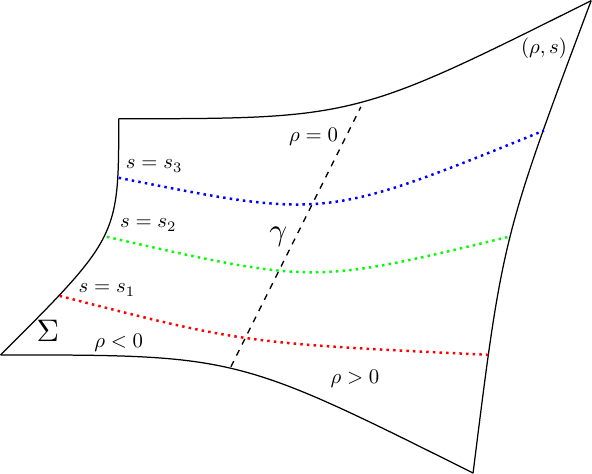}}
     \quad\quad
\subfloat[The evolved system]{\label{shiftm2}\includegraphics[width=6.4cm]{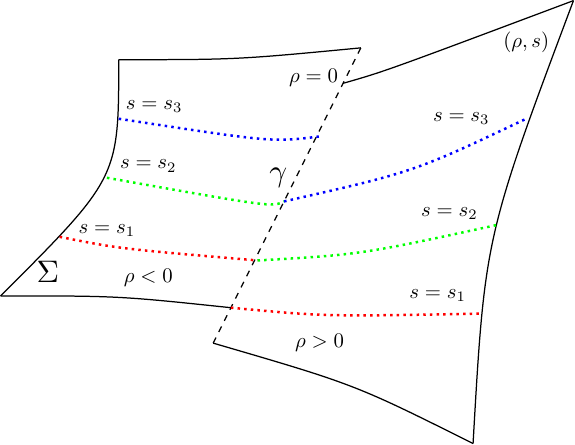}}
  \caption{The relative shift along the geodesic $\gamma$ generated by the twist. Fig.\ref{shiftm1} and Fig.\ref{shiftm2} represent the systems before and after the relative shift. As in figure \ref{kink}, these figures should be viewed in a pushforward perspective. Namely, in the evolution of the relative shift from Fig.\ref{shiftm1} to Fig.\ref{shiftm2}, we keep the three dimensional metric fixed and take a shift of one side of the Cauchy surface relative to the other side along the direction of the geodesic $\gamma$. (Here, the relative shift, though tangent to the geodesic $\gamma$, is not necessary parallel to the Cauchy surface $\Sigma$.) To manifest the relative shift, we introduce coordinates $(\rho,s)$ on the Cauchy surface $\Sigma$. Here, the dashed line denotes the geodesic $\gamma$ at $\rho=0$, and the dotted lines denote the curves with constant $s$ values. It is the evolution of the set of initial data in terms of the $(\rho,s)$ coordinates that equals to the system's evolution represented in (\ref{evotwist}). Moreover, in our mind, we should take an extra pullback on the evolved system in Fig.\ref{shiftm2} acting on both the metric and the Cauchy surface. The pullback maps the Cauchy surface $\Sigma$ back to the same location as the one in Fig.\ref{shiftm1} in terms of the three dimensional coordinates, maps the metric to be the evolved one, and keeps the set of initial data to be the one illustrated in Fig.\ref{shiftm2}. After acting this pullback on Fig.\ref{shiftm2}, the evolution of the set of initial data from Fig.\ref{shiftm1} to Fig.\ref{shiftm2} can indeed be interpreted as an evolution of the system.}\label{shiftgeo}
\end{figure}

We claim that the system's evolution (\ref{evotwist}) has a geometric interpretation, that is a relative shift along the geodesic $\gamma$ as in Fig.\ref{shiftgeo}.
Here, the relative shift is a diffeomorphism, which roughly speaking shifts the region in the one side of the geodesic relative to the region in the other side of the geodesic along the direction of the geodesic. And the system's evolution represented in (\ref{evotwist}) can be realized by this diffeomorphism.

The precise requirement of the diffeomorphism and a systematical discussion of the relative shift will be given in subsection \ref{seccovshift}.
In this subsection, we support this relative shift interpretation of the system's evolution (\ref{evotwist}) with a concrete example.
In this example, we directly give the system's evolution generated by the relative shift. And our goal here is to check its equivalence to the system's evolution represented in (\ref{evotwist}).

\begin{figure}
  \centering
  \includegraphics[width=7cm]{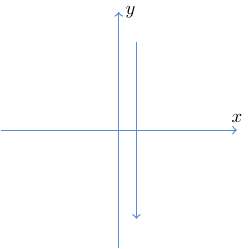}\\
  \caption{The relative shift along the $y$-axis in the flat metric. Here, we put the Cauchy surface at $t=0$. The diffeomorphism (\ref{flatshift}), when restricted on this Cauchy surface, indeed exhibits a relative shift along the $y$-axis for the two sides of the $y$-axis.}\label{flatshiftfig}
\end{figure}

We now introduce our example.
By focusing on the near geodesic region, we approximate the original system as a flat metric and the geodesic as a straight line.
In particular, we take the original system as the following flat metric
\be\label{3dgo} ds^{(o)2}=-dt^2+dx^2+dy^2, \ee
and we put the geodesic $\gamma$ at the $y$-axis.
We directly give the relative shift as the following diffeomorphism
\begin{align}\label{flatshift}
t\rightarrow &t \notag \\
x\rightarrow &x \notag \\
y\rightarrow &y-8\pi G \lambda \theta(x),
\end{align}
which, when restricted on a Cauchy surface containing the $y$-axis, indeed exhibits a relative shift along the $y$-axis, as in Fig.\ref{flatshiftfig}. By applying the diffeomorphism (\ref{flatshift}) to the metric of the original system (\ref{3dgo}), (which, to be more precise, is to take a pullback of the metric (\ref{3dgo}) by the diffeomorphism (\ref{flatshift}),) we get the metric of the evolved system as
\be\label{3dge} ds^{(e)2}=-dt^2+dx^2+dy^2-8\pi G \lambda \delta(x) (dxdy+dydx)+o(\lambda). \ee

Given the system's evolution generated by the relative shift from (\ref{3dgo}) to (\ref{3dge}), we can then check its equivalence to the system's evolution represented in (\ref{evotwist}). Here, we only need to focus on the evolution of the set of initial data on a Cauchy surface that contains the $y$-axis,\footnote{Remember that the system's evolution represented in (\ref{evotwist}) is only for the Cauchy surface that contains the geodesic.}
from which the evolution of the whole system is uniquely determined up to a non-physical diffeomorphism.
In the following, we will provide two choices of the Cauchy surface with zero and non-zero extrinsic curvature. And we will take the equivalence check respectively.

\subsubsection{The equivalence check on a Cauchy surface with zero extrinsic curvature}

We first put the Cauchy surface at $t=0$.
On such Cauchy surface, we can read out the set of initial data as
\begin{align}\label{in1o} &\sigma_{ab}^{(o)}d\bar{x}^ad\bar{x}^b=dx^2+dy^2, \notag \\
&K_{ab}^{(o)}=0,
\end{align}
for the original system (\ref{3dgo}), and
\begin{align}\label{in1e} &\sigma_{ab}^{(e)}d\bar{x}^a d\bar{x}^b=
dx^2+dy^2-8\pi G \lambda \delta(x) (dxdy+dydx)+o(\lambda)  \notag \\
&K_{ab}^{(e)}=0, \end{align}
for the evolved system (\ref{3dge}).
Here, we denote $\bar{x}^a=(x,y)$, and the vanishing of the extrinsic curvature is obtained from the time reversal symmetry of (\ref{3dgo}), (\ref{3dge}).
To compare with (\ref{evotwist}), we read out the $e_a$, $n_a$ for the original system from (\ref{in1o}) as
\begin{align}\label{eno}
&e_a=(0,1) \notag \\
&n_a=(1,0). \end{align}
And from (\ref{in1o}), (\ref{in1e}), (\ref{eno}), we can directly check that the system's evolution from (\ref{in1o}) to (\ref{in1e}) is precisely the system's evolution represented in (\ref{evotwist}).
This equivalence check therefore supports the relative shift interpretation of the system's evolution (\ref{evotwist}).

\subsubsection{The equivalence check on a Cauchy surface with non-zero extrinsic curvature}

We now take the equivalence check on a Cauchy surface with non-zero extrinsic curvature.
In particular, we consider the following Cauchy surface which contains the $y$-axis as
\footnote{Precisely speaking, the parametrization (\ref{rays}) only covers the near geodesic region of the Cauchy surface. Actually, for the $s$ with $f'(s)\neq 0$, the large $\rho$ region of (\ref{rays}) is even not spacelike, which can be seen from the expression of the induced metric (\ref{sKflato}).}
\footnote{See appendix \ref{Cauchycurvature} for the equivalence check on a general Cauchy surface containing the $y$-axis.}
\begin{align}\label{rays}
t=&\rho \sinh f(s) \notag \\
x=&\rho \cosh f(s) \notag \\
y=&s. \end{align}

We first compute the set of initial data for the original system (\ref{3dgo}).
Specifically, we compute the set of initial data with the following equations
\begin{align}\label{initialflat}
&\sigma_{ab}=\eta_{\mu\nu} \frac{\partial x^{\mu}}{\partial \bar{x}^a} \frac{\partial x^{\nu}}{\partial \bar{x}^b} \notag \\
&K_{ab}=\eta_{\mu\nu} \frac{\partial \tau^{\mu}}{\partial \bar{x}^a} \frac{\partial x^{\nu}}{\partial \bar{x}^b}.
\end{align}
Here, $x^{\mu}=(t,x,y)$ and $\bar{x}^a=(\rho,s)$. The first equation is by definition. And the second equation is derived as
\begin{align}
K_{ab}=&K_{\mu\nu} \frac{\partial x^{\mu}}{\partial \bar{x}^a} \frac{\partial x^{\nu}}{\partial \bar{x}^b} \notag \\
=& \sigma_{\mu}^{\phantom{\mu}\rho} \nabla_{\rho} \tau_{\nu}
\frac{\partial x^{\mu}}{\partial \bar{x}^a} \frac{\partial x^{\nu}}{\partial \bar{x}^b} \notag \\
=& \sigma_{\mu}^{\phantom{\mu}\rho} \partial_{\rho} \tau_{\nu}
\frac{\partial x^{\mu}}{\partial \bar{x}^a} \frac{\partial x^{\nu}}{\partial \bar{x}^b} \notag \\
=& \frac{\partial \tau_{\nu}}{\partial \bar{x}^a} \frac{\partial x^{\nu}}{\partial \bar{x}^b} \notag \\
=& \eta_{\mu\nu} \frac{\partial \tau^{\mu}}{\partial\bar{x}^a} \frac{\partial x^{\nu}}{\partial \bar{x}^b}.
\end{align}
We now compute the set of initial data with (\ref{initialflat}). We first compute the future-pointing normal vector $\tau^{\mu}$ of the Cauchy surface parameterized by (\ref{rays}). Here, the normal vector $\tau^{\mu}$ satisfies the following orthonormal conditions
\begin{align}
&\eta_{\mu\nu} \tau^{\mu} \frac{\partial x^{\nu}}{\partial \bar{x}^a}=0 \notag \\
&\eta_{\mu\nu} \tau^{\mu} \tau^{\nu}=-1,
\end{align}
and has the following expression
\be\label{normalo} \tau^{\mu}=\Big( \frac{\cosh f(s)}{[1-\rho^2 f'(s)^2]^{\frac{1}{2}}},
\frac{\sinh f(s)}{[1-\rho^2 f'(s)^2]^{\frac{1}{2}}},
\frac{\rho f'(s)}{[1-\rho^2 f'(s)^2]^{\frac{1}{2}}} \Big). \ee
By applying (\ref{rays}) and (\ref{normalo}) to (\ref{initialflat}), we then get the expression of the set of initial data for the original system as
\begin{align}\label{sKflato}
&\sigma_{\rho\rho}^{(o)}=1 \notag \\
&\sigma_{\rho s}^{(o)}=\sigma_{s \rho}^{(o)}=0 \notag \\
&\sigma_{ss}^{(o)}=1-\rho^2 f'(s)^2 \notag \\
&K_{\rho\rho}^{(o)}=0 \notag \\
&K_{\rho s}^{(o)}=K_{s \rho}^{(o)}=\frac{f'(s)}{[1-\rho^2 f'(s)^2]^{\frac{1}{2}}} \notag \\
&K_{ss}^{(o)}=\frac{\rho f''(s)}{[1-\rho^2 f'(s)^2]^{\frac{1}{2}}}.
\end{align}

We now compute the set of initial data for the evolved system (\ref{3dge}).
For convenience of the computation, we use a pushforward perspective to view the Cauchy surface embedded in the evolved system (\ref{3dge}).
The perspective is the following: instead of taking a pullback of the metric by the diffeomorphism (\ref{flatshift}), we keep the metric fixed and take a pushforward of the Cauchy surface by the diffeomorphism (\ref{flatshift}).
Under this perspective, we equivalently view the Cauchy surface embedded in the evolved system (\ref{3dge}) as the one parameterized by
\begin{align}\label{rayse}
t=&\rho \sinh f(s) \notag \\
x=&\rho \cosh f(s) \notag \\
y=&s-8\pi G \lambda \theta(\rho)+o(\lambda), \end{align}
and embedded in the flat metric (\ref{3dgo}).
We now compute the set of initial data for such Cauchy surface.
Following a similar computation in the previous paragraph, we first compute the future-pointing normal vector of the Cauchy surface (\ref{rayse}) as
\be\label{normale} \tau^{\mu}=\Big( \frac{\cosh f(s)}{[1-\rho^2 f'(s)^2]^{\frac{1}{2}}},
\frac{\sinh f(s)}{[1-\rho^2 f'(s)^2]^{\frac{1}{2}}},
\frac{\rho f'(s)}{[1-\rho^2 f'(s)^2]^{\frac{1}{2}}} \Big), \ee
which has the same expression as (\ref{normalo}).
And by applying (\ref{rayse}) and (\ref{normale}) to (\ref{initialflat}), we then get the set of initial data for the evolved system as
\begin{align}\label{sKflate}
&\sigma_{ss}^{(e)}=1-\rho^2 f'(s)^2+o(\lambda) \notag \\
&\sigma_{s\rho}^{(e)}=\sigma_{\rho s}^{(o)}=-8\pi G \lambda \delta(\rho)+o(\lambda) \notag \\
&\sigma_{\rho\rho}^{(e)}=1+o(\lambda) \notag \\
&K_{ss}^{(e)}=\frac{\rho f''(s)}{[1-\rho^2 f'(s)^2]^{\frac{1}{2}}}+o(\lambda) \notag \\
&K_{s\rho}^{(e)}=K_{\rho s}^{(e)}=\frac{f'(s)}{[1-\rho^2 f'(s)^2]^{\frac{1}{2}}}+o(\lambda) \notag \\
&K_{\rho\rho}^{(e)}=-8\pi G \lambda f'(s) \delta(\rho)+o(\lambda).
\end{align}

We now compare the evolution of the set of initial data from (\ref{sKflato}) to (\ref{sKflate}) with the one represented in (\ref{evotwist}).
To take the comparison, we read out the $e_a$, $n_a$ for the original system (\ref{sKflato}) as
\begin{align}\label{ennonflat}
&e_a=(0,\frac{1}{[1-\rho^2 f'(s)^2]^{\frac{1}{2}}}) \notag \\
&n_a=(1,0). \end{align}
From (\ref{sKflato}), (\ref{sKflate}), (\ref{ennonflat}), we then check that the system's evolution from (\ref{sKflato}) to (\ref{sKflate}) is precisely the system's evolution represented in (\ref{evotwist}).
The equivalence check again supports the relative shift interpretation of the system's evolution (\ref{evotwist}).

\section{The system's evolution generated by the twist with the covariant phase space formalism}\label{covariant}

So far, we have studied the system's evolution generated by the twist $\zeta$ with the canonical formalism.
In this section, we revisit the same problem with a different approach which is based on the covariant phase space formalism.

\subsection{An introduction of the approach}

We first introduce the approach, which is originally developed in \cite{Kaplan:2022orm} and is recast into the current form in \cite{ongoing2}.

The approach is based on the covariant phase space formalism \cite{Lee:1990nz,Iyer:1994ys,Wald:1999wa,Harlow:2019yfa}.
In the covariant phase space formalism, the phase space is viewed as the set of inequivalent solutions, the observables are reinterpreted as functions of the set of solutions, and the system's evolutions are represented as evolutions in the set of solutions.

\subsubsection{The main idea of the approach}

Based on the covariant phase space formalism, we now explain the main idea of the approach.

First, we represent the general solution under the Fefferman-Graham gauge as
\be\label{generalsol}
ds^2=\frac{dZ^2}{Z^2}-\frac{1}{Z^2} \Big(dU + \frac{6}{c}{Z^2}{T_{VV}}(V)dV \Big)
\Big(dV+\frac{6}{c}{Z^2}{T_{UU}}(U)dU \Big).
\ee
Here, we have taken use of the result in \cite{Banados:1998gg}.
$(Z,U,V)$ are a set of coordinates, where $Z>0$ is the radial coordinate and $U/V \in (-\infty,\infty)$ is the left/right moving null boundary coordinate. c is chosen to be
\be\label{cG}
c=\frac{3}{2G}. \ee
$T_{UU}$, $T_{VV}$ are the non-zero components of the boundary stress tensor.

Second, we reinterpret the observabes of our interest as functionals of $T_{UU}$, $T_{VV}$.
This is realized by evaluating the observables under the general solution (\ref{generalsol}).

Third, we compute the brackets of the observables from the brackets of $T_{UU}$, $T_{VV}$.
For example, given two observables $O_1$, $O_2$ which are already reinterpreted as functionals of $T_{UU}$, $T_{VV}$, their bracket can be computed by applying the chain rule as
\begin{align}\label{chainrule}
\{O_1,O_2\}=&\int d\widetilde{U}_1 d\widetilde{U}_2
\frac{\delta O_1} {\delta T_{UU}(\widetilde {U}_1)}
\frac{\delta O_2} {\delta T_{UU}(\widetilde {U}_2)}  \{T_{UU}(\widetilde{U}_1),T_{UU}(\widetilde{U}_2)\} \notag \\
&+\int d\widetilde {V}_1 d\widetilde{V}_2 \frac{\delta O_1}{\delta T_{VV}(\widetilde{V}_1)}
\frac{\delta O_2}{\delta T_{VV}(\widetilde{V}_2)} \{T_{VV}(\widetilde{V}_1),T_{VV}(\widetilde{V}_2)\}.
\end{align}
Here, the brackets of the boundary stress tensor have the following expressions
\begin{align}\label{TTbracket}
&\{ {T_{UU}}(U),{T_{UU}}(\widetilde U)\}
=-2\pi \Big[\frac{c}{12}\delta'''(U-\widetilde{U})
+2T_{UU}(U) \delta'(U-\widetilde{U})+T'_{UU}(U)\delta(U-\widetilde{U}) \Big] \notag \\
&\{T_{VV}(V),T_{VV}(\widetilde{V})\}
=-2\pi \Big[ \frac{c}{12} \delta'''(V-\widetilde{V})
+2T_{VV}(V)\delta'(V-\widetilde{V})+T'_{VV}(V)\delta(V-\widetilde{V}) \Big] \notag \\
&\{T_{UU}(U),T_{VV}(V)\}=0,
\end{align}
which can be read out from the results of the asymptotic symmetry analysis \cite{Brown:1986nw,Barnich:2001jy,Compere:2007az,Compere:2018aar}, and which are also derived in appendix \ref{deTTbracket} for completeness.

Fourth, we represent the system's evolution as the evolution of the general solution (\ref{generalsol}).
Precisely speaking, we apply the expression of the general solution (\ref{generalsol}) to the evolution equation (\ref{deltaW}) in the position of $W$.
And the system's evolution generated by a given observable $O$ is represented as
\be\label{Deltatildeg} \widetilde{\Delta} g_{\mu\nu}(X)=\lambda \{g_{\mu\nu}(X),O\}+o(\lambda). \ee
Here, $X^{\mu}=(Z,U,V)$. $g_{\mu\nu}(X)$ takes the expression of the general solution (\ref{generalsol}). The observable $O$ is reinterpreted as a functional of $T_{UU}$, $T_{VV}$. The bracket $\{g_{\mu\nu}(X),O\}$ can be computed by applying the chain rule (\ref{chainrule}).
The reason to use the symbol $\widetilde{\Delta}$ instead of $\Delta$ is to leave the symbol $\Delta$ for a transformed quantity below.

\subsubsection{A reformulated version of the approach}\label{approach}

So far, we have introduced the main idea of the approach. But, for the practical application, we take a further reformulation to simplify the computation and to manifest the geometric interpretation. We now introduce the reformulated version of the approach.

First, we reformulate the representation of the general solution (\ref{generalsol}) to a form appearing in \cite{Roberts:2012aq}, which we refer to as the pullback representation.
The pullback representation is to represent the general solution (\ref{generalsol}) as a pullback of the vacuum solution
\be\label{vacuum}
ds^2=\frac{dz^2-dudv}{z^2},
\ee
by the following diffeomorphism
\begin{align}\label{transform}
&z=Z\frac{4\big(u'_{(b)}(U)v'_{(b)}(V) \big)^{\frac{3}{2}}} {4u'_{(b)}(U)v'_{(b)}(V)-Z^2 u''_{(b)}(U) v''_{(b)}(V)} \notag \\
&u=u_{(b)}(U)+\frac{2Z^2 u'_{(b)}(U)^2 v''_{(b)}(V)} {4u'_{(b)}(U) v'_{(b)}(V)-Z^2 u''_{(b)}(U) v''_{(b)}(V)} \notag \\
&v=v_{(b)}(V)+\frac{2Z^2 u''_{(b)}(U) v'_{(b)}(V)^2} {4u'_{(b)}(U)v'_{(b)}(V)-Z^2u''_{(b)}(U)v''_{(b)}(V)}.
\end{align}
Here, we represent the vacuum solution in the $(z,u,v)$ coordinates and the general solution in the $(Z,U,V)$ coordinates.
Note that the functions $u_{(b)}$, $v_{(b)}$ in (\ref{transform}) should also be viewed as functionals of $T_{UU}$, $T_{VV}$, which are solved from
\begin{align}\label{Ttouv}
&T_{UU}(U)=\frac{c}{12} \bigg(\frac{u'''_{(b)}}{u'_{(b)}}-\frac{3}{2} \Big(\frac{{u''_{(b)}}}{{u'_{(b)}}}\Big)^2 \bigg) \notag \\
&T_{VV}(V)=\frac{c}{12} \bigg(\frac{v'''_{(b)}}{v'_{(b)}}-\frac{3}{2} \Big(\frac{{v''_{(b)}}}{{v'_{(b)}}} \Big)^2 \bigg),
\end{align}
and
\begin{align}\label{monotonic}
&u'_{(b)}(U)>0\nonumber\\
&v'_{(b)}(V)>0.
\end{align}
We assume that such $u_{(b)}$, $v_{(b)}$ always exist for the physical solutions. And we also point out that the choice of such $u_{(b)}$, $v_{(b)}$ has the following arbitrariness
\begin{align}\label{SL2}
&u_{(b)}(U)\rightarrow \frac{a_u u_{(b)}(U)+b_{u}}{c_u u_{(b)}(U)+d_{u}}
\notag \\
&v_{(b)}(V)\rightarrow \frac{a_v v_{(b)}(V)+b_{v}}{c_v v_{(b)}(V)+d_v},
\end{align}
where
\begin{align}
&a_ud_u-b_uc_u=1 \notag \\
&a_vd_v-b_vc_v=1. \end{align}
The pullback representation introduced here can be directly verified by taking a pullback of the vacuum solution (\ref{vacuum}) by the diffeomorphism (\ref{transform}) which gives precisely the general solution (\ref{generalsol}) after taking use of (\ref{Ttouv}).

Second, we express the observables of our interest as functionals of $u_{(b)}$, $v_{(b)}$.
This is realized by taking a pushforward of the observables by the diffeomorphism (\ref{transform}) to the vacuum solution (\ref{vacuum}) and computing their expressions there.
The observables of our interest usually have explicit expressions with respect to $u_{(b)}$, $v_{(b)}$.
But note that, as mentioned in the previous paragraph, the $u_{(b)}$, $v_{(b)}$ are in turn functionals of $T_{UU}$, $T_{VV}$.

Third, we compute the brackets of the observables by applying the chain rule recursively. Here, the chain rule still ends up at the brackets of the boundary stress tensor (\ref{TTbracket}). For example, given an observable $O$ which is already expressed with respect to $u_{(b)}$, $v_{(b)}$, the brackets with the observable $O$ can be computed as
\be\label{chainuvT}
\{\cdot,O\}=\int dU
\frac{\delta O}{\delta u_{(b)}(U)} \{\cdot,u_{(b)}(U)\}
+\int dV
\frac{\delta O}{\delta v_{(b)}(V)} \{\cdot,v_{(b)}(V)\},
\ee
together with
\begin{align}\label{uvbracket}
&\{\cdot,u_{(b)}(U)\}=\int d \widetilde{U} \frac{\delta u_{(b)}(U)}{\delta T_{UU}(\widetilde{U})}
\{\cdot, T_{UU}(\widetilde{U}) \} \notag \\
&\{\cdot,v_{(b)}(V)\}=\int d \widetilde{V} \frac{\delta v_{(b)}(V)}{\delta T_{VV}(\widetilde{V})}
\{\cdot, T_{VV}(\widetilde{V}) \}.
\end{align}
Here, the brackets $\{\cdot,T_{UU}(\widetilde{U})\}$, $\{\cdot,T_{UU}(\widetilde{V})\}$ can be computed by applying another chain rule.
$\frac{\delta u_{(b)}(U)}{\delta T_{UU}(\widetilde{U})}$, $\frac{\delta v_{(b)}(V)}{\delta T_{VV}(\widetilde{V})}$ are computed in the followings.
By taking a variation of (\ref{Ttouv}), we get the differential equations for $\delta u_{(b)}$, $\delta v_{(b)}$ as
\begin{align}\label{varTtouv}
&u'_{(b)}(U) \frac{\partial}{\partial U} \Big( \frac{1}{u'_{(b)}(U)}
\frac{\partial}{\partial U} \big( \frac{1}{u'_{(b)}(U)} \frac{\partial}{\partial U} \delta {u_{(b)}}(U) \big) \Big)
=\frac{12}{c} \delta T_{UU}(U) \notag \\
&v'_{(b)}(V) \frac{\partial}{\partial V} \Big( \frac{1}{v'_{(b)}(V)} \frac{\partial}{\partial V}
\big( \frac{1}{v'_{(b)}(V)} \frac{\partial}{\partial V} \delta v_{(b)}(V) \big) \Big)
=\frac{12}{c}\delta T_{VV}(V).
\end{align}
And by solving these differential equations (\ref{varTtouv}), we get
\begin{align}\label{varuv}
\delta u_{(b)}(U)=&-\int_U^{U_0} d{\widetilde U}_3 u'_{(b)}(\widetilde{U}_3)
\int_{\widetilde{U}_3}^{U_0} d\widetilde{U}_2 u'_{(b)}(\widetilde{U}_2)
\int_{\widetilde{U}_2}^{U_0} d\widetilde{U}_1 \frac{1}{u'_{(b)}(\widetilde{U}_1)}
\frac{12}{c} \delta T_{UU}(\widetilde{U}_1) \notag\\
&+a^{u,1}[u_{(b)},\delta T_{UU}]u_{(b)}(U)^2+a^{u,0}[u_{(b)},\delta T_{UU}]u_{(b)}(U)+a^{u, - 1}[u_{(b)},\delta T_{UU}]  \notag \\
\delta v_{(b)}(V)=&\int_{V_0}^V d\widetilde{V}_3 v'_{(b)}(\widetilde{V}_3)
\int_{V_0}^{\widetilde{V}_3} d\widetilde{V}_2 v'_{(b)}(\widetilde{V}_2)
\int_{V_0}^{\widetilde{V}_2} d\widetilde{V}_1 \frac{1}{v'_{(b)}(\widetilde{V}_1)}
\frac{12}{c} \delta T_{VV}(\widetilde{V}_1) \notag \\
&+a^{v,1}[v_{(b)},\delta T_{VV}]v_{(b)}(V)^2+a^{v,0}[v_{(b)},\delta T_{VV}] v_{(b)}(V) + a^{v,-1}[v_{(b)},\delta T_{VV}],
\end{align}
which are the integral form of $\frac{\delta u_{(b)}(U)}{\delta T_{UU}(\widetilde{U})}$, $\frac{\delta v_{(b)}(V)}{\delta T_{VV}(\widetilde{V})}$.
Here, $a^{u,i}$/$a^{v,i}$, with $i=-1,0,1$, are arbitrary functionals of $u_{b}$/$v_{(b)}$ and $\delta T_{UU}$/$\delta T_{VV}$, whose dependence on $\delta T_{UU}$/$\delta T_{VV}$ is linear; the arbitrariness of $a^{u,i}$, $a^{v,i}$ is the linearized version of the arbitrariness in (\ref{SL2}). ($U_0,V_0$) is a reference point, whose dependence can be absorbed into $a^{u,i}$, $a^{v,i}$. The integral in the form of $\int_a^b$ is interpreted as
\be
\int_a^b\equiv\left\{
\begin{array}{lcc}
\int_a^b & \mbox{for} & a<b \\
-\int_b^a & \mbox{for} & b<a
    \end{array} \right. ;
\ee
the reason to write the integral as if $U_0>U$, $V_0<V$ is because we will put the reference point $(U_0,V_0)$ to the left of all other relevant points in the discussion below.
The chain rule (\ref{chainuvT}), (\ref{uvbracket}) together with the expression (\ref{varuv}) is sufficient to compute the brackets of the observables.
While in practice, for a given observable $O$, we are most interested in the following two brackets
\be\label{bracketuvO} \{u_{(b)}(U),O\}~~~\{v_{(b)}(V),O\}, \ee
from which we can read out all of the other brackets with the observable $O$. For example, given another observable $W$ which is already expressed with respect to $u_{(b)}$, $v_{(b)}$, the bracket $\{W,O\}$ can be computed by applying the chain rule as
\be\label{bracketWO} \{W,O\}=\int dU \frac{\delta W}{\delta u_{(b)}(U)}\{u_{(b)}(U),O\}
+\int dV \frac{\delta W}{\delta v_{(b)}(V)} \{v_{(b)}(V),O\}.
\ee

Fourth, we represent the system's evolution as an infinitesimal diffeomorphism.
This can be derived by expressing the general solution with the pullback representation and applying the expression to the system's evolution equation (\ref{Deltatildeg}).
Precisely speaking, based on the pullback representation, we can represent the general solution in an abstract way as
\be\label{genpullback} g_{\mu\nu}(X)=g^{(0)}_{\alpha\beta}(x(X))
\frac{\partial x^{\alpha}}{\partial X^{\mu}} \frac{\partial x^{\beta}}{\partial X^{\nu}}, \ee
where we denote $X^{\mu}=(Z,U,V)$, $x^{\mu}=(z,u,v)$, $x(X)$ as the diffeomorphism (\ref{transform}), $g_{\mu\nu}(X)$ as the general solution, and $g^{(0)}_{\alpha\beta}(x)$ as the vacuum solution.
By applying (\ref{genpullback}) to (\ref{Deltatildeg}), we get the system's evolution generated by the observable $O$ as
\be\label{Deltagdiff} \widetilde{\Delta} g_{\mu\nu}={\cal{L}}_{\widetilde{\xi}} g_{\mu\nu}+o(\lambda),
\ee
where
\be\label{xitilde} \widetilde{\xi}^{\mu}=\frac{\partial X^{\mu}}{\partial x^{\alpha}} \cdot \Delta x^{\alpha}(X), \ee
and
\be\label{Deltax} \Delta x^{\alpha}(X)=\lambda \{x^{\alpha}(X),O\}. \ee
Here, we have ignored the $o(\lambda)$ terms in (\ref{xitilde}), (\ref{Deltax}) to simplify the convention, but we still keep the $o(\lambda)$ term in (\ref{Deltagdiff}).
The bracket in (\ref{Deltax}) can be computed by applying $x^{\alpha}(X)$ to (\ref{bracketWO}) in the position of $W$.
(Note that the diffeomorphism $x(X)$ represented in (\ref{transform}) is already expressed with respect to $u_{(b)}$, $v_{(b)}$.)
With some computation, we get the expression of $\widetilde{\xi}^{\mu}$ as
\begin{align}\label{xiFG}
\widetilde{\xi}^Z=&\frac{1}{2} Z \big(\eta^{U}{}'(U)+\eta^{V}{}'(V)\big) \notag \\
\widetilde{\xi}^U=& \eta^U(U)+\frac{\frac{1}{2}Z^2}{1-\frac{36}{c^2}Z^4 T_{UU}(U)T_{VV}(V)} \eta^{V}{}''(V)
-\frac{\frac{3}{c}Z^4T_{VV}(V)}{1-\frac{36}{c^2}Z^4T_{UU}(U)T_{VV}(V)} \eta^{U}{}''(U) \notag \\
\widetilde{\xi}^V=& \eta^V(V)+\frac{\frac{1}{2}Z^2}{1-\frac{36}{c^2}Z^4 T_{UU}(U) T_{VV}(V)} \eta^{U}{}''(U)
-\frac{\frac{3}{c}Z^4T_{UU}(U)}{1-\frac{36}{c^2}Z^4T_{UU}(U)T_{VV}(V)} \eta^{V}{}''(V),
\end{align}
where
\begin{align}\label{etaUV}
&\eta^U(U)=\frac{\Delta u_{(b)}(U)}{u_{(b)}'(U)} \notag \\
&\eta^V(V)=\frac{\Delta v_{(b)}(V)}{v_{(b)}'(V)},
\end{align}
and
\begin{align}\label{Deltauv}
&\Delta u_{(b)}(U)=\lambda \{u_{(b)}(U),O\}
\notag \\
&\Delta v_{(b)}(V)=\lambda \{v_{(b)}(V),O\}.
\end{align}
The infinitesimal diffeomorphism (\ref{Deltagdiff}) already describes the system's evolution.
But for some reason that will be clear below, we would like to allow an extra non-physical diffeomorphism and represent the system's evolution as
\be\label{nonphysical} \Delta g_{\mu\nu}=\widetilde{\Delta} g_{\mu\nu}+{\cal{L}}_{\Delta \xi}g_{\mu\nu}+o(\lambda), \ee
where the non-physical diffeomorphism parameter $\Delta \xi^{\mu}$ has the following asymptotic behavior\footnote{The asymptotic behavior of the non-physical diffeomorphism (\ref{gaugediff}) is under the asymptotic boundary conditions (\ref{asy}).}
\begin{align}\label{gaugediff}
&\Delta \xi^Z={\cal{O}}(Z^3) \notag \\
&\Delta \xi^U={\cal{O}}(Z^2) \notag \\
&\Delta \xi^V={\cal{O}}(Z^2).
\end{align}
By applying (\ref{Deltagdiff}) to (\ref{nonphysical}) and combining the two infinitesimal diffeomorphism, we get the final expression of the system's evolution as
\be\label{Deltagf} \Delta g_{\mu\nu}={\cal{L}}_{\xi} g_{\mu\nu}+o(\lambda), \ee
which again is an infinitesimal diffeomorphism whose diffeomorphism parameter $\xi^{\mu}$ can take any expression with the following asymptotic behavior 
\footnote{$\widetilde{\Delta} g_{\mu\nu}$ in (\ref{Deltagdiff}) and $\Delta g_{\mu\nu}$ in (\ref{Deltagf}) are both infinitesimal diffeomorphism. The only difference is on their diffeomorphism parameters $\widetilde{\xi}^{\mu}$ and $\xi^{\mu}$. The $\widetilde{\xi}^{\mu}$ for $\widetilde{\Delta} g_{\mu\nu}$ can only take the expression in (\ref{xiFG}) which preserves the Fefferman-Graham gauge; while the $\xi^{\mu}$ for $\Delta g_{\mu\nu}$ can take any expression satisfying the asymptotic behavior (\ref{xiZUV}).}
\begin{align}\label{xiZUV}
\xi^Z=&\frac{1}{2}Z \big(\eta^{U}{}'(U)+\eta^{V}{}'(V)\big)+{\cal{O}}(Z^3) \notag \\
\xi^U=&\eta^{U}(U)+{\cal{O}}(Z^2) \notag \\
\xi^V=&\eta^{V}(V)+{\cal{O}}(Z^2).
\end{align}
In the practical application, we would like to represent the diffeomorphism parameter $\xi^{\mu}$ in the $(z,u,v)$ coordinates, after a pushforward by the diffeomorphism (\ref{transform}).
Here, the diffeomorphism parameter $\xi^{\mu}$ in the $(z,u,v)$ coordinates has the following asymptotic behavior
\begin{align}\label{xifinalzuv}
\xi^z=&\frac{1}{2}z \big(\eta^{u}{}'(u)+\eta^{v}{}'(v)\big)+{\cal{O}}(z^3) \notag \\
\xi^u=&\eta^{u}(u)+{\cal{O}}(z^2) \notag \\
\xi^v=&\eta^{v}(v)+{\cal{O}}(z^2),
\end{align}
with
\begin{align}\label{etafinaluv}
&\eta^u(u)=\Delta u_{(b)} \circ u_{(b)}^{(-1)}(u) \notag \\
&\eta^v(v)=\Delta v_{(b)} \circ v_{(b)}^{(-1)}(v).
\end{align}

\subsubsection{Some useful equations}

So far, we have introduced the reformulated version of the approach. We now provide some useful equations for the application of the approach below.

First, we provide the following integral kernels
\begin{align}\label{kernel}
K^U(U,\widetilde{U}) &\equiv \int_{\widetilde{U}}^{U_0} d\widetilde{U}_3 u'_{(b)}(\widetilde{U}_3)
\int_{\widetilde{U}_3}^{U_0} d\widetilde{U}_2 u'_{(b)}(\widetilde{U}_2)
\int_{\widetilde{U}_2}^{U_0} d\widetilde{U}_1 \frac{1}{u'_{(b)}(\widetilde{U}_1)}
\delta(U-\widetilde{U}_1) \notag \\
&=\frac{1}{2u'_{(b)}(U)} \big(u_{(b)}(U)-u_{(b)}(\widetilde{U}) \big)^2
\big(\theta(U-\widetilde{U})-\theta(U-U_{0}) \big) \notag \\
K^V(V,\widetilde{V}) &\equiv  -\int_{V_0}^{\widetilde{V}} d\widetilde{V}_3 v'_{(b)}(\widetilde{V}_3)
\int_{V_0}^{\widetilde{V}_3} d\widetilde{V}_2 v'_{(b)}(\widetilde{V}_2)
\int_{V_0}^{\widetilde{V}_2} d\widetilde{V}_1 \frac{1}{v'_{(b)}(\widetilde{V}_1)}
\delta(V-\widetilde{V}_1) \notag \\
&=\frac{1}{2v'_{(b)}(V)} \big(v_{(b)}(V)-v_{(b)}(\widetilde{V}) \big)^2
\big(\theta(V-\widetilde{V})-\theta(V-V_{0}) \big).
\end{align}
Here, the integrals are extracted from the integral terms of (\ref{varuv}). And these integral kernels are used to compute the brackets with $u_{(b)}$, $v_{(b)}$ from the brackets with $T_{UU}$, $T_{VV}$ of the same observable.

Second, we compute the brackets $\{T_{UU}(U),u_{(b)}(\widetilde{U})\}$, $\{T_{VV}(V),v_{(b)}(\widetilde{V})\}$.
Specifically, by applying $T_{UU}$, $T_{VV}$ to (\ref{uvbracket}) and taking use of (\ref{TTbracket}), (\ref{kernel}), (\ref{Ttouv}), we get
\begin{align}\label{bracketTu}
&\{T_{UU}(U),u_{(b)}(\widetilde{U})\} \notag \\
=& 2\pi \left(\frac{\partial^3}{\partial U^3}+\frac{24}{c}T_{UU}(U)\frac{\partial}{\partial U}
+\frac{12}{c} T'_{UU}(U) \right) K^U(U,\widetilde{U}) \notag \\
&+\bar{a}^{u,1}(U)u_{(b)}(\widetilde{U})^2+\bar{a}^{u,0}(U)u_{(b)}(\widetilde{U})+\bar{a}^{u,-1}(U) \notag \\
=&2\pi u'_{(b)}(\widetilde{U}) \delta(U-\widetilde{U})+\widetilde{a}^{u,1}(U)u_{(b)}(\widetilde{U})^2
+\widetilde{a}^{u,0}(U)u_{(b)}(\widetilde{U})+\widetilde{a}^{u,-1}(U),
\end{align}
and
\begin{align}\label{bracketTv}
&\{T_{VV}(V),v_{(b)}(\widetilde{V})\} \notag \\
=& 2\pi\left(\frac{\partial^3}{\partial V^3}+\frac{24}{c}T_{VV}(V)\frac{\partial}{\partial V}
+\frac{12}{c}T'_{VV}(V) \right) K^V(V,\widetilde{V}) \notag \\
&+\bar{a}^{v,1}(V)v_{(b)}(\widetilde{V})^2+\bar{a}^{v,0}(V)v_{(b)}(\widetilde{V})+\bar{a}^{v,-1}(V) \notag \\
=&2\pi v'_{(b)}(\widetilde{V})\delta(V-\widetilde{V})+\widetilde{a}^{v,1}(V)v_{(b)}(\widetilde{V})^2
+\widetilde{a}^{v,0}(V)v_{(b)}(\widetilde{V})+\widetilde{a}^{v,-1}(V),
\end{align}
where
\begin{align}
&\bar{a}^{u,i}(U) \equiv a^{u,i}[u_{(b)},\delta T_{UU}] \big|_{\delta T_{UU}(U^*) \rightarrow \{T_{UU}(U),T_{UU}(U^*)\} } \notag \\
&\bar{a}^{v,i}(V) \equiv a^{v,i}[v_{(b)},\delta T_{VV}] \big|_{\delta T_{VV}(V^*) \rightarrow \{T_{VV}(V),T_{VV}(V^*)\} },
\end{align}
for $i=-1,0,1$, and $\widetilde{a}^{u,i}(U)/\widetilde{a}^{v,i}(V)$, with $i=-1,0,1$, are some $\widetilde{U}/\widetilde{V}$ independent quantities whose explicit expressions are not important.

\subsection{The expression of the twist}\label{kinematic}

Having introduced the approach, we now use it to study the system's evolution generated by the twist $\zeta$.
We divide the study into several subsections.
In this subsection, we compute the expression of the twist $\zeta$ with respect to $u_{(b)}$, $v_{(b)}$.

\begin{figure}
\centering
\subfloat[The geodesic and the IR cutoff surface in the $(Z,U,V)$ coordinates]{\label{cutofffigZUV}\includegraphics[width=7cm]{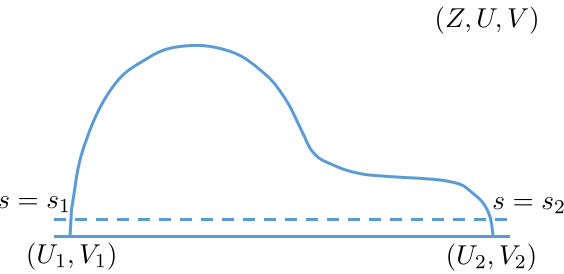}}
     \quad\quad\quad
\subfloat[The pushforward of the geodesic and the IR cutoff surface in the $(z,u,v)$ coordinates]{\label{cutofffigzuv}\includegraphics[width=7cm]{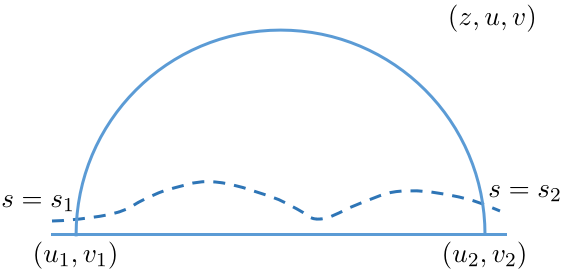}}
\caption{The geodesic and the IR cutoff surface in the $(Z,U,V)$ coordinates and in the $(z,u,v)$ coordinates. In both of the two figures, the solid curve denotes the geodesic and the dashed line/curve denotes the IR cutoff surface.}
\end{figure}

We first introduce the setup and also clarify a subtlety in the definition of the twist $\zeta$.
Here, the setup and the definition are all provided in the $(Z,U,V)$ coordinates where the general solution (\ref{generalsol}) is represented.
As in Fig.\ref{cutofffigZUV}, we consider a spacelike geodesic $\gamma$ with two endpoints $(U_1,V_1)$, $(U_2,V_2)$ on the asymptotic boundary $Z=0$, where we assume that $(U_1,V_1)$ is to the left of $(U_2,V_2)$ such that
\begin{align}
&U_1>U_2 \notag \\
&V_1<V_2.
\end{align}
For convenience of the following discussion, we equivalently represent the normal frames $(\hat{\tau}^{(1)},\hat{n}^{(1)})$, $(\hat{\tau}^{(2)},\hat{n}^{(2)})$ as $(\hat{l}^{(1)},\hat{r}^{(1)})$, $(\hat{l}^{(2)},\hat{r}^{(2)})$, respectively, which are defined as
\begin{align}
&\hat{l}^{(i)}=\hat{\tau}^{(i)}-\hat{n}^{(i)} \notag \\
&\hat{r}^{(i)}=\hat{\tau}^{(i)}+\hat{n}^{(i)},
\end{align}
for $i=1,2$, where $\hat{l}^{(i)}$/$\hat{r}^{(i)}$ is left/right future pointing null vectors from the viewpoint of an observer facing along the geodesic $\gamma$.
With the new representation of the normal frames $(\hat{l}^{(1)},\hat{r}^{(1)})$, $(\hat{l}^{(2)},\hat{r}^{(2)})$, we rewrite the parallel transport conditions (\ref{transport}) as
\be\label{parallellr} e^{\mu}\nabla_{\mu} l^{(i)\nu}\big|_{\gamma}=e^{\mu} \nabla_{\mu} r^{(i)\nu}\big|_{\gamma}=0, \ee
for $i=1,2$, the orthonormal conditions (\ref{orthonormal}) as
\begin{align}
&\hat{e}\cdot \hat{l}^{(i)} \big|_{\gamma}=\hat{e} \cdot \hat{r}^{(i)} \big|_{\gamma}=0 \notag \\
&\hat{l}^{(i)}\cdot \hat{l}^{(i)} \big|_{\gamma}=\hat{r}^{(i)}\cdot \hat{r}^{(i)} \big|_{\gamma}=0 \notag \\
&\hat{l}^{(i)}\cdot \hat{r}^{(i)} \big|_{\gamma}=-2,
\end{align}
for $i=1,2$, and the definition of the twist $\zeta$ (\ref{boost}) as
\begin{align}\label{lr21}
&\hat{l}^{(2)} \big|_{\gamma}=e^{\zeta} \hat{l}^{(1)} \big|_{\gamma} \notag \\
&\hat{r}^{(2)} \big|_{\gamma}=e^{-\zeta} \hat{r}^{(1)} \big|_{\gamma}.
\end{align}
As mentioned in section \ref{definition}, to seriously define the twist $\zeta$, we need to specify two points close to the respective geodesic endpoints and also adopt a proper value for the normal frames there.
Here, we introduce a cutoff surface at
\be\label{cutoffZUV} Z=\epsilon, \ee
as illustrated in Fig.\ref{cutofffigZUV}, where $\epsilon$ will be taken to zero at the end of the computation.
The cutoff surface intersects with the geodesic $\gamma$ at two points, denoted by their corresponding affine parameters of the geodesic as $s=s_1$, $s=s_2$.
These two intersections are the positions where we adopt the proper value for the normal frames by inheriting the frame from the asymptotic boundary.
Specifically, we adopt the following value for the normal frames at the two intersections $s=s_1$, $s=s_2$ as
\begin{align}\label{bframe}
&\hat{l}^{(1)} \big|_{s=s_1}={\cal{O}}(\epsilon^2)\frac{\partial }{\partial Z}
+\big(2\epsilon+{\cal{O}}(\epsilon^3) \big) \frac{\partial}{\partial U}+{\cal{O}}(\epsilon^3)\frac{\partial}{\partial V} \notag \\
&\hat{r}^{(1)} \big|_{s=s_1}={\cal{O}}(\epsilon^2)\frac{\partial}{\partial Z}
+{\cal{O}}(\epsilon^3)\frac{\partial}{\partial U}+\big(2\epsilon+{\cal{O}}(\epsilon^3) \big)\frac{\partial}{\partial V} \notag \\
&\hat{l}^{(2)} \big|_{s=s_2}={\cal{O}}(\epsilon^2)\frac{\partial}{\partial Z}
+{\cal{O}}(\epsilon^3)\frac{\partial}{\partial U}
+\big(2\epsilon+{\cal{O}}(\epsilon^3) \big) \frac{\partial}{\partial V} \notag \\
&\hat{r}^{(2)} \big|_{s=s_2}={\cal{O}}(\epsilon^2) \frac{\partial}{\partial Z}
+\big( 2\epsilon+{\cal{O}}(\epsilon^3) \big) \frac{\partial}{\partial U}
+{\cal{O}}(\epsilon^3)\frac{\partial}{\partial V}.
\end{align}

So far, we have clarified the definition of the twist for the general solution (\ref{generalsol}) in the $(Z,U,V)$ coordinates. And we now compute its expression by taking a pushforward of the whole kinematics by the diffeomorphism (\ref{transform}) to the vacuum solution (\ref{vacuum}) in the $(z,u,v)$ coordinates.

First, we study the geodesic $\gamma$, the cutoff surface, and their intersections in the $(z,u,v)$ coordinates, see Fig.\ref{cutofffigzuv} for an illustration.
We take a pushforward of the geodesic $\gamma$ by the diffeomorphism (\ref{transform}) to the $(z,u,v)$ coordinates, under which the geodesic endpoints $(U_1,V_1)$, $(U_2,V_2)$ are mapped to $(u_1,v_1)$, $(u_2,v_2)$ on the asymptotic boundary of the $(z,u,v)$ coordinates where
\begin{align}
&u_i=u_{(b)}(U_i) \notag \\
&v_i=v_{(b)}(V_i),
\end{align}
for $i=1,2$.
The geodesic $\gamma$ can therefore be expressed as
\begin{align}\label{geodesiczuv}
&z(s)=\frac{1}{2} \sqrt{(u_1-u_2)(v_2-v_1)} \frac{1}{\cosh s} \notag \\
&u(s)=\frac{u_2-u_1}{2} \frac{\sinh s}{\cosh s}+\frac{u_1+u_2}{2} \notag \\
&v(s)=\frac{v_2-v_1}{2} \frac{\sinh s}{\cosh s}+\frac{v_1+v_2}{2},
\end{align}
where $s$ is the proper length up to a shift.
We also take a pushforward of the cutoff surface (\ref{cutoffZUV}) to the $(z,u,v)$ coordinates, which can be represented as
\be\label{cutoffzuv} z=\epsilon \big( u_{(b)}'\circ u_{(b)}^{(-1)}(u) \big)^{\frac{1}{2}} \cdot
\big( v_{(b)}'\circ v_{(b)}^{(-1)}(v) \big)^{\frac{1}{2}}+{\cal{O}}(\epsilon^3), \ee
where we have also taken an expansion with respect to $\epsilon$.
By combining (\ref{geodesiczuv}) and (\ref{cutoffzuv}), we then solve the intersections between the geodesic and the cutoff surface represented by the affine parameters as
\begin{align}\label{s1s2}
&s_1=\log \epsilon-\frac{1}{2}\log (u_1-u_2)+\frac{1}{2} \log u_{(b)}'(U_1)
-\frac{1}{2} \log (v_2-v_1)+\frac{1}{2} \log v_{(b)}'(V_1)+{\cal{O}}(\epsilon^2) \notag \\
&s_2=-\log \epsilon+\frac{1}{2}\log(u_1-u_2)-\frac{1}{2}\log u_{(b)}'(U_2)
+\frac{1}{2}\log(v_2-v_1)-\frac{1}{2}\log v_{(b)}'(V_2)+{\cal{O}}(\epsilon^2).
\end{align}
Given the affine parameters of the intersections (\ref{s1s2}), we compute the geodesic's length as
\be\label{AHRT} A=s_2-s_1=-2\log\epsilon+A^{U}+A^{V}, \ee
with
\begin{align}\label{AUV}
&A^{U}=\log(u_1-u_2)-\frac{1}{2}\log u_{(b)}'(U_1)-\frac{1}{2}\log u_{(b)}'(U_2) \notag \\
&A^{V}=\log(v_2-v_1)-\frac{1}{2} \log v_{(b)}'(V_1)-\frac{1}{2}\log v_{(b)}'(V_2),
\end{align}
where we have ignored the terms which vanish under the $\epsilon\rightarrow 0$ limit in (\ref{AHRT}), (\ref{AUV}).
We also compute the value of the coordinates of intersections as
\begin{align}\label{zuvs1}
&z(s_1)=\epsilon u_{(b)}'(U_1)^{\frac{1}{2}}v_{(b)}'(V_1)^{\frac{1}{2}}+{\cal{O}}(\epsilon^3) \notag \\
&u(s_1)=u_1+{\cal{O}}(\epsilon^2) \notag \\
&v(s_1)=v_1+{\cal{O}}(\epsilon^2),
\end{align}
and
\begin{align}\label{zuvs2}
&z(s_2)=\epsilon u_{(b)}'(U_2)^{\frac{1}{2}} v_{(b)}'(V_2)^{\frac{1}{2}}+{\cal{O}}(\epsilon^3) \notag \\
&u(s_2)=u_2+{\cal{O}}(\epsilon^2) \notag \\
&v(s_2)=v_2+{\cal{O}}(\epsilon^2).
\end{align}
For the application below, we map the intersections (\ref{zuvs1}), (\ref{zuvs2}) back to the $(Z,U,V)$ coordinates by the reversal map of (\ref{transform}), which gives
\begin{align}\label{ZUVs1}
&Z|_{s=s_1}=\epsilon \notag \\
&U|_{s=s_1}=U_1+{\cal{O}}(\epsilon^2) \notag \\
&V|_{s=s_1}=V_1+{\cal{O}}(\epsilon^2),
\end{align}
and
\begin{align}\label{ZUVs2}
&Z|_{s=s_2}=\epsilon \notag \\
&U|_{s=s_2}=U_2+{\cal{O}}(\epsilon^2) \notag \\
&V|_{s=s_2}=V_2+{\cal{O}}(\epsilon^2).
\end{align}

Second, we study the normal frames $(\hat{l}^{(1)},\hat{r}^{(1)})$, $(\hat{l}^{(2)},\hat{r}^{(2)})$ and compute the twist $\zeta$ in the $(z,u,v)$ coordinates. We take a pushforward of the normal frames $(\hat{l}^{(1)},\hat{r}^{(1)})$, $(\hat{l}^{(2)},\hat{r}^{(2)})$ by the diffeomorphism (\ref{transform}) to the $(z,u,v)$ coordinates, under which their adopted value (\ref{bframe}) at the intersections $s=s_1$, $s=s_2$ are mapped to the following expression
\begin{align}\label{lrinzuv}
\hat{l}^{(1)} \big|_{s=s_1}=&{\cal{O}}(\epsilon^2) \frac{\partial}{\partial z}
+\big(2\epsilon u_{(b)}'(U_1)+{\cal{O}}(\epsilon^3)\big) \frac{\partial}{\partial u}
+{\cal{O}}(\epsilon^3) \frac{\partial}{\partial v} \notag \\
\hat{r}^{(1)} \big|_{s=s_1}=&{\cal{O}}(\epsilon^2) \frac{\partial}{\partial z}
+{\cal{O}}(\epsilon^3) \frac{\partial}{\partial u}
+\big(2\epsilon v_{(b)}'(V_1)+{\cal{O}}(\epsilon^3) \big) \frac{\partial}{\partial v} \notag \\
\hat{l}^{(2)}\big|_{s=s_2}=&{\cal{O}}(\epsilon^2)\frac{\partial}{\partial z}
+{\cal{O}}(\epsilon^3) \frac{\partial}{\partial u}
+\big(2\epsilon v_{(b)}'(V_2)+{\cal{O}}(\epsilon^3) \big) \frac{\partial}{\partial v} \notag \\
\hat{r}^{(2)}\big|_{s=s_2}=&{\cal{O}}(\epsilon^2)\frac{\partial}{\partial z}
+\big(2\epsilon u_{(b)}'(U_2)+{\cal{O}}(\epsilon^3)\big) \frac{\partial}{\partial u}
+{\cal{O}}(\epsilon^3) \frac{\partial}{\partial v},
\end{align}
where, in deriving (\ref{lrinzuv}), we have also used (\ref{ZUVs1}), (\ref{ZUVs2}).
We can in principle solve the expressions of $(\hat{l}^{(1)},\hat{r}^{(1)})$, $(\hat{l}^{(2)},\hat{r}^{(2)})$ on the whole geodesic $\gamma$ in the $(z,u,v)$ coordinates from (\ref{parallellr}) and (\ref{lrinzuv}) up to ${\cal{O}}(\epsilon^2)$ correction. But here we take the following shortcut.
We first provide a normal frame $(\hat{l}^{(0)},\hat{r}^{(0)})$ represented as
\begin{align}
\hat{l}^{(0)}\big|_{\gamma}=&
\frac{1}{2\cosh^2 s} \left( \sqrt{(u_1-u_2)(v_2-v_1)} \frac{\partial}{\partial z}
+(u_1-u_2) e^{-s} \frac{\partial}{\partial u}
+(v_2-v_1) e^{s} \frac{\partial}{\partial v} \right) \notag \\
\hat{r}^{(0)}\big|_{\gamma}=&
\frac{1}{2\cosh^2 s} \left( -\sqrt{(u_1-u_2)(v_2-v_1)} \frac{\partial}{\partial z}
+(u_1-u_2)e^{s} \frac{\partial}{\partial u}+(v_2-v_1) e^{-s} \frac{\partial}{\partial v}  \right),
\end{align}
where $\hat{l}^{(0)}$/$\hat{r}^{(0)}$ is a left/right future pointing null vector from the viewpoint of an observer facing along the geodesic, and $(\hat{l}^{(0)},\hat{r}^{(0)})$ satisfy the parallel transport conditions
\be e^{\mu} \nabla_{\mu} l^{(0)\nu} \big|_{\gamma}=e^{\mu} \nabla_{\mu} r^{(0)\nu} \big|_{\gamma}=0, \ee
and the orthonormal conditions
\begin{align}
&\hat{e}\cdot \hat{l}^{(0)} \big|_{\gamma}=\hat{e} \cdot \hat{r}^{(0)} \big|_{\gamma}=0 \notag \\
&\hat{l}^{(0)}\cdot \hat{l}^{(0)} \big|_{\gamma}=\hat{r}^{(0)}\cdot \hat{r}^{(0)} \big|_{\gamma}=0 \notag \\
&\hat{l}^{(0)}\cdot \hat{r}^{(0)} \big|_{\gamma}=-2.
\end{align}
Based on the same argument around (\ref{boost}), we can represent $(\hat{l}^{(1)},\hat{r}^{(1)})$, $(\hat{l}^{(2)},\hat{r}^{(2)})$ with $(\hat{l}^{(0)},\hat{r}^{(0)})$ as the following boost
\begin{align}\label{lr10}
&\hat{l}^{(1)}=e^{\zeta_1} \hat{l}^{(0)} \notag \\
&\hat{r}^{(1)}=e^{-\zeta_1} \hat{r}^{(0)}, \end{align}
and
\begin{align}\label{lr20}
&\hat{l}^{(2)}=e^{\zeta_2} \hat{l}^{(0)} \notag \\
&\hat{r}^{(2)}=e^{-\zeta_2} \hat{r}^{(0)},
\end{align}
where the rapidity $\zeta_1$, $\zeta_2$ are constant along the geodesic $\gamma$.
The constant rapidity $\zeta_1$, $\zeta_2$ can be furthermore fixed by comparing the value of $(\hat{l}^{(0)},\hat{r}^{(0)})$, $(\hat{l}^{(1)},\hat{r}^{(1)})$, $(\hat{l}^{(2)},\hat{r}^{(2)})$ at the intersections $s=s_1$, $s=s_2$. Specifically, the value of $(\hat{l}^{(0)},\hat{r}^{(0)})$ at the intersections $s=s_1$, $s=s_2$ is as
\begin{align}\label{lr0s1s2}
\hat{l}^{(0)}\big|_{s=s_1}=&{\cal{O}}(\epsilon^2) \frac{\partial}{\partial z}
+\bigg( 2\epsilon \Big(\frac{u_1-u_2}{v_2-v_1}\Big)^{\frac{1}{2}} u_{(b)}'(U_1)^{\frac{1}{2}} v_{(b)}'(V_1)^{\frac{1}{2}}
+{\cal{O}}(\epsilon^3) \bigg) \frac{\partial}{\partial u}
+{\cal{O}}(\epsilon^3) \frac{\partial}{\partial v} \notag \\
\hat{r}^{(0)}\big|_{s=s_1}=& {\cal{O}}(\epsilon^2) \frac{\partial}{\partial z}
+{\cal{O}}(\epsilon^3) \frac{\partial}{\partial u}
+\bigg( 2\epsilon \Big( \frac{v_2-v_1}{u_1-u_2} \Big)^{\frac{1}{2}} u_{(b)}'(U_1)^{\frac{1}{2}} v_{(b)}'(V_1)^{\frac{1}{2}}
+{\cal{O}}(\epsilon^3) \bigg) \frac{\partial}{\partial v} \notag \\
\hat{l}^{(0)} \big|_{s=s_2}=&
{\cal{O}}(\epsilon^2) \frac{\partial}{\partial z}+{\cal{O}}(\epsilon^3) \frac{\partial}{\partial u}
+\bigg( 2\epsilon \Big( \frac{v_2-v_1}{u_1-u_2} \Big)^{\frac{1}{2}} u_{(b)}'(U_2)^{\frac{1}{2}} v_{(b)}'(V_2)^{\frac{1}{2}}
+{\cal{O}}(\epsilon^3) \bigg) \frac{\partial}{\partial v} \notag \\
\hat{r}^{(0)} \big|_{s=s_2}=&
{\cal{O}}(\epsilon^2) \frac{\partial}{\partial z}
+\bigg(2\epsilon \Big( \frac{u_1-u_2}{v_2-v_1} \Big)^{\frac{1}{2}} u_{(b)}'(U_2)^{\frac{1}{2}} v_{(b)}'(V_2)^{\frac{1}{2}}
+{\cal{O}}(\epsilon^3) \bigg) \frac{\partial}{\partial u}
+{\cal{O}}(\epsilon^3) \frac{\partial}{\partial v}.
\end{align}
And by comparing (\ref{lrinzuv}) with (\ref{lr0s1s2}), we get the rapidity $\zeta_1$, $\zeta_2$ as
\begin{align}\label{zeta12}
&\zeta_1=-\frac{1}{2} \log(u_1-u_2) +\frac{1}{2} \log u_{(b)}'(U_1)
+\frac{1}{2}(v_2-v_1)-\frac{1}{2} \log v_{(b)}'(V_1)+{\cal{O}}(\epsilon^3) \notag \\
&\zeta_2=\frac{1}{2} \log(u_1-u_2)-\frac{1}{2} \log u_{(b)}'(U_2)
-\frac{1}{2}\log(v_2-v_1)+\frac{1}{2} \log v_{(b)}'(V_2)+{\cal{O}}(\epsilon^3).
\end{align}
(\ref{lr10}), (\ref{lr20}) together with (\ref{zeta12}) are in principle the expressions of $(\hat{l}^{(1)},\hat{r}^{(1)})$, $(\hat{l}^{(2)},\hat{r}^{(2)})$.
Moreover, by applying (\ref{lr10}), (\ref{lr20}) to (\ref{lr21}) and by taking use of (\ref{zeta12}), we get the expression of the twist $\zeta$ as
\be\label{twistexpression}
\zeta=\zeta_2-\zeta_1
=A^U-A^V,
\ee
where $A^{U}$, $A^{V}$ are given in (\ref{AUV}), and we have ignored the terms which vanish under the $\epsilon\rightarrow 0$ limit.

\subsection{The brackets with the twist}

In this subsection, we compute some relevant brackets with the twist $\zeta$.
First, by applying $T_{UU}$/$T_{VV}$ and $\zeta$ to (\ref{chainuvT}) with $T_{UU}$/$T_{VV}$ in the position of $\cdot$ and $\zeta$ in the position $O$, we get the brackets of $T_{UU}$, $T_{VV}$ with the twist $\zeta$ as
\begin{align}\label{bracketTzeta}
\{T_{UU}(U),\zeta\}=&2\pi
\Big( \frac{\partial\zeta}{\partial U_1} \delta(U-U_1)
+\frac{\partial\zeta}{\partial U_2} \delta(U-U_2)
-\frac{1}{2}\frac{\partial}{\partial U_1} \delta(U-U_1)-\frac{1}{2}\frac{\partial}{\partial U_2}\delta(U-U_2) \Big) \notag \\
\{T_{VV}(V),\zeta\}=&
2\pi \Big( \frac{\partial\zeta}{\partial V_1} \delta(V-V_1)
+\frac{\partial\zeta}{\partial V_2} \delta(V-V_2)
+\frac{1}{2}\frac{\partial}{\partial V_1}\delta(V-V_1)+\frac{1}{2}\frac{\partial}{\partial V_2}\delta(V-V_2) \Big),
\end{align}
where we have also used (\ref{bracketTu}), (\ref{bracketTv}), (\ref{AUV}), (\ref{twistexpression}).
Second, by applying $\zeta$ to (\ref{uvbracket}) in the position of $\cdot$, we get the brackets of $u_{(b)}$, $v_{(b)}$ with $\zeta$ as
\begin{align}\label{bracketuvzeta}
\{u_{(b)}(U),\zeta\}=&-\frac{12\pi}{c}\frac{1}{u_1-u_2} \big(u_{(b)}(U)-u_1\big) \big(u_{(b)}(U)-u_2 \big)
\big(-\theta(U-U_1)+\theta(U-U_2)\big) \notag \\
&+a_{\zeta}^{u,1}u_{(b)}(U)^2+a_{\zeta}^{u,0}u_{(b)}(U)+a_{\zeta}^{u,-1} \notag \\
\{v_{(b)}(V),\zeta\}=&\frac{12\pi}{c} \frac{1}{v_2-v_1} \big(v_{(b)}(V)-v_1\big) \big(v_{(b)}(V)-v_2\big)
\big(\theta(V-V_1)-\theta(V-V_2)\big) \notag \\
&+a_{\zeta}^{v,1}v_{(b)}(V)^2+a_{\zeta}^{v,0}v_{(b)}(V)+a_{\zeta}^{v,-1}.
\end{align}
Here, $a_{\zeta}^{u,i}$, $a_{\zeta}^{v,i}$, with $i=-1,0,1$, are defined from $a^{u,i}[u_{(b)},\delta T_{UU}]$, $a^{v,i}[v_{(b)},\delta T_{VV}]$, which appear in (\ref{varuv}), as
\begin{align}
&a^{u,i}_{\zeta}\equiv
a^{u,i}[u_{(b)},\delta T_{UU}] \big|_{\delta T_{UU}(U^{*})\rightarrow \{T_{UU}(U^{*}),\zeta\}} \notag \\
&a^{v,i}_{\zeta} \equiv
a^{v,i}[v_{(b)},\delta T_{VV}] \big|_{\delta T_{VV}(V^{*}) \rightarrow \{T_{VV}(V^{*}),\zeta\}},
\end{align}
which are independent of $U$, $V$.
In deriving (\ref{bracketuvzeta}), we have used (\ref{varuv}), (\ref{kernel}), (\ref{bracketTzeta}), and we have assumed that the reference point $(U_0,V_0)$ is to the left of $(U_1,V_1)$, $(U_2,V_2)$ such that
\begin{align}
&U_0>U_1>U_2 \notag \\
&V_0<V_1<V_2. \end{align}

\subsection{The system's evolution generated by the twist}\label{seccovshift}

In this subsection, we study the system's evolution generated by the twist $\zeta$.

In the approach applied in this section, the system's evolution generated by a given observable is represented as an infinitesimal diffeomorphism (\ref{Deltagf}).
Here, the diffeomorphism parameter $\xi^{\mu}$ is expressed as (\ref{xiZUV}) in the $(Z,U,V)$ coordinates, or equivalently as (\ref{xifinalzuv}) after a pushforward by the diffeomorphism (\ref{transform}) to the $(z,u,v)$ coordinates.
In the following discussion for the twist $\zeta$, we will express the corresponding diffeomorphism parameter $\xi^{\mu}$ in the $(z,u,v)$ coordinates.
But we will discuss the system's evolution in the $(Z,U,V)$ coordinates.

We now express the diffeomorphism parameter $\xi^{\mu}$ in the $(z,u,v)$ coordinates for the system's evolution generated by the twist $\zeta$.
Following the approach, the diffeomorphism parameter $\xi^{\mu}$ in the $(z,u,v)$ coordinates can be any one with the asymptotic behavior (\ref{xifinalzuv}), where the $\eta^u$, $\eta^v$ in (\ref{xifinalzuv}) take the following expression
\begin{align}\label{etauvforzeta}
&\eta^{u}(u)=\lambda \Big[ -\frac{12\pi}{c} \frac{1}{u_1-u_2}(u-u_1)(u-u_2)
\big(-\theta(u-u_1)+\theta(u-u_2)\big)
+a_{\zeta}^{u,1}u^2+a_{\zeta}^{u,0}u+a_{\zeta}^{u,-1} \Big] \notag \\
&\eta^{v}(v)=\lambda \Big[ \frac{12\pi}{c} \frac{1}{v_2-v_1}(v-v_1)(v-v_2)
\big(\theta(v-v_1)-\theta(v-v_2)\big)
+a^{v,1}_{\zeta}v^2+a^{v,0}_{\zeta}v+a^{v,-1}_{\zeta} \Big],
\end{align}
which are computed by applying (\ref{bracketuvzeta}) through (\ref{Deltauv}) to (\ref{etafinaluv}).
Note that all such diffeomorphism parameters are equivalent in describing the system's evolution.

Given the asymptotic behavior of the diffeomorphism parameter $\xi^{\mu}$, we would like to make a further requirement for its bulk behavior to simplify the geometric interpretation of the system's evolution.

\begin{figure}
  \centering
  \includegraphics[width=8cm]{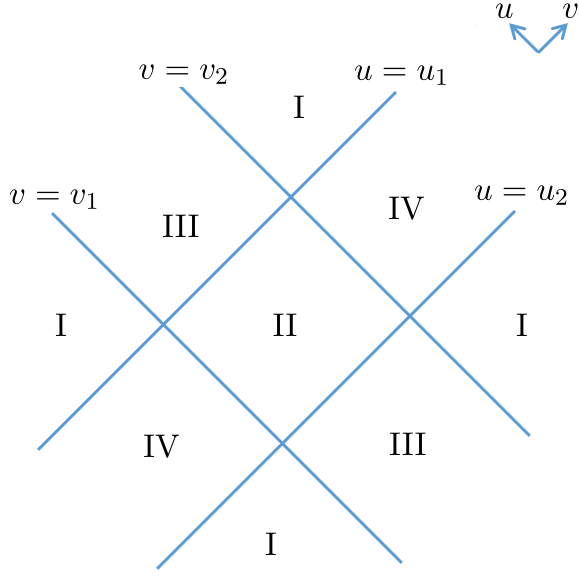}\\
  \caption{The subregions on the asymptotic boundary. We divide the asymptotic boundary into subregions labeled by the Roman numerals, where, in each subregion labeled by a Roman numeral, the $\eta^u$, $\eta^v$ in (\ref{etauvforzeta}) take uniform quadratic polynomials of $u$, $v$ respectively.}\label{subregions}
\end{figure}

The requirement is inspired from the observation that the $\eta^{u}$, $\eta^{v}$ in (\ref{etauvforzeta}) are piecewise quadratic polynomials of $u$, $v$ respectively, which implies a close relationship between the diffeomorphism parameter $\xi^{\mu}$ and the Killing fields (\ref{xiu}), (\ref{xiv}).
To be more precise, we divide the asymptotic boundary into subregions as in Fig.\ref{subregions}, such that, in each subregion, $\eta^u$, $\eta^v$ take uniform quadratic polynomials of $u$, $v$ respectively.
Each of these subregions on the asymptotic boundary corresponds to a near boundary region in the bulk.
And in each of these near boundary regions, the asymptotic behavior of the diffeomorphism parameter $\xi^{\mu}$ in (\ref{xifinalzuv}) coincides with the one of a Killing field. Here, the Killing field can be represented by its corresponding global conformal Killing field as
\be\label{eta} \hat{\eta}=\eta^u(u) \frac{\partial}{\partial u}+\eta^v(v) \frac{\partial}{\partial v}, \ee
where we have used the correspondence between the Killing fields (\ref{xiu}), (\ref{xiv}) and the global conformal Killing fields (\ref{etauv}), and the fact that the $\eta^u$, $\eta^v$ take uniform quadratic polynomials respectively in the corresponding subregion on the asymptotic boundary.
For the following application, we explicitly represent the Killing fields
$\xi_{\mbox{\scriptsize I}}^{\phantom{\mbox{\scriptsize I}}\mu}$, $\xi_{\mbox{\scriptsize II}}^{\phantom{\mbox{\scriptsize II}}\mu}$
corresponding to the subregion I, II on the asymptotic boundary by their corresponding global conformal Killing fields $\hat{\eta}_{\mbox{\scriptsize I}}$, $\hat{\eta}_{\mbox{\scriptsize II}}$ as
\begin{align}\label{eta12}
\hat{\eta}_{\mbox{\scriptsize I}}=&\lambda(a_{\zeta}^{u,1}u^2+a_{\zeta}^{u,0}u+a_{\zeta}^{u,-1}) \frac{\partial}{\partial u}
+\lambda(a_{\zeta}^{v,1}v^2+a_{\zeta}^{v,0}v+a^{v,-1})\frac{\partial}{\partial v} \notag \\
\hat{\eta}_{\mbox{\scriptsize II}}=&\lambda \Big(-\frac{12\pi}{c}\frac{1}{u_1-u_2}(u-u_1)(u-u_2)
+a_{\zeta}^{u,1}u^2+a_{\zeta}^{u,0}u+a_{\zeta}^{u,-1} \Big) \frac{\partial}{\partial u} \notag \\
&+\lambda \Big(\frac{12\pi}{c}\frac{1}{v_2-v_1}(v-v_1)(v-v_2)
+a_{\zeta}^{v,1}v^2+a_{\zeta}^{v,0}v+a_{\zeta}^{v,-1} \Big) \frac{\partial}{\partial v}.
\end{align}

\begin{figure}
  \centering
  \subfloat[The Cauchy surface $\Sigma$]{\includegraphics[width=6cm]{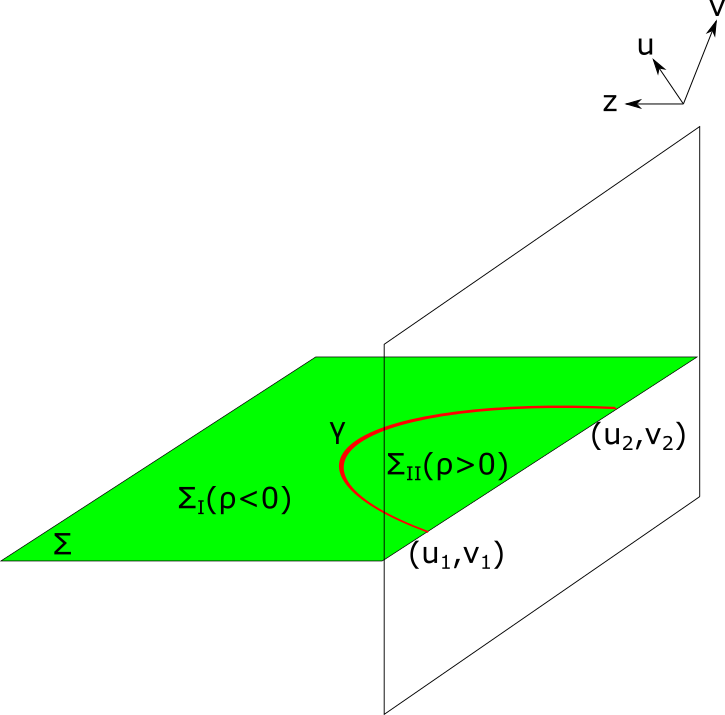}\label{Cauchytot}}
  \quad\quad\quad
  \subfloat[The intersection of the Cauchy surface $\Sigma$ with the asymptotic boundary]{\includegraphics[width=5.5cm]{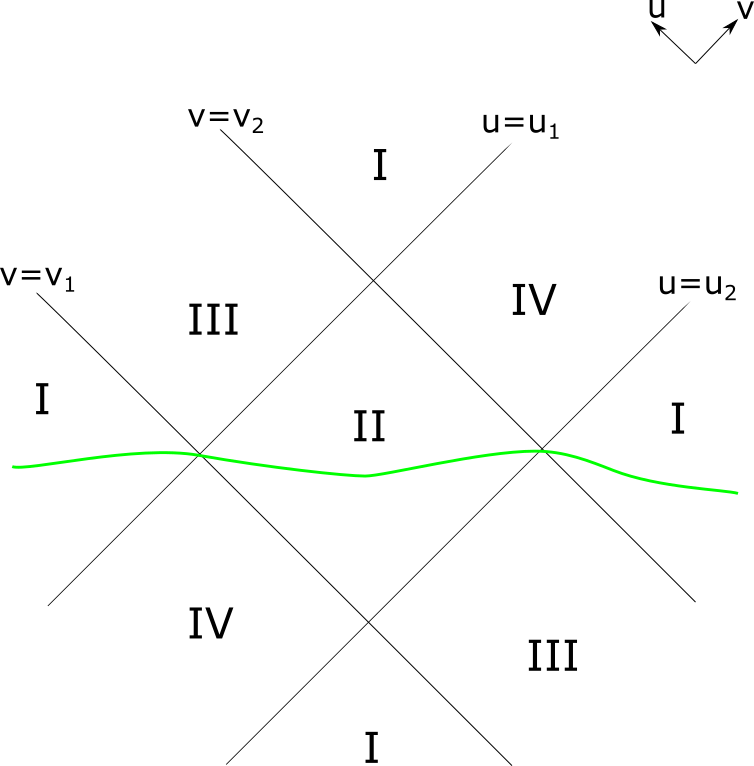}\label{bCauchy}}
  \caption{The Cauchy surface $\Sigma$ that contains the geodesic $\gamma$. (a) illustrates the Cauchy surface in the bulk denoted by the green plane. (b) illustrates the intersection of the Cauchy surface with the asymptotic boundary denoted by the green curve.}\label{Cauchy}
\end{figure}

We now explicitly provide the requirement for the diffeomorphism parameter $\xi^{\mu}$.
We first introduce a Cauchy surface $\Sigma$ containing and being divided by the geodesic $\gamma$ as in Figure.\ref{Cauchytot}. Here, we denote the piece of the Cauchy surface $\Sigma$ outside/inside the geodesic $\gamma$ as $\Sigma_{\mbox{\scriptsize I}}$/$\Sigma_{\mbox{\scriptsize II}}$. And by comparing Fig.\ref{Cauchytot} and Fig.\ref{bCauchy}, we see that the intersection of $\Sigma_{\mbox{\scriptsize I}}$/$\Sigma_{\mbox{\scriptsize II}}$ with the asymptotic boundary is in the subregion I/II.
Given this setup, we make the following requirement for the restriction of the diffeomorphism parameter $\xi^{\mu}$ on the Cauchy surface $\Sigma$ as
\be\label{xiSigma}
\xi^{\mu}|_{\Sigma}=\big( \theta(-\rho)\xi_{\mbox{\scriptsize I}}^{\phantom{\mbox{\scriptsize I}}\mu}
+\theta(\rho)\xi_{\mbox{\scriptsize II}}^{\phantom{\mbox{\scriptsize II}}\mu} \big) \big|_{\Sigma}.
\ee
Here, $\rho$ takes the definition below (\ref{deltaA}).
$\xi_{\mbox{\scriptsize I}}^{\phantom{\mbox{\scriptsize I}}\mu}$, $\xi_{\mbox{\scriptsize II}}^{\phantom{\mbox{\scriptsize II}}\mu}$ are Killing fields presented in (\ref{eta12}) by the corresponding global conformal Killing fields.
And the requirement (\ref{xiSigma}) means that the diffeomorphism parameter $\xi^{\mu}$, when restricted on the piece of the Cauchy surface $\Sigma_{\mbox{\scriptsize I}}$/$\Sigma_{\mbox{\scriptsize II}}$, coincides with the Killing field
$\xi_{\mbox{\scriptsize I}}^{\phantom{\mbox{\scriptsize I}}\mu}$/$\xi_{\mbox{\scriptsize II}}^{\phantom{\mbox{\scriptsize II}}\mu}$ that corresponds to the subregion I/II on the asymptotic boundary.
It is clear that the requirement (\ref{xiSigma}) of the diffeomorphism parameter $\xi^{\mu}$ is consistent with its asymptotic behavior. 
\footnote{Note that the requirement (\ref{xiSigma}) is only compatible with the asymptotic behavior (\ref{xiZUV}) but not with the one (\ref{xiFG}) that preserves the Fefferman-Graham gauge. This is the reason that we introduce $\Delta g_{\mu\nu}$ beyond $\widetilde{\Delta} g_{\mu\nu}$.}
For the following application, we also derive the relative difference between
$\xi_{\mbox{\scriptsize I}}^{\phantom{\mbox{\scriptsize I}}\mu}$, $\xi_{\mbox{\scriptsize II}}^{\phantom{\mbox{\scriptsize II}}\mu}$ as
\be\label{xidiff}
\xi_{\mbox{\scriptsize II}}^{\phantom{\mbox{\scriptsize II}}\mu}
-\xi_{\mbox{\scriptsize I}}^{\phantom{\mbox{\scriptsize I}}\mu}
=-\lambda \frac{6\pi}{c} \xi_{(u,e)}^{\phantom{(u,e)}\mu}
-\lambda\frac{6\pi}{c}\xi_{(v,e)}^{\phantom{(v,e)}\mu},
\ee
and its behavior near the geodesic $\gamma$ as
\begin{align}\label{xidifference}
&(\xi _{\mbox{\scriptsize II}}^{\phantom{\mbox{\scriptsize II}}\mu}
-\xi_{\mbox{\scriptsize I}}^{\phantom{\mbox{\scriptsize I}}\mu}) \big|_{\gamma}
=-\lambda\frac{12\pi}{c}e^\mu \big|_{\gamma} \notag \\
&\nabla_{\mu}(\xi_{\mbox{\scriptsize II}}^{\phantom{\mbox{\scriptsize II}}\nu}
-\xi_{\mbox{\scriptsize I}}^{\phantom{\mbox{\scriptsize I}}\nu}) \big|_{\gamma}=0.
\end{align}
Here, (\ref{xidiff}) is derived by equivalently mapping to the following equation for the corresponding global conformal Killing fields
\be \hat{\eta}_{\mbox{\scriptsize II}}-\hat{\eta}_{\mbox{\scriptsize I}}=
-\lambda \frac{6\pi}{c} \hat{\eta}_{(u,e)}-\lambda \frac{6\pi}{c} \hat{\eta}_{(v,e)}, \ee
which in turn can be verified with (\ref{eta12}) and (\ref{etae}).
And (\ref{xidifference}) is derived by restricting (\ref{xidiff}) and its derivative on the geodesic $\gamma$ and taking use of (\ref{Killingequ}) and (\ref{derivativexi}).

So far we have expressed the diffeomorphism parameter $\xi^{\mu}$ in the $(z,u,v)$ coordinates.
But remember that the system's evolution (\ref{Deltagf}) should be discussed in the $(Z,U,V)$ coordinates, so we need to take a pullback by the diffeomorphism (\ref{transform}) from the $(z,u,v)$ coordinates to the $(Z,U,V)$ coordinates.
Specifically, we take a pullback of the Cauchy surface $\Sigma$, the Killing fields
$\xi_{\mbox{\scriptsize I}}^{\phantom{\mbox{\scriptsize I}}\mu}$, $\xi_{\mbox{\scriptsize II}}^{\phantom{\mbox{\scriptsize II}}\mu}$, and
the diffeomorphism parameter $\xi^{\mu}$ to the $(Z,U,V)$ coordinates.
For the corresponding objects in the $(Z,U,V)$ coordinates, the Cauchy surface $\Sigma$ still contains the geodesic $\gamma$,
$\xi_{\mbox{\scriptsize I}}^{\phantom{\mbox{\scriptsize I}}\mu}$, $\xi_{\mbox{\scriptsize II}}^{\phantom{\mbox{\scriptsize II}}\mu}$ are still the Killing fields of the corresponding metric, and the equations (\ref{xiSigma}), (\ref{xidifference}) still hold.

We are now ready to discuss the system's evolution (\ref{Deltagf}) in the $(Z,U,V)$ coordinates.
Here, we already have all relevant quantities, and we only need to make two explanations.
First, up to a non-physical diffeomorphism, the system's evolution (\ref{Deltagf}) is completely captured by the restriction of the diffeomorphism parameter $\xi^{\mu}$ on the Cauchy surface $\Sigma$ which is represented in (\ref{xiSigma}).\footnote{This statement can be seen from the action of a diffeomorphism on the set of initial data on a given Cauchy surface represented in (\ref{diff}). Here, the transformation of the set of initial data only depends on the value of the diffeomorphism parameter restricted on the Cauchy surface. And the system's evolution is completely captured by the evolution of the set of initial data up to a non-physical diffeomorphism.}
Second, since the
$\xi_{\mbox{\scriptsize I}}^{\phantom{\mbox{\scriptsize I}}\mu}$, $\xi_{\mbox{\scriptsize II}}^{\phantom{\mbox{\scriptsize II}}\mu}$ in (\ref{xiSigma}) are both Killing fields, the system's evolution is finally captured by the relative difference between
$\xi_{\mbox{\scriptsize I}}^{\phantom{\mbox{\scriptsize I}}\mu}$ and $\xi_{\mbox{\scriptsize II}}^{\phantom{\mbox{\scriptsize II}}\mu}$ for the region near the geodesic $\gamma$ which is given in (\ref{xidifference}).
Based on the two explanations, we now give the final expression of the system's evolution: it is the infinitesimal diffeomorphism (\ref{Deltagf}) whose diffeomorphism parameter $\xi^{\mu}$ satisfy (\ref{xiSigma}) on the Cauchy surface $\Sigma$, where the
$\xi_{\mbox{\scriptsize I}}^{\phantom{\mbox{\scriptsize I}}\mu}$, $\xi_{\mbox{\scriptsize II}}^{\phantom{\mbox{\scriptsize II}}\mu}$ in (\ref{xiSigma}) are the Killing fields of the unevolved metric and satisfy (\ref{xidifference}) on the geodesic $\gamma$.
From (\ref{xiSigma}), (\ref{xidifference}), we see that the system's evolution is again a relative shift along the geodesic $\gamma$. And by taking use of (\ref{cG}), we see that this relative shift is the same as the one we mentioned in subsection \ref{twistill}.

We can also compare the system's evolution with the one derived with the canonical formalism (\ref{evotwist}).
Specifically, by applying (\ref{xiSigma}) to (\ref{diff}), we get the evolution of the set of initial data on the Cauchy surface $\Sigma$ as
\begin{align}\label{codeltaini}
&\sigma_{\alpha}^{\phantom{\alpha}\mu} \sigma_{\beta}^{\phantom{\beta}\nu}
\Delta \sigma_{\mu\nu}|_{\Sigma}=-\lambda \frac{12\pi}{c} \delta(\rho)(n_{\alpha}e_{\beta}+e_{\alpha}n_{\beta})+o(\lambda) \notag \\
&\sigma_{\alpha}^{\phantom{\alpha}\mu} \sigma_{\beta}^{\phantom{\beta}\nu}
\Delta K_{\mu\nu}|_{\Sigma}=-\lambda\frac{12\pi}{c} \delta (\rho) K_{\mu\nu}e^{\mu}n^{\nu} n_{\alpha}n_{\beta}+o(\lambda),
\end{align}
which is precisely the same as (\ref{evotwist}) after the replacement of (\ref{cG}).\footnote{More precisely, we should also take a non-physical diffeomorphism mapping the Cauchy surface to the given one whose boundary contains the geodesic's endpoints.}
In deriving (\ref{codeltaini}), we have used the fact that
$\xi_{\mbox{\scriptsize I}}^{\phantom{\mbox{\scriptsize I}}\mu}$, $\xi_{\mbox{\scriptsize II}}^{\phantom{\mbox{\scriptsize II}}\mu}$ are Killing fields, so the terms proportional to $\theta(\rho)$, $\theta(-\rho)$ automatically vanish, and the remaining terms can be combined into the ones only in terms of
$\xi_{\mbox{\scriptsize II}}^{\phantom{\mbox{\scriptsize II}}\mu}-\xi_{\mbox{\scriptsize I}}^{\phantom{\mbox{\scriptsize I}}\mu}$.
We have also used the following equations
\begin{align}\label{ADMxidiff}
&\tau_{\mu} (\xi_{\mbox{\scriptsize II}}^{\phantom{\mbox{\scriptsize II}}\mu}
-\xi_{\mbox{\scriptsize I}}^{\phantom{\mbox{\scriptsize I}}\mu}) \big|_{\gamma}=0 \notag \\
&\sigma_{\alpha\mu} (\xi_{\mbox{\scriptsize II}}^{\phantom{\mbox{\scriptsize II}} \mu}
-\xi_{\mbox{\scriptsize I}}^{\phantom{\mbox{\scriptsize I}} \mu}) \big|_{\gamma}
=-\lambda \frac{12\pi}{c} e_{\alpha} \big|_{\gamma} \notag \\
&D_{\alpha} \big(\tau_{\mu} (\xi_{\mbox{\scriptsize II}}^{\phantom{\mbox{\scriptsize II}} \mu}
-\xi_{\mbox{\scriptsize I}}^{\phantom{\mbox{\scriptsize I}} \mu}) \big) \big|_{\gamma}
=-\lambda \frac{12\pi}{c} K_{\alpha \mu} e^\mu \big|_{\gamma},
\end{align}
the equation (\ref{nadrho}), and the last equation of (\ref{relationongeo}),
where (\ref{ADMxidiff}) are derived from an ADM decomposition of (\ref{xidifference}).

\section{The bracket between the geodesic length and the twist}\label{bracket}

We now switch to the second topic, the brackets with the twist $\zeta$. In particular, we only study the bracket between the geodesic length $A$ and the twist $\zeta$ of the same geodesic $\gamma$.
Remember that our original motivation to study the twist $\zeta$ is that it may be a candidate element in the center.
But being an element in the center requires the twist $\zeta$ at least to be commutative with the HRT-area which is the geodesic length $A$ in our setup.
In the following, we will compute the bracket between the geodesic length $A$ and the twist $\zeta$ with the canonical formalism and the covariant phase space formalism respectively. Both of the two approaches give the same result, that is the twist $\zeta$ indeed commutes with the geodesic length $A$.

We first compute the bracket $\{A,\zeta\}$ with the canonical formalism.
Following the treatment in section \ref{evocan}, we choose a Cauchy surface $\Sigma$ whose boundary $\partial\Sigma$ contains the geodesic's endpoints. We also restrict the computation to the subset of the solutions such that the geodesic $\gamma$ is contained in the Cauchy surface $\Sigma$. Here, this restriction makes no difference to the final result, since all solutions can be mapped to this subset by a non-physical diffeomorphism.
Based on this setup, we compute the bracket $\{A,\zeta\}$ with the chain rule as
\begin{align}\label{bracketchain}
\{A,\zeta\}=&
\int d^2 x \int d^2 y
\bigg[ \frac{\delta A}{\delta \sigma_{ab}(t_0,x)} \frac{\delta \zeta}{\delta \sigma_{cd}(t_0,y)}
\{\sigma_{ab}(t_0,x),\sigma_{cd}(t_0,y)\} \notag \\
&+\frac{\delta A}{\delta \sigma_{ab}(t_0,x)} \frac{\delta \zeta}{\delta K_{cd}(t_0,y)} \{\sigma_{ab}(t_0,x),K_{cd}(t_0,y)\} \notag \\
&+\frac{\delta A}{\delta K_{ab}(t_0,x)} \frac{\delta \zeta}{\delta \sigma_{cd}(t_0,y)} \{K_{ab}(t_0,x),\sigma_{cd}(t_0,y)\} \notag \\
&+\frac{\delta A}{\delta K_{ab}(t_0,x)} \frac{\delta \zeta}{\delta K_{cd}(t_0,y)} \{K_{ab}(t_0,x),K_{cd}(t_0,y)\} \bigg].
\end{align}
Here, we have introduced a coordinate system $(t,x^a)$ such that the Cauchy surface $\Sigma$ is at $t=t_0$.
The functional derivatives of $A$ and $\zeta$ with respect to the set of initial data can be read out from (\ref{deltaA}) and (\ref{deltazetafinal}). And the brackets between the set of initial data are given in (\ref{Poisson2}).\footnote{There is no difference between Poisson bracket and Dirac bracket when computing the brackets between diffeomorphism invariant observables.}
In the first sight, we may expect a divergent result in the bracket $\{A,\zeta\}$, since both $\delta A$ in (\ref{deltaA}) and $\delta \zeta$ in (\ref{deltazetafinal}) support only on the geodesic $\gamma$.
However, by applying (\ref{deltaA}), (\ref{deltazetafinal}), (\ref{Poisson2}) to (\ref{bracketchain}), we can explicitly compute the bracket $\{A,\zeta\}$ and verify its vanishing as
\begin{align}\label{braAzeta}
\{A,\zeta\}=&(-1)\int_{\Sigma}d^2x \sqrt{\sigma(x)} \frac{1}{2} e^a(x)e^b(x)\delta\big(\rho(x)\big)
\int_{\Sigma} d^2y \sqrt{\sigma(y)} e^{c}(y) n^d(y) \delta \big(\rho(y)\big) \notag \\
&\cdot 8\pi G \Big( \sigma_{ac}(x)\sigma_{bd}(x)+\sigma_{ad}(x)\sigma_{bc}(x)-2\sigma_{ab}(x)\sigma_{cd}(x) \Big)
\frac{1}{\sqrt{\sigma(x)}} \delta^2(x-y)
\notag \\
=&(-1) \int_{\Sigma} d^2x \sqrt{\sigma(x)} \frac{1}{2} e^a(x)e^b(x)e^c(x)n^d(x) \notag \\
&\cdot 8\pi G \Big( \sigma_{ac}(x)\sigma_{bd}(x)+\sigma_{ad}(x)\sigma_{bc}(x)-2\sigma_{ab}(x)\sigma_{cd}(x) \Big)
\delta\big(\rho(x)\big)^2 \notag \\
=&\int_{\Sigma} d^2x \sqrt{\sigma(x)} \cdot 0 \cdot \delta\big( \rho(x)\big)^2 \notag \\
=&0.
\end{align}
Here, in the second line from the end, the $\delta(\rho)^2$ reflects the potential divergence that we just mentioned.
Nevertheless, the $0$ is a $0$ everywhere on the Cauchy surface, so we believe it beats the $\delta(\rho)^2$ and leads to a zero result in the integral.
We admit that this argument is not solid enough and requires further regularization. See subsubsection \ref{bracketAzeta} for an attempt of the regularization.

We now provide a more rigorous computation of the bracket $\{A,\zeta\}$ with the covariant phase space formalism namely the approach used in section \ref{covariant}. We list the expressions of the geodesic length (\ref{AHRT}) and the twist (\ref{twistexpression}) as
\begin{align}
&A=-2\log\epsilon+A^{U}+A^{V} \notag \\
&\zeta=A^{U}-A^{V},
\end{align}
with $A^U$, $A^V$ given in (\ref{AUV}).
Naively, if we accept that $A^U$, $A^V$ are self commutative
\be \{A^{U},A^{U}\}=\{A^{V},A^{V}\}=0, \ee
we can directly show that the geodesic length $A$ and the twist $\zeta$ are commutative
\be \{A,\zeta\}=0. \ee
However, we point out that the brackets $\{A^{U},A^{U}\}$, $\{A^{V},A^{V}\}$ are ill-defined in our approach, and the whole computation needs a regularization.

We first explain why the brackets $\{A^{U},A^{U}\}$, $\{A^{V},A^{V}\}$ are ill-defined. In our approach, these brackets are computed in two steps. First, we compute the brackets of $u_{(b)}$, $v_{(b)}$ with $A^{U}$, $A^{V}$ respectively as
\begin{align}\label{bracketuvAUV}
\{u_{(b)}(U),A^{U}\}=&
-\frac{12\pi}{c} \frac{1}{u_1-u_2} \big(u_{(b)}(U)-u_1\big) \big(u_{(b)}(U)-u_2\big)
\big(-\theta(U-U_1)+\theta(U-U_2) \big) \notag \\
&+a_{A^{U}}^{u,1} u_{(b)}(U)^2+a_{A^{U}}^{u,0} u_{(b)}(U) +a_{A^U}^{u,-1} \notag \\
\{v_{(b)}(V),A^V\}=&
-\frac{12\pi}{c} \frac{1}{v_2-v_1} \big( v_{(b)}(V)-v_1 \big) \big(v_{(b)}(V)-v_2\big)
\big( \theta(V-V_1) -\theta(V-V_2) \big) \notag \\
&+a_{A^V}^{v,1} v_{(b)}(V)^2+a_{A^V}^{v,0} v_{(b)}(V) +a_{A^V}^{v,-1},
\end{align}
where
\begin{align}
&a_{A^{U}}^{u,i} \equiv a^{u,i}[u_{(b)},\delta T_{UU}] \big|_{\delta T_{UU}(U^{*}) \rightarrow \{T_{UU}(U^{*}),A^U\}} \notag \\
&a_{A^{V}}^{v,i} \equiv a^{v,i}[v_{(b)},\delta T_{VV}] \big|_{\delta T_{VV}(V^{*}) \rightarrow \{T_{VV}(V^{*}),A^V\}},
\end{align}
for $i=-1,0,1$.
Second, we compute the brackets $\{A^U,A^U\}$, $\{A^V,A^V\}$ by applying (\ref{bracketuvAUV}) to the following chain rule
\begin{align}\label{selfbracket}
&\{A^U,A^U\}=\int dU \frac{\delta A^U}{\delta u_{(b)}(U)}\{u_{(b)}(U),A^{U}\} \notag \\
&\{A^V,A^V\}=\int dV \frac{\delta A^V}{\delta v_{(b)}(V)} \{v_{(b)}(V),A^{V}\}.
\end{align}
We point out that the computation in (\ref{selfbracket}) lead to the brackets $\{A^U,A^U\}$, $\{A^V,A^V\}$ being ill-defined.
In more detail, the bracket $\{u_{(b)}(U),A^U\}$/$\{v_{(b)}(V),A^V\}$ computed in (\ref{bracketuvAUV}) has discontinuity in its first order derivative at $U_1$/$V_1$, $U_2$/$V_2$. While, by taking use of the expressions (\ref{AUV}), the functional derivative $\frac{\delta A^{U}}{\delta u_{(b)}(U)}$/$\frac{\delta A^V}{\delta v_{(b)}(V)}$ contains terms $\delta'(U-U_1)$/$\delta'(V-V_1)$, $\delta'(U-U_2)$/$\delta'(V-V_2)$. These two facts together lead to the computation (\ref{selfbracket}) being ill-defined.
Technically, both of the two facts are from the appearance of $u_{(b)}'$, $v_{(b)}'$ in the expressions of $A^U$, $A^V$ in (\ref{AUV}), which in turn is because the geodesic $\gamma$ ends up at the asymptotic boundary.

\begin{figure}
  \centering
  \includegraphics[width=10cm]{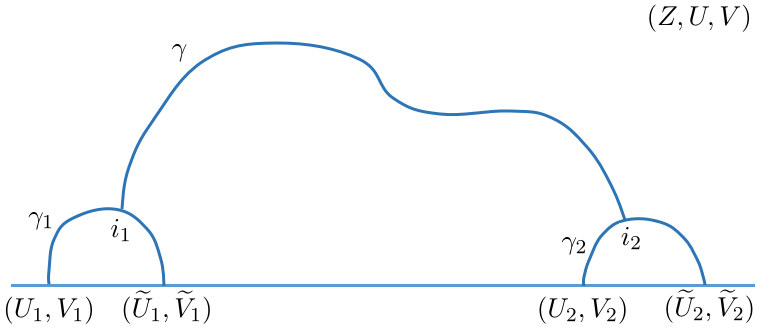}\\
  \caption{The geodesic network for the regularization. Here, the geodesic $\gamma_1$ connects the boundary points $(U_1,V_1)$, $(\widetilde{U}_1,\widetilde{V}_1)$, the geodesic $\gamma_2$ connects the boundary points $(U_2,V_2)$, $(\widetilde{U}_2,\widetilde{V}_2)$, and the geodesic $\gamma^{(r)}$ orthogonally intersects with the geodesics $\gamma_1$, $\gamma_2$ at $i_1$, $i_2$ respectively.}\label{EWCS}
\end{figure}

We now provide a regularization for the computation of the bracket $\{A,\zeta\}$, where we also take a regularization for the geodesic $\gamma$, the geodesic length $A$, and the twist $\zeta$, and where the main idea is to regularize the geodesic $\gamma$ to one which no long ends up at the asymptotic boundary.
In particular, we consider the geodesic network in Fig.\ref{EWCS}, where the geodesic $\gamma_1$ connects the boundary points $(U_1,V_1)$, $(\widetilde{U}_1,\widetilde{V}_1)$, the geodesic $\gamma_2$ connects the boundary points $(U_2,V_2)$, $(\widetilde{U}_2,\widetilde{V}_2)$, the geodesic $\gamma^{(r)}$ orthogonally intersects with the geodesics $\gamma_1$, $\gamma_2$ at $i_1$, $i_2$ respectively, and we choose
\begin{align}\label{UVtilde}
&\widetilde{U}_1=U_1-2\epsilon \notag \\
&\widetilde{V}_1=V_1+2\epsilon \notag \\
&\widetilde{U}_2=U_2-2\epsilon \notag \\
&\widetilde{V}_2=V_2+2\epsilon,
\end{align}
for the reason that will be clear below.
In this geodesic network, we view the geodesics $\gamma_1$, $\gamma_2$ as the regularized version of the original geodesic's endpoints, and we view the geodesic $\gamma^{(r)}$ as the regularized version of the original geodesic $\gamma$.
Given this geodesic network, we define the regularized version of the geodesic length and the twist through the geodesic $\gamma^{(r)}$.
Here, we define the regularized version of the geodesic length, denoted by $A^{(r)}$, as the length of $\gamma^{(r)}$.
And we define the regularized version of the twist as the followings.
Similar as the original definition, we introduce two normal frames $(\hat{\tau}^{(r,1)},\hat{n}^{(r,1)})$, $(\hat{\tau}^{(r,2)},\hat{n}^{(r,2)})$ to the geodesic $\gamma^{(r)}$ by the following two steps. First, we construct the normal frames $(\hat{\tau}^{(r,1)},\hat{n}^{(r,1)})$, $(\hat{\tau}^{(r,2)},\hat{n}^{(r,2)})$ at the intersections $i_1$, $i_2$  respectively as
\begin{align}
&n^{(r,1)\mu} \big|_{i_1}=e_1^{\phantom{1}\mu} \big|_{i_1} \notag \\
&\tau^{(r,1)\mu} \big|_{i_1}=-g^{\mu\lambda}\epsilon_{\lambda\nu\rho} e_1^{\phantom{1}\nu} e^{(r)\rho} \big|_{i_1},
\end{align}
and
\begin{align}
&n^{(r,2)\mu} \big|_{i_2}=-e_2^{\phantom{2}\mu} \big|_{i_2} \notag \\
&\tau^{(r,2)\mu} \big|_{i_2}=g^{\mu\lambda} \epsilon_{\lambda\nu\rho} e_2^{\phantom{2}\nu} e^{(r)\rho} \big|_{i_2},
\end{align}
where $e_1^{\phantom{1}\mu}$, $e_2^{\phantom{2}\mu}$, $e^{(r)\mu}$ are the unit tangent vector of $\gamma_1$, $\gamma_2$, $\gamma^{(r)}$ pointing from left to right.
Second, by taking a parallel transport, we extend the definition of the normal frames
$(\hat{\tau}^{(r,1)},\hat{n}^{(r,1)})$, $(\hat{\tau}^{(r,2)},\hat{n}^{(r,2)})$ to the whole geodesic $\gamma^{(r)}$.
Based on the same argument around (\ref{boost}), the two normal frames are different by a relative boost
\begin{align}
&\tau^{(r,2)\mu}-n^{(r,2)\mu} \big|_{\gamma}=e^{\zeta^{(r)}}(\tau^{(r,1)\mu}-n^{(r,1)\mu}) \big|_{\gamma}
\notag \\
&\tau^{(r,2)\mu}+n^{(r,2)\mu} \big|_{\gamma} =
e^{-\zeta^{(r)}} (\tau^{(r,1)\mu}+n^{(r,1)\mu}) \big|_{\gamma},
\end{align}
with a constant rapidity $\zeta^{(r)}$. And we define the regularized version of the twist as this constant rapidity $\zeta^{(r)}$.
Here, we directly give the expressions of $A^{(r)}$, $\zeta^{(r)}$ as\footnote{See \cite{Held:2024bgs,ongoing2} for the derivation of (\ref{Azetar}), (\ref{ArUV}).}
\begin{align}\label{Azetar}
&A^{(r)}=A^{(r)U}+A^{(r)V} \notag \\
&\zeta^{(r)}=A^{(r)U}-A^{(r)V},
\end{align}
where
\begin{align}\label{ArUV}
&A^{(r)U}=\log \left[ \left( \frac{(u_1-u_2)(\tilde{u}_1-\tilde{u}_2)}{(u_1-\tilde{u}_1)(u_2-\tilde{u}_2)} \right)^{\frac{1}{2}}
+\left( \frac{(u_1-\tilde{u}_2)(\tilde{u}_1-u_2)}{(u_1-\tilde{u}_1)(u_2-\tilde{u}_2)} \right)^{\frac{1}{2}} \right] \notag \\
&A^{(r)V}=\log \left[ \left( \frac{(v_2-v_1)(\tilde{v}_2-\tilde{v}_1)}{(\tilde{v}_1-v_1)(\tilde{v}_2-v_2)} \right)^{\frac{1}{2}}
+\left( \frac{(\tilde{v}_2-v_1)(v_2-\tilde{v}_1)}{(\tilde{v}_1-v_1)(\tilde{v}_2-v_2)} \right)^{\frac{1}{2}} \right],
\end{align}
and
\be
\left\{ \begin{array}{l}
u_1\equiv u_{(b)}(U_1) \\
v_1\equiv v_{(b)}(V_1) \end{array} \right.~~~
\left\{ \begin{array}{l}
\tilde{u}_1\equiv u_{(b)}(\widetilde{U}_1) \\
\tilde{v}_1\equiv v_{(b)}(\widetilde{V}_1) \end{array} \right.~~~
\left\{ \begin{array}{l}
u_2\equiv u_{(b)}(U_2) \\
v_2 \equiv v_{(b)}(V_2) \end{array} \right. ~~~
\left\{ \begin{array}{l}
\tilde{u}_2\equiv u_{(b)}(\widetilde{U}_2) \\
\tilde{v}_2\equiv v_{(b)}(\widetilde{V}_2) \end{array} \right. ,
\ee
and remember that $(\widetilde{U}_1,\widetilde{V}_1)$, $(\widetilde{U}_2,\widetilde{V}_2)$ are set as (\ref{UVtilde}).
With these expressions, we can get the following two results.
First, we verify the regularization by showing that
\begin{align}
&A^{(r)}=A \notag \\
&\zeta^{(r)}=\zeta, \end{align}
under the $\epsilon \rightarrow 0$ limit, where $A$, $\zeta$ are given in (\ref{AHRT}), (\ref{twistexpression}) with $A^U$, $A^V$ given in (\ref{AUV}).
Second, we show that the brackets $\{A^{(r)U},A^{(r)U}\}$, $\{A^{(r)V},A^{(r)V}\}$ are well defined and vanish as
\be \{A^{(r)U},A^{(r)U}\}=\{A^{(r)V},A^{(r)V}\}=0.
\ee
Based on these two results, we then regularize the bracket $\{A,\zeta\}$ and show its vanishing as
\be \{A,\zeta\}\equiv \{A^{(r)},\zeta^{(r)}\}=0. \ee

\section{Conclusion and discussion}\label{discussion}

In this paper, we studied the twist along a geodesic for its properties in the Hamiltonian formalism. Specifically, we got the following two results
\begin{itemize}
\item First, we studied the system's evolution generated by the twist, which, under a proper gauge choice, exhibits a relative shift along the geodesic.
\item Second, we showed that the twist commutes with the geodesic length of the same geodesic, which supports that the twist can be a candidate element in the center.
\end{itemize}
In the following, we will give some discussions related to our study.

\subsection{Operators defined in geodesic networks}

We first discuss the relation between our study with our original motivation, the construction of the Hilbert space of gravity.
A generally used approach to construct the Hilbert space in other theories is to take use of the properties of operators.
(For example, in oscillator or in free field theories, the Hilbert space can be constructed in a straightforward way after writing down the creation and annihilation operators.)
However, in gravity, our knowledge of the properties of operators are very limited.
Therefore, before constructing the Hilbert space of gravity, we need to first study some properties of diffeomorphism invariant operators.
The study of HRT-area \cite{Kaplan:2022orm} and the study in this paper are indeed some exploratory attempts in this direction. However, to pursue further, we need to introduce more operators and study their properties.

Partially inspired from the tensor network model of holography \cite{Swingle:2009bg,Swingle:2012wq,Pastawski:2015qua,Hayden:2016cfa}, we would like to focus on operators defined in a geodesic network, where the geodesic network can be viewed as the dual network of the tensor network \cite{Held:2024bgs,Bao:2018pvs,Bao:2019fpq}.
(See Fig.\ref{EWCS} for an example of the geodesic network.)
Given a geodesic network, we can define diffeomorphism invariant operators by measuring the length and the twist along each geodesic segment.
We can furthermore study the properties of these operators including their system's evolutions and their commutators as what we did in this paper.

Holding the commutators of the operators, we may then study the Hilbert space of gravity.
Among all operators, we may have a set of operators commuting with each other. We view these operators as elements in the center and use them to take Hilbert space decomposition. We may also have sets of operators, where the operators of each set form a noncommutative subalgebra. We can study the representations of these subalgebras and use them to construct the Hilbert space.

The final goal in this direction is to study a sufficient fine-grained geodesic network and use it to approximate the gravity in the algebraic aspects.

\subsection{The asymptotic boundary conditions of the pure AdS$_3$ gravity}\label{arguebd}

We now discuss the compatibility between the system's evolutions (\ref{evolutionA}), (\ref{evotwist}) and the asymptotic boundary conditions of the pure AdS$_3$ gravity.
We notice that the system's evolutions (\ref{evolutionA}) and (\ref{evotwist}), generated by the geodesic length $A$ and the twist $\zeta$ respectively, have non-zero behaviors which extend to the asymptotic boundary without an apparent suppression.
We therefore come up with the question whether these system's evolutions (\ref{evolutionA}) and (\ref{evotwist}) are compatible with the asymptotic boundary conditions of the pure AdS$_3$ gravity.\footnote{\label{gaugecompatible}Note that even if we have shown that the system's evolutions (\ref{evolutionA}) and (\ref{evotwist}) are not compatible with a given choice of the asymptotic boundary conditions, it doesn't imply that the system's evolutions generated by the geodesic length $A$ and the twist $\zeta$ are not well defined in the pure AdS$_3$ gravity adopted the given asymptotic boundary conditions.
The system's evolutions represented in (\ref{evolutionA}) and (\ref{evotwist}) are only the ones under a special gauge choice, where we require that in the initial system the Cauchy surface $\Sigma$ contains the geodesic $\gamma$ and we also require that the evolutions on the set of initial data on the Cauchy surface $\Sigma$ only support on the geodesic $\gamma$. The previous incompatibility only means that the gauge choice which leads to (\ref{evolutionA}) and (\ref{evotwist}) is not compatible with the asymptotic boundary conditions. However, we can still use some other gauge choice consistent with the asymptotic boundary conditions, for example the Fefferman-Graham gauge, to describe the system's evolutions generated by the geodesic's length $A$ and the twist $\zeta$. Under such gauge choice, the system's evolutions are still well defined but don't exhibit as (\ref{evolutionA}) and (\ref{evotwist}).}

The widely adopted asymptotic boundary conditions are the ones provided in \cite{Brown:1986nw} as
\begin{align}\label{asy0}
&g_{ZZ}=\frac{1}{Z^2}+{\cal{O}}(Z^0) \notag \\
&g_{ZA}={\cal{O}}(Z) \notag \\
&g_{AB}=\frac{1}{Z^2}\eta_{AB}+{\cal{O}}(Z^0),
\end{align}
where $A$, $B$ run through the boundary coordinates. However, we find that the asymptotic boundary conditions (\ref{asy0}) are not compatible with the system's evolutions (\ref{evolutionA}) and (\ref{evotwist}).
In particular, we do the following computation. We start from the vacuum solution which is compatible with the asymptotic boundary conditions (\ref{asy0}). We act the system's evolutions (\ref{evolutionA}) and (\ref{evotwist}) on the vacuum solution. And we find that the evolutions on $g_{ZA}$ have the following asymptotic behavior
\be\label{DeltagZA} \Delta g_{ZA}={\cal{O}}(\frac{1}{Z}), \ee
which is not compatible with the asymptotic boundary conditions of the corresponding components, namely the second equation of (\ref{asy0}).
(See appendix \ref{incompatibility} for more detail.)

To discuss the system's evolutions (\ref{evolutionA}) and (\ref{evotwist}) in a rigorous way, we suggest to weaken the asymptotic behavior of $g_{ZA}$ and adopt the following asymptotic boundary conditions
\begin{align}\label{asy}
&g_{ZZ}=\frac{1}{Z^2}+{\cal{O}}(Z^0) \notag \\
&g_{ZA}={\cal{O}}(\frac{1}{Z}) \notag \\
&g_{AB}=\frac{1}{Z^2} \eta_{AB}+{\cal{O}}(Z^0).
\end{align}
Here, the asymptotic boundary conditions (\ref{asy}) are equivalent to (\ref{asy0}) after removing the extra gauge redundancies by taking some non-physical diffeomorphism.
Up to now, we haven't found any incompatibility between the system's evolutions (\ref{evolutionA}), (\ref{evotwist}) and the asymptotic boundary conditions (\ref{asy}).\footnote{There is a potential question here: since the two choices of the asymptotic boundary conditions (\ref{asy0}), (\ref{asy}) are equivalent to each other, how can the system's evolutions (\ref{evolutionA}), (\ref{evotwist}) being compatible with only one of the asymptotic boundary condition but not the other one? The answer to this question is in footnote \ref{gaugecompatible}.}
\footnote{There is another choice of the asymptotic boundary conditions provided in \cite{Compere:2007az} as
\begin{align}\label{asy1}
&g_{ZZ}\sim \frac{1}{Z^2}-\frac{1}{Z}g_1+{\cal{O}}(Z^0) \notag \\
&g_{ZA}\sim {\cal{O}}(\frac{1}{Z}) \notag \\
&g_{AB}\sim \frac{1}{Z^2} \eta_{AB}+\frac{1}{Z}g_1\eta_{AB}+{\cal{O}}(Z^0),
\end{align}
where $g_1$ is an arbitrary function of the boundary coordinates. The asymptotic boundary conditions (\ref{asy1}), though even weaker than (\ref{asy}), are still equivalent to (\ref{asy0}) and (\ref{asy}) after removing some gauge redundancies. We believe that the asymptotic boundary conditions (\ref{asy1}) are also compatible with the system's evolutions (\ref{evolutionA}), (\ref{evotwist}). But for simplicity, we still prefer to adopt the asymptotic boundary conditions (\ref{asy}) in the future application.}

\subsection{Some technical questions for the canonical formalism}\label{discussioncan}

We now raise some technical questions for the canonical formalism.
Comparing the two approaches used in this paper, we believe that the canonical formalism has its advantages in illustrating the geometric interpretations and in generalizing to other systems.
However, we admit that there are still some technical issues in the canonical formalism.
In particular, we raise two technical questions in the following and leave them for the future study.

\subsubsection{The brackets with the boundary stress tensor}

The first question is about the computation of the brackets of the boundary stress tensor with the geodesic length $A$ and with the twist $\zeta$.
In the covariant phase space formalism, after we get the expressions of the geodesic length $A$ and the twist $\zeta$, the computation is straightforward. (See (\ref{bracketTzeta}) for an example.)
However, in the canonical formalism, we haven't achieved such a computation.
In principle, the computation can be done by applying the system's evolutions (\ref{evolutionA}) and (\ref{evotwist}), generated by the geodesic length $A$ and the twist $\zeta$, on the boundary stress tensor.
However, to perform such a computation in practice, we need to first figure out the expression of the boundary stress tensor under the asymptotic boundary conditions (\ref{asy}), which to our knowledge is not totally clear.\footnote{We may try to evaluate the boundary stress tensor with the following expression
\be\label{BrownYork} T_{AB}=\lim_{\epsilon\rightarrow 0} \frac{1}{4G} (\bar{K}_{AB}-\bar{K}\gamma_{AB}+\gamma_{AB}), \ee
where $\gamma_{AB}$ and $\bar{K}_{AB}$ denote the induced metric and the extrinsic curvature on the cutoff surface $Z=\epsilon$.
Indeed the expression (\ref{BrownYork}) gives the correct result under the Fefferman-Graham gauge.
But its validity for the general solution with the asymptotic boundary conditions (\ref{asy}) requires an explicit check.}
Regarding the value of this question, we admit that it is only to reproduce a known result derived with the covariant phase space formalism.
But we still think the question is worth studying: we view it as a stress test of the capability of the canonical formalism approach.

\subsubsection{The bracket $\{A,\zeta\}$}\label{bracketAzeta}

The second question is about the regularization in the computation of the bracket $\{A,\zeta\}$.
In the computation of the bracket $\{A,\zeta\}$ in (\ref{braAzeta}), the last equation is not very rigorous and requires a further regularization.
Here, the main problem is that both $\delta A$ in (\ref{deltaA}) and $\delta \zeta$ in (\ref{deltazetafinal}) support on the same lower dimensional submanifold of the Cauchy surface namely the geodesic $\gamma$.
And we suggest to regularize $\delta A$ and $\delta \zeta$ by adding terms proportional to the variation of the constraints, such that the regularized version of $\delta A$ and $\delta \zeta$ are two dimensional integrals with finite integrands.
\footnote{This suggestion is inspired from the derivations of the ADM energy with the covariant phase space formalism \cite{Iyer:1994ys,Wald:1999wa,Harlow:2019yfa} and with the canonical formalism \cite{Brown:1986nw,Regge:1974zd}. Following the derivations step by step, the ADM energy, in an intermediate step, can be expressed as a summation of a boundary term and a bulk term. The boundary term is the widely used surface integral for the expression of the ADM energy. And the bulk term is a bulk integral of the equations of motion in the covariant phase formalism and a bulk integral of the constraints in the canonical formalism. If we are only interested in evaluating the ADM energy for a given solution, we can freely ignore the bulk term without losing anything. But if we want to take a variation of the ADM energy, for example to check the derivation or to compute the brackets of the ADM energy with other given observables, we'd better include the bulk term. The variation of the ADM energy with the bulk term included has a more regular behavior. In particular, such variation can be expressed purely as a bulk integral under a proper gauge choice at the near boundary region.} We will provide more details in the following.

Here, we directly give the regularized observables corresponding to $A$ and $\zeta$ as
\begin{align}
&\widetilde{A}=A-8\pi G \int_{\Sigma} d^2 x \rho \theta(\rho) {\cal{H}} \notag \\
&\widetilde{\zeta}=\zeta-8\pi G  \int_{\Sigma} d^2 x \theta(-\rho) {\cal{H}}_m e^m
+8\pi G \int_{\Sigma} d^2 x \rho \theta(-\rho) e^m n^n K_{mn} {\cal{H}},
\end{align}
where ${\cal{H}}$, ${\cal{H}}_a$ are constraints given in (\ref{constraints}). 
And we now explain in which sense the observables $\widetilde{A}$ and $\widetilde{\zeta}$ regularize $A$ and $\zeta$. 
First, since $A$ and $\zeta$ are diffeomorphism invariant observables, the bracket of the regularized observables $\{\widetilde{A},\widetilde{\zeta}\}$ is the same as the bracket of the original ones $\{A,\zeta\}$, namely
\be \{\widetilde{A},\widetilde{\zeta}\}=\{A,\zeta\}. \ee
We can therefore use the bracket of the regularized observables $\{\widetilde{A},\widetilde{\zeta}\}$ to represent the bracket of the original ones $\{A,\zeta\}$. 
Second, the observables $\widetilde{A}$ and $\widetilde{\zeta}$ indeed have more regular behaviors. 
Specifically, by taking a variation of $\widetilde{A}$ and $\widetilde{\zeta}$, we get the following expressions
\begin{align}\label{dtilde}
\delta \widetilde{A}=&
\int_{\Sigma} d^2x \sqrt{\sigma}
\bigg\{ \theta(\rho) \Big[\rho (K^{a}_{\phantom{a}m} K^{bm}-KK^{ab}-\frac{1}{4}\widetilde{R} \sigma^{ab})
+(\frac{1}{2} D^a D^b \rho-\frac{1}{2} D^mD_{m} \rho \sigma^{ab}) \Big] \delta \sigma_{ab} \notag \\
&+\rho \theta(\rho) (-K^{ab}+K\sigma^{ab}) \delta K_{ab} \notag \\
&+D_{m}\Big[ \theta(\rho) \Big(
\frac{1}{2} \rho \sigma^{am} D^b \delta \sigma_{ab}
-\frac{1}{2} D^a \rho \sigma^{bm} \delta \sigma_{ab}
-\frac{1}{2} \rho D^m(\sigma^{ab} \delta \sigma_{ab})
+\frac{1}{2} D^m \rho \sigma^{ab} \delta \sigma_{ab} \Big) \Big] \bigg\} \notag \\
\delta \widetilde{\zeta}=&\int_{\Sigma} d^2 x\sqrt{\sigma}
\bigg\{\theta(-\rho) \Big[
\big( \frac{1}{2}e^mD_mK^{ab}-\frac{1}{2}D^nK_{mn}e^m\sigma^{ab}
-\frac{1}{2}D_me^m K^{ab}+D^ae^m K^{b}_{\phantom{b}m} -\frac{1}{2}K_{mn}D^me^n \sigma^{ab} \big) \notag \\
&+\rho e^m n^n K_{mn} \big(-K^{ap}K^b_{\phantom{b}p}+KK^{ab}+\frac{1}{4} \widetilde{R}\sigma^{ab} \big) \notag \\
&+\big(-\frac{1}{2}D^aD^b(\rho e^m n^n K_{mn})+\frac{1}{2}D^pD_p(\rho e^mn^nK_{mn}) \sigma^{ab} \big) \Big] \delta \sigma_{ab} \notag \\
&+\theta(-\rho) \Big[ (-D^ae^b+D_me^m\sigma^{ab})+\rho e^mn^nK_{mn}(K^{ab}-K\sigma^{ab}) \Big] \delta K_{ab} \notag \\
&+D_m \Big[ \theta(-\rho) \Big(
-\delta \sigma_{ab}K^a_{\phantom{a}n} e^n \sigma^{bm}+\frac{1}{2}\sigma^{ab} \delta \sigma_{ab}K^m_{\phantom{m}n}e^n
+\frac{1}{2}K^{ab}\delta \sigma_{ab}e^m \notag \\
&-\frac{1}{2} \rho e^n n^{p} K_{np}D^b \delta \sigma_{ab}\sigma^{am}
+\frac{1}{2} D^a(\rho e^nn^pK_{np})\delta \sigma_{ab} \sigma^{bm} \notag \\
&+\frac{1}{2}\rho e^n n^p K_{np} D^m(\sigma^{ab} \delta \sigma_{ab})
-\frac{1}{2} D^m(\rho e^n n^p K_{np}) \sigma^{ab} \delta \sigma_{ab} \notag \\
&+\delta K_{ab}e^a \sigma^{bm}-\sigma^{ab} \delta K_{ab} e^m \Big) \Big]  \bigg\},
\end{align}
where we have taken several integration by parts and used
(\ref{constraintsequ}), (\ref{constraints}), (\ref{Riemann}), (\ref{relationongeo}). 
We see that, up to total derivative terms, the expressions of $\delta \widetilde{A}$, $\delta \widetilde{\zeta}$ in (\ref{dtilde}) are indeed two dimensional integrals of $\delta \sigma_{ab}$, $\delta K_{ab}$ with finite coefficients.

We now try to use the expressions (\ref{dtilde}) to argue that the bracket $\{\widetilde{A},\widetilde{\zeta}\}$ should vanish.
In an intuitive level, we make the following argument.
By taking an integration by parts, we first rewrite the total derivative terms in (\ref{dtilde}) as boundary terms. After this integration by parts, we see that $\delta \widetilde{A}$ only supports at the region with $\rho>0$ and $\delta \zeta$ only supports at the region with $\rho<0$.
And we then get the conclusion that the bracket $\{\widetilde{A},\widetilde{\zeta}\}$ vanishes.
However, there is a subtlety related to the boundary terms. 
Based on the experience in studying the ADM energy in \cite{Brown:1986nw,Regge:1974zd}, we may further require that the boundary terms from the integration by parts vanishes. 
For the part of the boundary away from the geodesic's endpoints, this requirement may be achieved by properly choosing the near boundary behavior of $\rho$, $e^a$, $n^a$. (So far, we have only made requirement for $\rho$, $e^a$, $n^a$ at the near geodesic region, which is to represent $\delta A$ (\ref{deltaA}) and $\delta \zeta$ (\ref{deltazetafinal}). We therefore have a lot of freedom for their near boundary behavior.) 
While, for the part of the boundary around the geodesic's endpoints, we may need a further regularization, for example to regularize the geodesic $\gamma$ to Fig.\ref{EWCS}. We leave this subtlety to the future study.

As for the value of this question, we admit that regularizing a zero result may not be very interesting.
But the regularization technique may have more applications, for example to compute the brackets between the segment lengths in the geodesic network in Fig.\ref{EWCS}.
(It was shown in \cite{Held:2024bgs,ongoing2}, with the covariant phase space formalism, that some of these brackets are non-zero.)

\subsection{The generalization to the gravitational systems with bulk degrees of freedom}

We now comment about the generalization of our study to the gravitational systems with bulk degrees of freedom.
In this paper, as an exploratory attempt, we restrict the study to the pure AdS$_3$ gravity, where the twist may be the only non-trivial observable which is defined on a geodesic and different from the geodesic length.
While, in future, we should generalize the discussion to the gravitational systems with bulk degrees of freedom (for example the systems with matter fields coupled or the systems in higher dimension) and study the operators/observables defined on the HRT-surface.
The difficulty in the generalization is that there are too many independent operators/observables defined on the HRT-surface, and it is not clear how to exhaust them or which ones of them are more important.
We don't have concrete idea to deal with this difficulty. But one attempt may be to start from some easily constructed operators/observables and use their commutators/brackets to generate the other operators/observables.

\subsection{The generalization to the gravitational systems with boundaries at finite distance}

We now comment about the generalization to the gravitational systems with boundaries at finite distance.
The gravitational systems with boundaries at finite distance \cite{Marolf:2012dr,Andrade:2015qea,Andrade:2015gja} are interesting models, especially considering the recent developments on their field theory correspondences
\cite{McGough:2016lol,Kraus:2018xrn,Guica:2019nzm,Caputa:2020lpa,Kraus:2021cwf,Hartman:2018tkw}.
Given this fact, it is therefore interesting to generalize the previous study on the action of the HRT-area \cite{Bousso:2019dxk,Bousso:2020yxi,Kaplan:2022orm} and our current study to the gravitational systems with boundaries at finite distance.

One observation in such generalization is an extra discontinuity in the set of initial data on the Cauchy surface at the intersection of the HRT-surface and the spatial boundary. We now explain it in the pure AdS$_3$ gravity with the results of this paper. We notice that the evolutions (\ref{evolutionA}), (\ref{evotwist}), for the set of initial data on the Cauchy surface $\Sigma$ embedded in the bulk, have singular behaviors on the geodesic extending to the spatial boundary.
While, the induced metric and the extrinsic curvature of the boundary of the Cauchy surface $\partial \Sigma$ embedded in the spatial boundary are regular and fixed.
Though the former and the latter are not precisely the same quantity, their different behaviors still indicate a discontinuity in the set of initial data on the Cauchy surface $\Sigma$ at the intersection of the geodesic with the spatial boundary.

Given the previous observation, one interesting question is to figure out the origin of the discontinuity of the set of initial data.
Here, the set of initial data with discontinuity is generated by an observable. And the more precise question is to figure out which specific property of the observable generates this discontinuity.

\subsection{The field theory dual of the twist}

We now comment about the field theory dual of the twist $\zeta$.
From the JLMS formula \cite{Jafferis:2014lza,Jafferis:2015del,Dong:2016eik}, we know that the HRT-area is dual to the modular Hamiltonian in field theory, in the sense of taking the classical limit and restricting to the amplitudes of the code subspace \cite{Harlow:2016vwg,Almheiri:2014lwa}.
Comparing the expressions of the geodesic length $A$ in (\ref{AHRT}) and the twist $\zeta$ in (\ref{twistexpression}), we suggest that the twist $\zeta$ may correspond to a conjugation of the modular Hamiltonian, where the conjugation is a specific action on the left moving modes which changes the signature of $A^V$, and the correspondence is in the same sense of the HRT-area.
However, we don't know any further properties of the conjugation or if the conjugation really exists. We leave these questions for the future study.

\subsection{The twist along a closed geodesic}

We now discuss the generalization to the twist along a closed geodesic.
So far, we have only focused on the twist along a geodesic anchored on the asymptotic boundary.
However, it is interesting to generalize the discussion to the twist along a closed geodesic.
In particular, we can consider a system with two asymptotic boundaries. And we consider the set of solutions with the two asymptotic boundaries being connected, for example the eternal wormhole or the fluctuations beyond it.
For such solutions, we do have a closed geodesic going around the non-contractible cycle. (For the eternal wormhole, this curve is the bifurcated horizon.) And we can then define the twist along this closed geodesic.
Specifically, we can introduce a normal frame $(\hat{\tau},\hat{n})$ on the geodesic, which satisfies the parallel transport conditions and is multi-valued. With the multi-valued normal frame, we can then define the twist in the following equation
\begin{align}
&\hat{\tau}(s+A)-\hat{n}(s+A) \big|_{\gamma}=e^{\zeta}(\hat{\tau}(s)-\hat{n}(s)) \big|_{\gamma}
\notag \\
&\hat{\tau}(s+A)+\hat{n}(s+A) \big|_{\gamma} =
e^{-\zeta} (\hat{\tau}(s)+\hat{n}(s)) \big|_{\gamma},
\end{align}
where $A$ is the length of the closed geodesic, and $\zeta$ is the twist along the closed geodesic.
We can also study the properties of the twist along the closed geodesic.
Actually, our canonical formalism approach can be directly applied to this study. And our two results work the same for the twist along the closed geodesic.

\section*{Acknowledgments}

We would like to thank for the useful discussions with Bin Chen, Jin Chen, Zhe Feng, Zongkuan Guo, Molly Kaplan, Jianxin Lu, Donald Marolf, Rongxin Miao, Tao Shi, Hao Song, Wei Song, Yu Tian, Huajia Wang, Shao-jiang Wang, Qiang Wen, Zhenbin Yang, Hong Zhang, Hongbao Zhang, Yang Zhou.
J.q.W would like to thank the organizers and the participants of the workshop ``Black hole and quantum entanglement'' held by International Centre for Theoretical Physics Asia-Pacific, University of Chinese Academy of Sciences, the workshop ``Entanglement and emergent spacetime'' held by Tianjin University, the workshop ``Frontiers of quantum gravity'' in Institute for Advanced Study at Tsinghua University, the ``Beijing Osaka String/Gravity/Black Hole Workshop'' in Kavli Institute for Theoretical Sciences, Chinese Academy of Sciences, and the ``East Asia Joint Workshop on Fields and Strings 2023'' held by the Interdisciplinary Center for Theoretical Study at the University of Science and Technology of China and the Peng Huanwu Center for Fundamental Theory. J.q.W would like to thank for the hospitality from the Interdisciplinary Center for Theoretical Study at the University of Science and Technology China and the Peng Huanwu Center for Fundamental Theory.

X.s.W. and J.q.W. are supported by the National Natural Science Foundation of China (NSFC) Project No.12047503.


\appendix

\section{Two results in terms of the set of initial data}\label{ADMframe}

In this appendix, we introduce two results used in the main context in terms of the set of initial data $(\sigma_{ab},K_{ab})$.

In order to introduce these two results, we first review the definitions of the induced metric, the extrinsic curvature, and some other relevant quantities appearing in these two results. Here, the definitions are in the framework of the ADM decomposition. \footnote{See \cite{Gourgoulhon:2007ue} for a pedagogical review of the ADM decomposition.}
In more detail, we foliate the spacetime by Cauchy surfaces $\Sigma_{t}$ parameterized by a time parameter $t$.
We introduce a coordinate system $x^{\mu}=(t,x^a)$ adapted to the foliation which specifies a time shift vector
\be \hat{T}=\frac{\partial}{\partial t}. \ee
We define the future-pointing normal vector $\tau^{\mu}$, the induced metric $\sigma_{\mu\nu}$, the lapse $N$, and the shift $\beta^{\mu}$ as
\begin{align}\label{defADM} \tau_{\mu}=&-N\partial_{\mu}t \notag \\
\sigma_{\mu\nu}=&g_{\mu\nu}+\tau_{\mu}\tau_{\nu} \notag \\
N=&\frac{1}{\sqrt{-g^{\alpha\beta}\partial_{\alpha}t\partial_{\beta}t}} \notag \\
\beta^{\mu}=&T^{\mu}-N\tau^{\mu}.
\end{align}
And we also define the extrinsic curvature and the acceleration as
\begin{align}\label{defKa} &K_{\mu\nu}\equiv \sigma_{\mu}^{\phantom{\mu}\rho}\nabla_{\rho}\tau_{\nu}
=\sigma_{\nu}^{\phantom{\nu}\rho}\nabla_{\rho}\tau_{\mu} \notag \\
&a_{\mu}\equiv \tau^{\rho}\nabla_{\rho}\tau_{\mu}=\frac{1}{N}D_{\mu}N. \end{align}
For clarity, we provide the expressions of the metric $g_{\mu\nu}$ and the normal vector $\tau^{\mu}$ in component formalism as
\be\label{ADMmetric} g_{\mu\nu}=\left( \begin{array}{cc}
-N^2+\sigma_{ij}\beta^i\beta^j & \beta_b \\
\beta_a & \sigma_{ab} \end{array} \right)~~~
g^{\mu\nu}=\left( \begin{array}{cc}
-\frac{1}{N^2} & \frac{\beta^b}{N^2} \\
\frac{\beta^a}{N^2} & \sigma^{ab}-\frac{\beta^a\beta^b}{N^2} \end{array} \right),
\ee
and
\be \tau_{\mu}=(-N,0)~~~\tau^{\mu}=(\frac{1}{N},-\frac{\beta^a}{N}). \ee

Having reviewed the previous definitions, we now introduce the two results in terms of the set of initial data $(\sigma_{ab},K_{ab})$. Note that, when we refer to the induced metric and the extrinsic curvature in the set of initial data, we only mean their spatial components.

The first result is about the variation of the set of initial data with respect to the variation of the metric.
In practice, we compute the following two quantities
\be \sigma_{\alpha}^{\phantom{\alpha}\mu} \sigma_{\beta}^{\phantom{\beta}\nu} \delta \sigma_{\mu\nu}
~~~\sigma_{\alpha}^{\phantom{\alpha}\mu} \sigma_{\beta}^{\phantom{\beta}\nu} \delta K_{\mu\nu}, \ee
whose spatial components are exactly what we want
\begin{align}
&\delta \sigma_{ab}=\sigma_{a}^{\phantom{a}\mu} \sigma_{b}^{\phantom{b}\nu} \delta \sigma_{\mu\nu} \notag \\
&\delta K_{ab}=\sigma_{a}^{\phantom{a}\mu} \sigma_{b}^{\phantom{b}\nu} \delta K_{\mu\nu}. \end{align}
With a concrete computation, we get
\begin{align}\label{deltasigmaK}
\sigma_{\alpha}^{\phantom{\alpha}\mu} \sigma_{\beta}^{\phantom{\beta}\nu} \delta \sigma_{\mu\nu}
=&\sigma_{\alpha}^{\phantom{\alpha}\mu} \sigma_{\beta}^{\phantom{\beta}\nu} \delta g_{\mu\nu} \notag \\
\sigma_{\alpha}^{\phantom{\alpha}\mu} \sigma_{\beta}^{\phantom{\beta}\nu} \delta K_{\mu\nu}=&
\frac{1}{2} \sigma_{\alpha}^{\phantom{\alpha}\mu} \sigma_{\beta}^{\phantom{\beta}\nu} \tau^{\rho}
\nabla_{\rho}(\sigma_{\mu}^{\phantom{\mu}\sigma} \sigma_{\nu}^{\phantom{\nu}\lambda} \delta g_{\sigma\lambda})
+\frac{1}{2} K_{\alpha}^{\phantom{\alpha}\mu} \sigma_{\beta}^{\phantom{\beta}\nu} \delta g_{\mu\nu}
+\frac{1}{2} K_{\beta}^{\phantom{\beta}\mu} \sigma_{\alpha}^{\phantom{\alpha} \nu} \delta g_{\mu\nu} \notag \\
&-\frac{1}{2} D_{\alpha}(\sigma_{\beta}^{\phantom{\beta}\mu} \tau^{\nu} \delta g_{\mu\nu})
-\frac{1}{2} D_{\beta}(\sigma_{\alpha}^{\phantom{\alpha}\mu} \tau^{\nu} \delta g_{\mu\nu}) \notag \\
&-\frac{1}{2N}D_{\alpha}N \sigma_{\beta}^{\phantom{\beta}\mu} \tau^{\nu} \delta g_{\mu\nu}
-\frac{1}{2N} D_{\beta}N \sigma_{\alpha}^{\phantom{\alpha} \mu} \tau^{\nu} \delta g_{\mu\nu}
+\frac{1}{2} K_{\alpha\beta} \tau^{\mu} \tau^{\nu} \delta g_{\mu\nu}.
\end{align}
Here, we have represented the expressions in the form of the ADM decomposition.
In deriving (\ref{deltasigmaK}), we have used (\ref{defADM}), (\ref{defKa}), and
\be \delta \tau_{\mu}=-\frac{1}{2} \tau_{\mu} \tau^{\alpha}\tau^{\beta} \delta g_{\alpha\beta}, \ee
which can be read out from $\delta (g^{\mu\nu}\tau_{\mu}\tau_{\nu})=0$.

The second result is about the action of a diffeomorphism on the set of initial data $(\sigma_{ab},K_{ab})$.
Here, we consider the following infinitesimal diffeomorphism
\be \Delta_{\xi} g_{\mu\nu}=\nabla_{\mu}\xi_{\nu}+\nabla_{\nu}\xi_{\mu}, \ee
but we keep the foliation being fixed.
Under such a diffeomorphism, we would like to compute the transformation of the set of initial data $(\Delta_{\xi}\sigma_{ab},\Delta_{\xi}K_{ab})$.
Again, we transform the computation to the following two quantities
\be \sigma_{\alpha}^{\phantom{\alpha}\mu} \sigma_{\beta}^{\phantom{\beta}\nu} \Delta_{\xi} \sigma_{\mu\nu}~~~
\sigma_{\alpha}^{\phantom{\alpha}\mu} \sigma_{\beta}^{\phantom{\beta}\nu} \Delta_{\xi} K_{\mu\nu}, \ee
whose spatial components are exactly what we want
\begin{align} &\Delta_{\xi} \sigma_{ab}=\sigma_{a}^{\phantom{a}\mu} \sigma_{b}^{\phantom{b}\nu} \Delta_{\xi} \sigma_{\mu\nu} \notag \\
&\Delta_{\xi} K_{ab}=\sigma_{a}^{\phantom{a}\mu} \sigma_{b}^{\phantom{b}\nu} \Delta_{\xi} K_{\mu\nu}.
\end{align}
With a concrete computation, we get
\begin{align}\label{diff}
\sigma_{\alpha}^{\phantom{\alpha} \mu} \sigma_{\beta}^{\phantom{\beta}\nu} \Delta_{\xi} \sigma_{\mu\nu}
=&D_{\alpha}(\sigma_{\beta\mu}\xi^{\mu})+D_{\beta}(\sigma_{\alpha\mu}\xi^{\mu})-2K_{\alpha\beta}\tau_{\mu}\xi^{\mu} \notag \\
\sigma_{\alpha}^{\phantom{\alpha}\mu} \sigma_{\beta}^{\phantom{\beta}\nu} \Delta_{\xi} K_{\mu\nu}=&
K_{\alpha\nu} D_{\beta}(\sigma^{\nu}_{\phantom{\nu}\mu}\xi^{\mu})
+K_{\beta\nu} D_{\alpha}(\sigma^{\nu}_{\phantom{\nu}\mu}\xi^{\mu})+D_{\nu}K_{\alpha\beta} \sigma^{\nu}_{\phantom{\nu}\mu}\xi^{\mu}
 \notag \\
&-D_{\alpha}D_{\beta}(\tau_{\mu}\xi^{\mu})
+\left(-\sigma_{\alpha}^{\phantom{\alpha}\nu}\sigma_{\beta}^{\phantom{\beta}\rho}
\tau^{\sigma}\nabla_{\sigma}K_{\nu\rho}
-2K_{\alpha\nu}K_{\beta}^{\phantom{\beta}\nu}
+\frac{1}{N}D_{\alpha}D_{\beta}N \right) \tau_{\mu}\xi^{\mu}.
\end{align}
Here, we have again represented the expressions in the form of the ADM decomposition. In deriving (\ref{diff}), we have used (\ref{defADM}), (\ref{defKa}). We have used
\be\label{deltaxi} \Delta_{\xi} \tau_{\mu}
=\left(\frac{1}{N} D_{\nu}N \sigma^{\nu}_{\phantom{\nu}\rho}\xi^{\rho}-\tau^{\nu}\nabla_{\nu}(\tau_{\rho}\xi^{\rho}) \right) \tau_{\mu},
\ee
which can be read out from $\Delta_{\xi} (g_{\mu\nu}\tau^{\mu}\tau^{\nu})=0$.
We have used
\begin{align}\label{deltaGamma}
\Delta_{\xi} \Gamma_{\mu\nu}^{\rho}=&
\frac{1}{2} g^{\rho\sigma} \big( \nabla_{\mu}\Delta_{\xi} g_{\nu\sigma}+\nabla_{\nu} \Delta_{\xi} g_{\mu\sigma}
-\nabla_{\sigma} \Delta_{\xi} g_{\mu\nu} \big) \notag \\
=& \frac{1}{2} g^{\rho\sigma}
\big( \nabla_{\mu}\nabla_{\nu}\xi_{\sigma}+\nabla_{\nu}\nabla_{\mu}\xi_{\sigma}
+R_{\mu\sigma\nu\lambda}\xi^{\lambda}
+R_{\nu\sigma\mu\lambda}\xi^{\lambda} \big).
\end{align}
And we have also used the ADM decomposition of the Riemann tensor
\begin{align}\label{ADM}
\sigma_{\alpha}^{\phantom{\alpha}\mu} \sigma_{\beta}^{\phantom{\beta}\nu} \sigma_{\gamma}^{\phantom{\gamma}\rho}
\sigma_{\delta}^{\phantom{\delta}\sigma} R_{\mu\nu\rho\sigma}=&
\widetilde{R}_{\alpha\beta\gamma\delta}+K_{\alpha\gamma}K_{\beta\delta}-K_{\beta\gamma}K_{\alpha\delta} \notag \\
\sigma_{\alpha}^{\phantom{\alpha}\mu} \sigma_{\beta}^{\phantom{\beta}\nu} \sigma_{\gamma}^{\phantom{\gamma}\rho}
\tau^{\sigma} R_{\mu\nu\rho\sigma}=&D_{\alpha}K_{\beta\gamma}-D_{\beta}K_{\alpha\gamma} \notag \\
\sigma_{\alpha}^{\phantom{\alpha}\mu} \tau^{\nu} \sigma_{\gamma}^{\phantom{\gamma}\rho} \tau^{\sigma}
R_{\mu\nu\rho\sigma}=&
-K_{\alpha}^{\phantom{\alpha}\mu} K_{\gamma\mu}
-\sigma_{\alpha}^{\phantom{\alpha}\mu} \sigma_{\gamma}^{\phantom{\gamma}\rho} \tau^{\lambda}
\nabla_{\lambda}K_{\mu\rho} +\frac{1}{N} D_{\alpha}D_{\gamma}N,
\end{align}
where $\widetilde{R}_{\alpha\beta\gamma\delta}$ denotes the Riemann curvature of the induced metric.

\section{Canonical formalism}\label{can}

In this appendix, we recast the pure AdS$_3$ gravity into the canonical formalism. During this procedure, we introduce the set of initial data, the Poisson bracket, and the constraints, which are used in the main context and in other appendices.
\footnote{See \cite{Dirac,Henneaux:1992ig} for a pedagogical introduction of the canonical formalism for the systems with constraints.}

To apply the canonical formalism, we first rewrite the action in a proper form.
Specifically, we represent the metric in the framework of the ADM decomposition as in (\ref{ADMmetric}).
And by applying (\ref{ADMmetric}) to the action (\ref{action}), by taking use of (\ref{ADM}), and by taking a proper integration by parts, we can rewrite the action as
\be\label{actioncanonical} S=\int dtd^2x {\cal{L}}+\mbox{boundary terms},
\ee
with the Lagrangian density ${\cal{L}}$ being
\be\label{Ldensity} {\cal{L}}=\frac{1}{16\pi G}
N \sqrt{\sigma} \left( \widetilde{R}+K_{ab}K^{ab}-K^2+2 \right),
\ee
where $\widetilde{R}$ is the Ricci scalar of the induced metric $\sigma_{ab}$, and $K_{ab}$ is the spatial components of the extrinsic curvature with the following expression
\be\label{Kab} K_{ab}=\frac{1}{2N}\left( \partial_t \sigma_{ab}-D_a\beta_b-D_b\beta_a \right). \ee
The action (\ref{actioncanonical}) is now in a proper form to apply the canonical formalism.
Specifically, the Lagrangian density (\ref{Ldensity}) depends on $\sigma_{ab}$, $\dot{\sigma}_{ab}$, $N$, $\beta^a$. And we thus view $\sigma_{ab}$ as a dynamical field, and $N$, $\beta^a$ as Lagrangian multipliers.

We now recast the action (\ref{actioncanonical}) into the canonical formalism.
Following the canonical formalism, we define the conjugate momentum of $\sigma_{ab}$ as
\be\label{piab} \pi^{ab}\equiv \frac{\partial {\cal{L}}}{\partial \dot{\sigma}_{ab}(t,x)}=\frac{1}{16\pi G} \sqrt{\sigma}(K^{ab}-K\sigma^{ab}). \ee
We view the canonical pair $(\sigma_{ab},\pi^{ab})$ together as a set of initial data.
We introduce the following Poisson bracket for $(\sigma_{ab},\pi^{ab})$ as
\begin{align}\label{Poisson1}
&\{\sigma_{ab}(t,x_1),\sigma_{cd}(t,x_2)\}=0 \notag \\
&\{\sigma_{ab}(t,x_1),\pi^{cd}(t,x_2)\}=
\frac{1}{2}\left( \delta_a^{\phantom{a}c} \delta_{b}^{\phantom{b}d}
+\delta_a^{\phantom{a}d}\delta_b^{\phantom{b}c} \right) \delta^2(x_1-x_2) \notag \\
&\{\pi^{ab}(t,x_1),\pi^{cd}(t,x_2)\}=0.
\end{align}
We define the Hamiltonian as
\begin{align}\label{Hamiltonian}
H=&\int d^2x (\pi^{ab}\dot{\sigma}_{ab}-{\cal{L}})+\mbox{boundary terms} \notag \\
=&\int d^2x N{\cal{H}}+\beta^a{\cal{H}}_a+\mbox{boundary terms},
\end{align}
with
\begin{align}\label{constraints0}
&{\cal{H}}=16\pi G \frac{1}{\sqrt{\sigma}}(\pi^{ab}\pi_{ab}-\pi^2)-
\frac{1}{16\pi G} \sqrt{\sigma} (\widetilde{R}+2)  \notag \\
&{\cal{H}}_a=-2\sqrt{\sigma}D^b(\frac{1}{\sqrt{\sigma}}\pi_{ab}),
\end{align}
where $\pi=\sigma_{ab}\pi^{ab}$, and the covariant derivative is acted as if $\frac{1}{\sqrt{\sigma}}\pi_{ab}$ is a tensor.
And we point out that the ${\cal{H}}$, ${\cal{H}}_a$ appearing in (\ref{Hamiltonian}) are actually constraints which satisfy
\begin{align}\label{constraintsequ}
&{\cal{H}}=0 \notag \\
&{\cal{H}}_a=0. \end{align}
So far, we have recast the pure $AdS_3$ gravity into the canonical formalism, which is specified by the set of initial data $(\sigma_{ab},\pi^{ab})$, the Poisson bracket (\ref{Poisson1}), the Hamiltonian (\ref{Hamiltonian}), and the constraints (\ref{constraints0}), (\ref{constraintsequ}).

For the applications in other places, we equivalently represent the set of initial data as $(\sigma_{ab},K_{ab})$.
With this set of initial data $(\sigma_{ab},K_{ab})$, we represent the Poisson bracket (\ref{Poisson1}) as
\begin{align}\label{Poisson2}
\{\sigma_{ab}(t,x_1),\sigma_{cd}(t,x_2)\}&=0 \notag \\
\{\sigma_{ab}(t,x_1),K_{cd}(t,x_2)\}&=
8\pi G (\sigma_{ac}\sigma_{bd}+\sigma_{ad}\sigma_{bc}-2\sigma_{ab}\sigma_{cd})
\frac{1}{\sqrt{\sigma}} \delta^2(x_1-x_2) \notag \\
\{K_{ab}(t,x_1),K_{cd}(t,x_2)\}&=
8\pi G (\sigma_{ab}K_{cd}-K_{ab}\sigma_{cd}) \frac{1}{\sqrt{\sigma}} \delta^2(x_1-x_2), \end{align}
and the constraints (\ref{constraints0}) as
\begin{align}\label{constraints}
&{\cal{H}}=\frac{1}{16\pi G} \sqrt{\sigma} (K^{ab}K_{ab}-K^2-\widetilde{R}-2)  \notag \\
&{\cal{H}}_a=-\frac{1}{8\pi G} \sqrt{\sigma} D^b (K_{ab}-K\sigma_{ab}).
\end{align}

\section{The derivation for the equations of the variation of the geodesic and the normal frames}\label{deriving}

In this appendix, we provide a detailed derivation for the equations of the variation of the geodesic and the normal frames.

First, we derive the equation of the variation of the geodesic.
As we know, the geodesic equation is
\be\label{geoAppendix} \frac{d^2 x^{\mu}}{ds^2}+\Gamma^{\mu}_{\nu\rho}(x(s))\frac{dx^{\nu}}{ds} \frac{dx^{\rho}}{ds}=0. \ee
By taking a variation, we get
\be\label{geovar0} \frac{d^2}{ds^2} \delta x^{\mu}
+\partial_{\lambda}\Gamma^{\mu}_{\nu\rho}(x(s)) \delta x^{\lambda}(s) \frac{dx^{\nu}}{ds} \frac{dx^{\rho}}{ds}
+\delta_{g} \Gamma^{\mu}_{\nu\rho}(x(s)) \frac{dx^{\nu}}{ds} \frac{dx^{\rho}}{ds}
+2\Gamma^{\mu}_{\nu\rho}(x(s)) \frac{dx^{\nu}}{ds} \frac{d}{ds} \delta x^{\rho}=0, \ee
where $\delta_g \Gamma^{\mu}_{\nu\rho}$ denotes the variation of the Christoffel symbol generated by the variation of the metric, which equals to
\be \delta_g \Gamma^{\mu}_{\nu\rho}=\frac{1}{2} g^{\mu\sigma}
(\nabla_{\nu}\delta g_{\rho\sigma}+\nabla_{\rho} \delta g_{\nu\sigma}-\nabla_{\sigma} \delta g_{\nu\rho}). \ee
We now rewrite (\ref{geovar0}) into a covariant form.
We need the following results, which are the action of covariant derivatives on $\delta x^{\mu}$ along the tangent direction, as
\be e^{\alpha}\nabla_{\alpha} \delta x^{\mu}=\frac{d}{ds} \delta x^{\mu}
+\Gamma^{\mu}_{\nu\rho} \frac{dx^{\nu}}{ds} \delta x^{\rho}, \ee
and
\begin{align}\label{twoderivative}
e^{\alpha}\nabla_{\alpha}(e^{\beta} \nabla_{\beta} \delta x^{\mu})=&
\frac{d^2}{ds^2} \delta x^{\mu}+2\Gamma^{\mu}_{\nu\rho} \frac{dx^{\nu}}{ds} \frac{d}{ds} \delta x^{\rho}
+\partial_{\sigma}\Gamma^{\mu}_{\nu\rho} \frac{d x^{\nu}}{ds} \frac{d x^{\sigma}}{ds} \delta x^{\rho}  \notag \\
&+\Gamma^{\mu}_{\nu\rho} \frac{d^2 x^{\nu}}{ds^2} \delta x^{\rho}
+\Gamma^{\mu}_{\nu\lambda} \Gamma^{\lambda}_{\rho\sigma} \frac{dx^{\nu}}{ds} \frac{dx^{\rho}}{ds}\delta x^{\sigma}.
\end{align}
By applying (\ref{twoderivative}) to (\ref{geovar0}) and taking use of (\ref{geoAppendix}), we get the equation of the variation of the geodesic as
\be e^{\alpha}\nabla_{\alpha}(e^{\beta}\nabla_{\beta} \delta x^{\mu})
+R_{\nu\alpha\phantom{\mu}\beta}^{\phantom{\nu\alpha}\mu} \delta x^{\nu} e^{\alpha}e^{\beta}
+\delta_g \Gamma^{\mu}_{\alpha\beta} e^{\alpha}e^{\beta}=0. \ee

Second, we derive the equation of the variation of the normal frames.
By definition, the basis of the normal frames satisfy the parallel transport equation
\be \frac{dx^{\alpha}}{ds}\nabla_{\alpha} V^{\mu}=0, \ee
or equivalently written as
\be\label{tranAppendix} \frac{d}{ds} V^{\mu}+\Gamma^{\mu}_{\nu\rho}(x(s)) \frac{dx^{\nu}}{ds} V^{\rho}(s)=0, \ee
where $V^{\mu}$ represents $\tau^{(1)\mu}$, $n^{(1)\mu}$, $\tau^{(2)\mu}$, $n^{(2)\mu}$.
By taking a variation of (\ref{tranAppendix}), we get
\begin{align}\label{varframe0}
&\frac{d}{ds} \delta V^{\mu}
+\partial_{\sigma} \Gamma^{\mu}_{\nu\rho}(x(s)) \delta x^{\sigma}(s) \frac{dx^{\nu}}{ds} V^{\rho}(s)
+\delta_g \Gamma^{\mu}_{\nu\rho}(x(s)) \frac{dx^{\nu}}{ds} V^{\rho}(s) \notag \\
&+\Gamma^{\mu}_{\nu\rho}(x(s)) \frac{d}{ds} \delta x^{\nu} V^{\rho}(s)
+\Gamma^{\mu}_{\nu\rho}(x(s)) \frac{dx^{\nu}}{ds} \delta V^{\rho}(s)=0. \end{align}
We now try to rewrite (\ref{varframe0}) into a covariant form.
We first notice that $V^{\mu}(s)+\delta V^{\mu}(s)$ and $V^{\mu}(s)$ are defined at different locations, which are at $x^{\mu}(s)+\delta x^{\mu}(s)$ and $x^{\mu}(s)$, respectively.
Therefore, it is reasonable to introduce the following definition
\be \delta^{(c)} V^{\mu}(s)\equiv \delta V^{\mu}(s)+\Gamma^{\mu}_{\nu\rho}(x(s)) \delta x^{\nu}(s) V^{\rho}(s), \ee
which measures the relative difference between $V^{\mu}(s)+\delta V^{\mu}(s)$ and the parallel transported $V^{\mu}(s)$ from $x^{\mu}(s)$ to $x^{\mu}(s)+\delta x^{\mu}(s)$.
Moreover, as the derivation of the equation of the variation of the geodesic, we also act a covariant derive on $\delta^{(c)}V^{\mu}$ along the tangent direction and get
\begin{align}\label{derivative}
e^{\alpha}\nabla_{\alpha} \delta^{(c)}V^{\mu}=&
\frac{d}{ds} \delta V^{\mu}+\partial_{\sigma}\Gamma^{\mu}_{\nu\rho} \frac{dx^{\sigma}}{ds} \delta x^{\nu} V^{\rho}+
\Gamma^{\mu}_{\nu\rho} \frac{d}{ds} \delta x^{\nu} V^{\rho}+\Gamma^{\mu}_{\nu\rho} \delta x^{\nu} \frac{d}{ds}V^{\rho}  \notag \\ &+\Gamma^{\mu}_{\nu\rho} \frac{d x^{\nu}}{ds} \delta V^{\rho}
+\Gamma^{\mu}_{\nu\lambda} \Gamma^{\lambda}_{\rho\sigma} \frac{dx^{\nu}}{ds} \delta x^{\rho} V^{\sigma}.
\end{align}
By applying (\ref{derivative}) to (\ref{varframe0}) and taking use of (\ref{tranAppendix}), we get the equation of the variation of the normal frames as
\be e^{\alpha}\nabla_{\alpha} \delta^{(c)}V^{\mu}
+R_{\nu\alpha\phantom{\mu}\beta}^{\phantom{\nu\alpha}\mu} \delta x^{\nu} e^{\alpha}V^{\beta}
+\delta_g \Gamma^{\mu}_{\alpha\beta} e^{\alpha}V^{\beta}=0. \ee

\section{The solution for the equations of the variation of the geodesic and the normal frames}\label{solequations}

In this appendix, we solve the differential equations (\ref{vargeo}), (\ref{varframe}) for the variation of the geodesic $\delta x^{\mu}$ and the variation of the normal frames $\delta^{(c)}\tau^{(1)\mu}$, $\delta^{(c)} n^{(1)\mu}$, $\delta^{(c)} \tau^{(2)\mu}$, $\delta^{(c)} n^{(2)\mu}$.

We first solve the differential equation (\ref{vargeo}) for $\delta x^{\mu}$.
For convenience, we take a decomposition with respect to the orthonormal frames $(\hat{e},\hat{\tau}^{(i)},\hat{n}^{(i)})$ as
\be\label{deltaxdecom} \delta x^{\mu}(s)=C^{e}(s)e^{\mu}(s)+C^{\tau^{(i)}}(s) \tau^{(i)\mu}(s)+C^{n^{(i)}}(s) n^{(i)\mu}(s), \ee
for $i=1,2$. Here, because of the relation (\ref{boost}), the $C$s satisfy the following relation
\begin{align}\label{boostinC}
&C^{\tau^{(2)}}(s)=\cosh \zeta \cdot C^{\tau^{(1)}}(s)+\sinh \zeta \cdot C^{n^{(1)}}(s) \notag \\
&C^{n^{(2)}}(s)=\sinh \zeta \cdot C^{\tau^{(1)}}(s)+\cosh \zeta \cdot C^{n^{(1)}}(s).
\end{align}
By applying (\ref{deltaxdecom}) to the differential equation (\ref{vargeo}) and also taking a decomposition for the differential equation (\ref{vargeo}) with respect to $(\hat{e},\hat{\tau}^{(i)},\hat{n}^{(i)})$, for $i=1,2$, we get the following differential equations
\begin{align}\label{dxdecom}
&\frac{d^2}{ds^2} C^e+\frac{1}{2}\frac{d}{ds} (\delta g_{\alpha\beta} e^{\alpha}e^{\beta})=0 \notag \\
&\frac{d^2}{ds^2} C^{\tau^{(i)}}-{C^{{\tau ^{(i)}}}}
-\frac{d}{ds} (\delta {g_{\alpha \beta }}{e^\alpha }{\tau ^{(i)\beta }}) + \frac{1}{2}{\tau ^{(i)\gamma }}{\nabla _\gamma }\delta {g_{\alpha \beta }}{e^\alpha }{e^\beta } = 0  \notag \\
&\frac{d^2}{ds^2} C^{n^{(i)}} - {C^{{n^{(i)}}}} + \frac{d}{{ds}}(\delta {g_{\alpha \beta }}{e^\alpha }{n^{(i)\beta }}) - \frac{1}{2}{n^{(i)\gamma }}{\nabla _\gamma }\delta {g_{\alpha \beta }}{e^\alpha }{e^\beta } = 0,
\end{align}
for $i=1,2$. Here, the differential equations in (\ref{dxdecom}) are decoupled from each other, and can be solved as
\begin{align}\label{soldeltax}
C^e(s)=&-\frac{s_2-s}{2({s_2} - {s_1})}
\int_{s_1}^s d\widetilde{s} \delta g_{\alpha \beta}(x(\widetilde s)) e^{\alpha}e^{\beta}
+\frac{s-s_1}{2(s_2-s_1)} \int_s^{s_2} d\widetilde{s} \delta {g_{\alpha \beta }}(x(\widetilde s)) e^{\alpha}e^{\beta} \notag \\
&+\frac{s_2-s}{s_2-s_1} C^{e}(s_1)+\frac{s-s_1}{s_2-s_1} C^{e}(s_2) \notag \\
C^{\tau^{(i)}}(s)=&
\frac{\sinh(s_2-s)}{\sinh(s_2-s_1)} \int_{{s_1}}^s d\widetilde s
\big(\cosh (\widetilde{s} - {s_1})\delta {g_{\alpha \beta }}(x(\widetilde s)){e^\alpha }{\tau ^{(i)\beta }}+
\frac{1}{2}\sinh(\widetilde{s} - {s_1})
\tau ^{(i)\gamma} \nabla_{\gamma} \delta g_{\alpha \beta}(x(\widetilde s)){e^\alpha }{e^\beta } \big) \notag \\
&+\frac{\sinh(s-s_1)}{\sinh(s_2-s_1)}
\int_s^{{s_2}} d\widetilde{s}
\big( -\cosh(s_2-\widetilde{s})\delta {g_{\alpha \beta }}(x(\widetilde s)){e^\alpha }{\tau ^{(i)\beta }}
+\frac{1}{2}\sinh (s_2-\widetilde{s}){\tau ^{(i)\gamma }}
\nabla_{\gamma} \delta g_{\alpha \beta}(x(\widetilde{s})) e^{\alpha}e^{\beta} \big)
\notag \\
&+\frac{\sinh(s_2-s)}{\sinh(s_2-s_1)} C^{\tau^{(i)}}(s_1)
+\frac{\sinh(s-s_1)}{\sinh(s_2-s_1)} C^{\tau^{(i)}}(s_2) \notag \\
{C^{{n^{(i)}}}}(s)=&-\frac{\sinh(s_2-s)}{\sinh ({s_2} - {s_1})}
\int_{s_1}^s d\widetilde{s}
\big(\cosh (\widetilde{s}-s_1) \delta {g_{\alpha \beta }}(x(\widetilde s)){e^\alpha }{n^{(i)\beta }}
+\frac{1}{2}\sinh (\widetilde{s}-s_1)
n^{(i)\gamma} \nabla_{\gamma} \delta {g_{\alpha \beta }}(x(\widetilde s)){e^\alpha }{e^\beta } \big) \notag \\
&+ \frac{{\sinh (s - {s_1})}}{{\sinh ({s_2} - {s_1})}}
\int_s^{s_2} d\widetilde{s} \big(\cosh(s_2-\widetilde{s}) \delta g_{\alpha \beta}(x(\widetilde{s})){e^\alpha }{n^{(i)\beta }}
-\frac{1}{2}\sinh(s_2-\widetilde{s}) n^{(i)\gamma} \nabla_{\gamma} \delta {g_{\alpha \beta }}(x(\widetilde s)){e^\alpha }{e^\beta } \big) \notag \\
&+\frac{\sinh(s_2-s)}{\sinh(s_2-s_1)} C^{n^{(i)}}(s_1)
+\frac{\sinh(s-s_1)}{\sinh(s_2-s_1)} C^{n^{(i)}}(s_2).
\end{align}
Here, the $C$s appear in the right hand side of the equations are the boundary values at $s_1$, $s_2$ of the $C$s in the left hand side.
They also satisfy (\ref{boostinC}), which can be explicitly written as
\begin{align}\label{cobrel}
&C^{\tau^{(2)}}(s_1)=\cosh \zeta \cdot C^{\tau^{(1)}}(s_1)+\sinh \zeta \cdot C^{n^{(1)}}(s_1) \notag \\
&C^{n^{(2)}}(s_1)=\sinh \zeta \cdot C^{\tau^{(1)}}(s_1)+\cosh \zeta \cdot C^{n^{(1)}}(s_1) \notag \\
&C^{\tau^{(2)}}(s_2)=\cosh \zeta \cdot C^{\tau^{(1)}}(s_2)+\sinh \zeta \cdot C^{n^{(1)}}(s_2) \notag \\
&C^{n^{(2)}}(s_2)=\sinh \zeta \cdot C^{\tau^{(1)}}(s_2)+\cosh \zeta \cdot C^{n^{(1)}}(s_2).
\end{align}
These boundary values of the $C$s are nothing else but some linear combinations of the components of $\delta x^{\mu}(s_1)$, $\delta x^{\mu}(s_2)$, which, as mentioned in subsection \ref{cdeltazeta}, are allowed to appear in the solution.

We next solve the differential equations (\ref{varframe}) for $\delta^{(c)} \tau^{(1)\mu}$, $\delta^{(c)} n^{(1)\mu}$, $\delta^{(c)} \tau^{(2)\mu}$, $\delta^{(c)} n^{(2)\mu}$. We also take a decomposition as
\begin{align}\label{decomdtaudn}
&\delta^{(c)} \tau^{(i)\mu}(s)=C_{(\tau^{(i)})}^{e}(s) e^{\mu}(s)
+C_{(\tau^{(i)})}^{\tau^{(i)}}(s) \tau^{(i)\mu}(s)+C_{(\tau^{(i)})}^{n^{(i)}}(s) n^{(i)\mu}(s) \notag \\
&\delta^{(c)} n^{(i)\mu}(s)=C_{(n^{(i)})}^{e}(s) e^{\mu}(s)
+C_{(n^{(i)})}^{\tau^{(i)}}(s) \tau^{(i)\mu}(s)+C_{(n^{(i)})}^{n^{(i)}}(s) n^{(i)\mu}(s),
\end{align}
for $i=1,2$.
By applying (\ref{decomdtaudn}) in (\ref{varframe}), we then get
\begin{align}\label{dframe}
&\frac{d}{{ds}}C_{(\tau^{(i)} )}^e - {C^{\tau^{(i)}} } + \frac{1}{2}{\tau ^{(i)\gamma} }{\nabla _\gamma }\delta {g_{\alpha \beta }}{e^\alpha }{e^\beta } = 0 \notag \\
&\frac{d}{ds}C_{(\tau^{(i)} )}^{\tau^{(i)}}-\frac{1}{2}\frac{d}{ds}(\delta g_{\alpha\beta} \tau ^{(i)\alpha} \tau^{(i)\beta})=0 \notag \\
&\frac{d}{{ds}}C_{(\tau^{(i)} )}^{n^{(i)}}+\frac{1}{2}\frac{d}{ds}(\delta g_{\alpha \beta} \tau^{(i)\alpha} n^{(i)\beta})
+\frac{1}{2}{\tau ^{(i)\gamma} }{\nabla _\gamma }\delta g_{\alpha \beta} e^{\alpha} n^{(i)\beta}
-\frac{1}{2}n^{(i)\gamma} \nabla_{\gamma} \delta g_{\alpha \beta} e^{\alpha} \tau ^{(i)\beta}= 0 \notag \\
&\frac{d}{{ds}}C_{(n^{(i)})}^e+C^{n^{(i)}}+
\frac{1}{2}{n^{(i)\gamma} }{\nabla _\gamma }\delta {g_{\alpha \beta }}{e^\alpha }{e^\beta } = 0 \notag\\
&\frac{d}{{ds}}C_{(n^{(i)})}^{\tau^{(i)}}  - \frac{1}{2}\frac{d}{ds}(\delta {g_{\alpha \beta }}{\tau ^{(i)\alpha} }{n^{(i)\beta} })
+\frac{1}{2}{\tau ^{(i)\gamma} }{\nabla _\gamma }\delta {g_{\alpha \beta }}{e^\alpha }{n^{(i)\beta} }
-\frac{1}{2}{n^{(i)\gamma} }{\nabla _\gamma }\delta {g_{\alpha \beta }} e^{\alpha} \tau ^{(i)\beta}=0 \notag \\
&\frac{d}{{ds}}C_{(n^{(i)})}^{n^{(i)}} + \frac{1}{2}\frac{d}{ds}(\delta {g_{\alpha \beta }}{n^{(i)\alpha} }{n^{(i)\beta} }) = 0,
\end{align}
for $i=1,2$. We can solve these differential equations (\ref{dframe}) as
\begin{align}\label{Cftau10}
C_{(\tau^{(1)})}^{e}(s)=&
\int_{s_1}^{s} d\widetilde{s}
\bigg( \frac{-\cosh(s_2-s)\cosh(\widetilde{s}-s_1)+\cosh(s_2-\widetilde{s})}{\sinh(s_2-s_1)}
\delta g_{\alpha\beta}(x(\widetilde{s})) e^{\alpha} \tau^{(1)\beta} \notag \\
&-\frac{\cosh(s_2-s)\sinh(\widetilde{s}-s_1)+\sinh(s_2-\widetilde{s})}{2\sinh(s_2-s_1)}
\tau^{(1)\gamma}\nabla_{\gamma} \delta g_{\alpha\beta}(x(\widetilde{s})) e^{\alpha}e^{\beta} \bigg) \notag \\
&+\frac{\cosh(s-s_1)-1}{\sinh(s_2-s_1)} \int_s^{s_2} d\widetilde{s}
\big(-\cosh(s_2-\widetilde{s}) \delta g_{\alpha\beta}(x(\widetilde{s}))e^{\alpha}\tau^{(1)\beta} \notag \\
&+\frac{1}{2} \sinh(s_2-\widetilde{s}) \tau^{(1)\gamma}\nabla_{\gamma}\delta g_{\alpha\beta}(x(\widetilde{s}))e^{\alpha}e^{\beta} \big)
\notag \\
&+C_{(\tau^{(1)})}^e(s_1)
+\frac{\cosh(s_2-s_1)-\cosh(s_2-s)}{\sinh(s_2-s_1)} C^{\tau^{(1)}}(s_1)
+\frac{\cosh(s-s_1)-1}{\sinh(s_2-s_1)} C^{\tau^{(1)}}(s_2) \notag \\
C_{(\tau^{(1)})}^{\tau^{(1)}}(s)=&
\frac{1}{2} \delta g_{\alpha\beta}(x(s))\tau^{(1)\alpha}\tau^{(1)\beta}
-\frac{1}{2} \delta g_{\alpha\beta}(x(s_1))\tau^{(1)\alpha}\tau^{(1)\beta}
+C_{(\tau^{(1)})}^{\tau^{(1)}}(s_1) \notag \\
C_{(\tau^{(1)})}^{n^{(1)}}(s)=&
\int_{s_1}^{s} d\widetilde{s}
\big(-\frac{1}{2}\tau^{(1)\gamma}\nabla_{\gamma} \delta g_{\alpha\beta}(x(\widetilde{s})) e^{\alpha} n^{(1)\beta}
+\frac{1}{2} n^{(1)\gamma}\nabla_{\gamma} \delta g_{\alpha\beta} (x(\widetilde{s})) e^{\alpha} \tau^{(1)\beta} \big) \notag \\
&-\frac{1}{2} \delta g_{\alpha\beta}(x(s))\tau^{(1)\alpha} n^{(1)\beta}
+\frac{1}{2} \delta g_{\alpha\beta}(x(s_1))\tau^{(1)\alpha} n^{(1)\beta}
+C_{(\tau^{(1)})}^{n^{(1)}}(s_1),
\end{align}
for $\delta^{(c)} \tau^{(1)\mu}$,
\begin{align}\label{Cfn10}
C_{(n^{(1)})}^{e}(s)=&
\int_{s_1}^s d\widetilde{s}
\bigg( \frac{-\cosh(s_2-s)\cosh(\widetilde{s}-s_1)+\cosh(s_2-\widetilde{s})}{\sinh(s_2-s_1)}
\delta g_{\alpha\beta}(x(\widetilde{s})) e^{\alpha} n^{(1)\beta} \notag \\
&-\frac{\cosh(s_2-s)\sinh(\widetilde{s}-s_1)+\sinh(s_2-\widetilde{s})}{2\sinh(s_2-s_1)}
n^{(1)\gamma}\nabla_{\gamma} \delta g_{\alpha\beta}(x(\widetilde{s}))e^{\alpha}e^{\beta} \bigg)  \notag \\
&+\frac{\cosh(s-s_1)-1}{\sinh(s_2-s_1)}
\int_{s}^{s_2} d\widetilde{s}
\big( -\cosh(s_2-\widetilde{s}) \delta g_{\alpha\beta}(x(\widetilde{s})) e^{\alpha} n^{(1)\beta} \notag \\
&+\frac{1}{2} \sinh(s_2-\widetilde{s}) n^{(1)\gamma}\nabla_{\gamma} \delta g_{\alpha\beta} (x(\widetilde{s})) e^{\alpha} e^{\beta} \big) \notag \\
&+C_{(n^{(1)})}^{e}(s_1)
-\frac{\cosh(s_2-s_1)-\cosh(s_2-s)}{\sinh(s_2-s_1)} C^{n^{(1)}}(s_1)
-\frac{\cosh(s-s_1)-1}{\sinh(s_2-s_1)} C^{n^{(1)}}(s_2) \notag \\
C_{(n^{(1)})}^{\tau^{(1)}}(s)=&
\int_{s_1}^{s} d\widetilde{s}
\big(-\frac{1}{2} \tau^{(1)\gamma} \nabla_{\gamma} \delta g_{\alpha\beta} (x(\widetilde{s})) e^{\alpha} n^{(1)\beta}
+\frac{1}{2} n^{(1)\gamma} \nabla_{\gamma} \delta g_{\alpha\beta} (x(\widetilde{s})) e^{\alpha} \tau^{(1)\beta} \big) \notag \\
&+\frac{1}{2} \delta g_{\alpha\beta}(x(s)) \tau^{(1)\alpha} n^{(1)\beta}
-\frac{1}{2} \delta g_{\alpha\beta}(x(s_1)) \tau^{(1)\alpha} n^{(1)\beta}
+C_{(n^{(1)})}^{\tau^{(1)}}(s_1) \notag \\
C_{(n^{(1)})}^{n^{(1)}}(s)=&
-\frac{1}{2} \delta g_{\alpha\beta}(x(s)) n^{(1)\alpha} n^{(1)\beta}
+\frac{1}{2} \delta g_{\alpha\beta}(x(s_1)) n^{(1)\alpha} n^{(1)\beta}
+C_{(n^{(1)})}^{n^{(1)}}(s_1),
\end{align}
for $\delta^{(c)}n^{(1)\mu}$,
\begin{align}\label{Cftau20}
C_{(\tau^{(2)})}^{e}(s)=&
-\frac{\cosh(s_2-s)-1}{\sinh(s_2-s_1)}
\int_{s_1}^{s} d\widetilde{s}
\big( \cosh(\widetilde{s}-s_1) \delta g_{\alpha\beta}(x(\widetilde{s})) e^{\alpha} \tau^{(2)\beta} \notag \\
&+\frac{1}{2} \sinh(\widetilde{s}-s_1) \tau^{(2)\gamma} \nabla_{\gamma} \delta g_{\alpha\beta} (x(\widetilde{s})) e^{\alpha} e^{\beta} \big)
\notag \\
&+\int_{s}^{s_2} d\widetilde{s}
\bigg( \frac{\cosh(\widetilde{s}-s_1)-\cosh(s-s_1)\cosh(s_2-\widetilde{s})}{\sinh(s_2-s_1)}
\delta g_{\alpha\beta}(x(\widetilde{s})) e^{\alpha} \tau^{(2)\beta} \notag \\
&+\frac{\sinh(\widetilde{s}-s_1)+\cosh(s-s_1)\sinh(s_2-\widetilde{s})}{2\sinh(s_2-s_1)}
\tau^{(2)\gamma} \nabla_{\gamma} \delta g_{\alpha\beta} (x(\widetilde{s})) e^{\alpha} e^{\beta} \bigg) \notag \\
&+C_{(\tau^{(2)})}^{e}(s_2)
-\frac{\cosh(s_2-s)-1}{\sinh(s_2-s_1)} C^{\tau^{(2)}}(s_1)
-\frac{\cosh(s_2-s_1)-\cosh(s-s_1)}{\sinh(s_2-s_1)} C^{\tau^{(2)}}(s_2) \notag \\
C_{(\tau^{(2)})}^{\tau^{(2)}}(s)=&
\frac{1}{2} \delta g_{\alpha\beta}(x(s))\tau^{(2)\alpha} \tau^{(2)\beta}
-\frac{1}{2} \delta g_{\alpha\beta}(x(s_2)) \tau^{(2)\alpha} \tau^{(2)\beta}
+C_{(\tau^{(2)})}^{\tau^{(2)}}(s_2) \notag \\
C_{(\tau^{(2)})}^{n^{(2)}}(s)=&
\int_{s}^{s_2} d\widetilde{s}
\big( \frac{1}{2} \tau^{(2)\gamma} \nabla_{\gamma} \delta g_{\alpha\beta}(x(\widetilde{s})) e^{\alpha} n^{(2)\beta}
-\frac{1}{2} n^{(2)\gamma} \nabla_{\gamma} \delta g_{\alpha\beta} (x(\widetilde{s})) e^{\alpha} \tau^{(2)\beta} \big) \notag \\
&-\frac{1}{2} \delta g_{\alpha\beta}(x(s)) \tau^{(2)\alpha} n^{(2)\beta}
+\frac{1}{2} \delta g_{\alpha\beta}(x(s_2)) \tau^{(2)\alpha} n^{(2)\beta}
+C_{(\tau^{(2)})}^{n^{(2)}}(s_2),
\end{align}
for $\delta^{(c)}\tau^{(2)\mu}$, and
\begin{align}\label{Cfn20}
C_{(n^{(2)})}^{e}(s)=&
-\frac{\cosh(s_2-s)-1}{\sinh(s_2-s_1)}
\int_{s_1}^{s} d\widetilde{s}
\big( \cosh(\widetilde{s}-s_1) \delta g_{\alpha\beta}(x(\widetilde{s}))e^{\alpha} n^{(2)\beta} \notag \\
&+\frac{1}{2} \sinh(\widetilde{s}-s_1) n^{(2)\gamma} \nabla_{\gamma} \delta g_{\alpha\beta} (x(\widetilde{s})) e^{\alpha} e^{\beta} \big) \notag \\
&+\int_{s}^{s_2} d\widetilde{s}
\bigg( \frac{\cosh(\widetilde{s}-s_1)-\cosh(s-s_1)\cosh(s_2-\widetilde{s})}{\sinh(s_2-s_1)}
\delta g_{\alpha\beta}(x(\widetilde{s})) e^{\alpha} n^{(2)\beta} \notag \\
&+\frac{\sinh(\widetilde{s}-s_1)+\cosh(s-s_1)\sinh(s_2-\widetilde{s})}{2\sinh(s_2-s_1)}
n^{(2)\gamma} \nabla_{\gamma} \delta g_{\alpha\beta} (x(\widetilde{s})) e^{\alpha} e^{\beta} \bigg) \notag \\
&+C_{(n^{(2)})}^{e}(s_2)+\frac{\cosh(s_2-s)-1}{\sinh(s_2-s_1)} C^{n^{(2)}}(s_1)
+\frac{\cosh(s_2-s_1)-\cosh(s-s_1)}{\sinh(s_2-s_1)} C^{n^{(2)}}(s_2) \notag \\
C_{(n^{(2)})}^{\tau^{(2)}}(s)=&
\int_{s}^{s_2} d\widetilde{s} \big( \frac{1}{2} \tau^{(2)\gamma} \nabla_{\gamma} \delta g_{\alpha\beta} (x(\widetilde{s})) e^{\alpha} n^{(2)\beta}
-\frac{1}{2} n^{(2)\gamma} \nabla_{\gamma} \delta g_{\alpha\beta} (x(\widetilde{s})) e^{\alpha} \tau^{(2)\beta} \big) \notag \\
&+\frac{1}{2} \delta g_{\alpha\beta}(x(s)) \tau^{(2)\alpha} n^{(2)\beta}
-\frac{1}{2} \delta g_{\alpha\beta}(x(s_2)) \tau^{(2)\alpha} n^{(2)\beta}
+C_{(n^{(2)})}^{\tau^{(2)}}(s_2) \notag \\
C_{(n^{(2)})}^{n^{(2)}}(s)=&
-\frac{1}{2} \delta g_{\alpha\beta}(x(s)) n^{(2)\alpha} n^{(2)\beta}
+\frac{1}{2} \delta g_{\alpha\beta}(x(s_2)) n^{(2)\alpha} n^{(2)\beta} +C_{(n^{(2)})}^{n^{(2)}}(s_2),
\end{align}
for $\delta^{(c)} n^{(2)\mu}$.
Here, the boundary values of the $C$s, that appear in the right hand side of the equations in (\ref{Cftau10}), (\ref{Cfn10}), (\ref{Cftau20}), (\ref{Cfn20}), are some linear combinations of the components of $\delta x^{\mu}(s_1)$, $\delta x^{\mu}(s_2)$, $\delta^{(c)} \tau^{(1)\mu}(s_1)$, $\delta^{(c)} n^{(1)\mu}(s_1)$, $\delta^{(c)} \tau^{(2)\mu}(s_2)$, $\delta^{(c)} n^{(2)\mu}(s_2)$, which, as mentioned in subsection \ref{cdeltazeta}, are allowed to appear in the solution. Moreover, by applying the orthonormal conditions (\ref{orthonormal}), we can get some relations for these boundary values of the $C$s, which we will explain in the following two paragraphs.

We will need $\delta^{(c)} e^{\mu}$ in deriving the relations for the boundary values of the $C$s. So we now compute $\delta^{(c)}e^{\mu}$ from $\delta x^{\mu}$.
We first read out $\delta^{(c)} \frac{d x^{\mu}}{ds}$ from $\delta x^{\mu}$ as
\begin{align} \label{deltadxds0} \delta^{(c)} \frac{dx^{\mu}}{ds}=&
\delta \frac{d}{ds}  x^{\mu}+\Gamma^{\mu}_{\nu\rho} \delta x^{\nu} \frac{dx^{\rho}}{ds} \notag \\
=&e^{\nu}\partial_{\nu} \delta x^{\mu}+\Gamma^{\mu}_{\nu\rho}\delta x^{\nu} e^{\rho} \notag \\
=&e^{\nu} \nabla_{\nu} \delta x^{\mu} \notag \\
=&\frac{d}{ds} C^{e} e^{\mu}+\frac{d}{ds} C^{\tau^{(i)}} \tau^{(i)\mu}+\frac{d}{ds} C^{n^{(i)}} n^{(i)\mu},
\end{align}
where, in the second equation, we have used
\be e^{\mu}=\frac{dx^{\mu}}{ds}. \ee
We can then read out $\delta^{(c)}e^{\mu}$ from $\delta^{(c)} \frac{dx^{\mu}}{ds}$.
Remember that we have chosen the parameter $s$ to be the proper length up to a shift for the geodesic $x^{\mu}(s)$ in the unvaried metric $g_{\mu\nu}$ and to be the affine parameter for the geodesic $x^{\mu}(s)+\delta x^{\mu}(s)$ in the varied metric $g_{\mu\nu}+\delta g_{\mu\nu}$, so we have
\begin{align}\label{dxds}
&\frac{d}{ds}x^{\mu}=e^{\mu} \notag \\
&\frac{d}{ds}(x^{\mu}+\delta x^{\mu})=(1+\delta A_0) (e^{\mu}+\delta e^{\mu}),
\end{align}
where $\delta A_0$ is a constant along the geodesic. From the two equations in (\ref{dxds}), we can derive
\be\label{deltadxds} \delta^{(c)} \frac{dx^{\mu}}{ds}=\delta A_0 e^{\mu}+\delta^{(c)} e^{\mu}. \ee
The decomposition in the right hand side of (\ref{deltadxds}) can be uniquely fixed by the normalization of the tangent vector.
Specifically, by taking a variation on the normalization condition of the tangent vector
\be g_{\mu\nu}(x(s))e^{\mu}(s)e^{\nu}(s)=1, \ee
we have
\begin{align}\label{variationunit}
0=&\delta g_{\mu\nu}(x(s))e^{\mu}(s)e^{\nu}(s)
+\delta x^{\rho}(s) \partial_{\rho}  g_{\mu\nu} (x(s)) e^{\mu}(s) e^{\nu}(s)
+2g_{\mu\nu}(x(s)) \delta e^{\mu}(s) e^{\nu}(s) \notag \\
=&\delta g_{\mu\nu}(x(s))e^{\mu}(s)e^{\nu}(s)+2\delta^{(c)}e^{\mu}(s) e_{\mu}(s).
\end{align}
From (\ref{deltadxds0}), (\ref{deltadxds}), (\ref{variationunit}) together with (\ref{soldeltax}), we then read out
\begin{align}
\delta A_0=&
\left(\frac{1}{2(s_2-s_1)} \int_{s_1}^{s_2}d\widetilde{s} \delta g_{\alpha\beta}(x(\widetilde{s})) e^{\alpha} e^{\beta}\right)
-\frac{1}{s_2-s_1}C^{e}(s_1)+\frac{1}{s_2-s_1}C^{e}(s_2),
\end{align}
and
\be \delta^{(c)}e^{\mu}=C_{(e)}^e e^{\mu}+C_{(e)}^{\tau^{(i)}} \tau^{(i)\mu}+C_{(e)}^{n^{(i)}} n^{(i)\mu}, \ee
where
\begin{align}\label{Cfe}
C_{(e)}^{e}(s)=&-\frac{1}{2} \delta g_{\alpha\beta}(x(s)) e^{\alpha} e^{\beta} \notag \\
C_{(e)}^{\tau^{(i)}}(s)=&
-\frac{\cosh(s_2-s)}{\sinh(s_2-s_1)}
\int_{s_1}^{s} d\widetilde{s}
\big( \cosh(\widetilde{s}-s_1) \delta g_{\alpha\beta}(x(\widetilde{s})) e^{\alpha} \tau^{(i)\beta}
+\frac{1}{2} \sinh(\widetilde{s}-s_1) \tau^{(i)\gamma} \nabla_{\gamma} \delta g_{\alpha\beta} (x(\widetilde{s})) e^{\alpha}e^{\beta} \big)
\notag \\
&+\frac{\cosh(s-s_1)}{\sinh(s_2-s_1)}
\int_{s}^{s_2} d\widetilde{s}
\big( -\cosh(s_2-\widetilde{s}) \delta g_{\alpha\beta}(x(\widetilde{s})) e^{\alpha} \tau^{(i)\beta}
+\frac{1}{2} \sinh(s_2-\widetilde{s}) \tau^{(i)\gamma} \nabla_{\gamma} \delta g_{\alpha\beta}(x(\widetilde{s})) e^{\alpha} e^{\beta} \big) \notag \\
&+\delta g_{\alpha\beta}(x(s)) e^{\alpha} \tau^{(i)\beta}
-\frac{\cosh(s_2-s)}{\sinh(s_2-s_1)} C^{\tau^{(i)}}(s_1)
+\frac{\cosh(s-s_1)}{\sinh(s_2-s_1)} C^{\tau^{(i)}}(s_2) \notag \\
C_{(e)}^{n^{(i)}}(s)=&
\frac{\cosh(s_2-s)}{\sinh(s_2-s_1)}
\int_{s_1}^{s} d\widetilde{s}
\big( \cosh(\widetilde{s}-s_1) \delta g_{\alpha\beta}(x(\widetilde{s})) e^{\alpha} n^{(i)\beta}
+\frac{1}{2} \sinh(\widetilde{s}-s_1) n^{(i)\gamma} \nabla_{\gamma} \delta g_{\alpha\beta}(x(\widetilde{s})) e^{\alpha} e^{\beta} \big) \notag \\
&+\frac{\cosh(s-s_1)}{\sinh(s_2-s_1)} \int_{s}^{s_2} d\widetilde{s}
\big( \cosh(s_2-\widetilde{s}) \delta g_{\alpha\beta}(x(\widetilde{s})) e^{\alpha} n^{(i)\beta}
-\frac{1}{2} \sinh(s_2-\widetilde{s}) n^{(i)\gamma} \nabla_{\gamma} \delta g_{\alpha\beta}(x(\widetilde{s})) e^{\alpha} e^{\beta} \big) \notag \\
&-\delta g_{\alpha\beta}(x(s))e^{\alpha} n^{(i)\beta}
-\frac{\cosh(s_2-s)}{\sinh(s_2-s_1)} C^{n^{(i)}}(s_1)+\frac{\cosh(s-s_1)}{\sinh(s_2-s_1)} C^{n^{(i)}}(s_2).
\end{align}

We are now ready to derive the boundary values of the $C$s from the orthonormal conditions (\ref{orthonormal}), where, similar as equation (\ref{variationunit}), we need a variational version of the orthonormal conditions.
By taking a variation for the orthonormal conditions
\begin{align}
&\hat{e}\cdot \hat{\tau}^{(i)}=\hat{e} \cdot \hat{n}^{(i)}=\hat{\tau}^{(i)}\cdot \hat{n}^{(i)}=0 \notag \\
&-\hat{\tau}^{(i)2}=\hat{n}^{(i)2}=1,
\end{align}
we get
\begin{align}\label{varorthonormal0}
&\delta g_{\mu\nu}(x(s))e^{\mu} \tau^{(i)\nu}+\delta^{(c)} e^{\mu} \tau^{(i)}_{\phantom{(i)}\mu}
+\delta^{(c)} \tau^{(i)\mu} e_{\mu}=0 \notag \\
&\delta g_{\mu\nu}(x(s))e^{\mu} n^{(i)\nu}+\delta^{(c)} e^{\mu} n^{(i)}_{\phantom{(i)}\mu}
+\delta^{(c)} n^{(i)\mu} e_{\mu}=0 \notag \\
&\delta g_{\mu\nu}(x(s))\tau^{(i)\mu} n^{(i)\nu}+\delta^{(c)} \tau^{(i)\mu} n^{(i)}_{\phantom{(i)}\mu}
+\delta^{(c)} n^{(i)\mu} \tau^{(i)}_{\phantom{(i)}\mu}=0 \notag \\
&\delta g_{\mu\nu}(x(s)) \tau^{(i)\mu} \tau^{(i)\nu}+2\delta^{(c)}\tau^{(i)\mu} \tau^{(i)}_{\phantom{(i)}\mu}=0 \notag \\
&\delta g_{\mu\nu}(x(s)) n^{(i)\mu} n^{(i)\nu}+2\delta^{(c)}n^{(i)\mu} n^{(i)}_{\phantom{(i)}\mu}=0,
\end{align}
for $i=1,2$, and, by taking use of (\ref{deltaxdecom}), (\ref{decomdtaudn}), we equivalently represent (\ref{varorthonormal0}) as
\begin{align}\label{varorthonormal}
&\delta g_{\alpha\beta}(x(s))e^{\alpha}\tau^{(i)\beta}-C_{(e)}^{\tau^{(i)}}+C_{(\tau^{(i)})}^{e}=0 \notag \\
&\delta g_{\alpha\beta}(x(s))e^{\alpha} n^{(i)\beta}+C_{(e)}^{n^{(i)}}+C_{(n^{(i)})}^{e}=0 \notag \\
&\delta g_{\alpha\beta}(x(s))\tau^{(i)\alpha} n^{(i)\beta}+C_{(\tau^{(i)})}^{n^{(i)}}-C_{(n^{(i)})}^{\tau^{(i)}}=0 \notag \\
&\delta g_{\alpha\beta}(x(s)) \tau^{(i)\alpha} \tau^{(i)\beta}-2C_{(\tau^{(i)})}^{\tau^{(i)}}=0 \notag \\
&\delta g_{\alpha\beta}(x(s)) n^{(i)\alpha} n^{(i)\beta}+2C_{(n^{(i)})}^{n^{(i)}}=0,
\end{align}
for $i=1,2$.
By applying (\ref{Cftau10}), (\ref{Cfn10}), (\ref{Cftau20}), (\ref{Cfn20}), (\ref{Cfe}) in (\ref{varorthonormal}), we get the relations for the boundary values of the $C$s as
\begin{align}\label{Cb1}
C_{(\tau^{(1)})}^{e}(s_1)=&
\frac{1}{\sinh(s_2-s_1)}
\int_{s_1}^{s_2} d\widetilde{s}
\big( -\cosh(s_2-\widetilde{s}) \delta g_{\alpha\beta}(x(\widetilde{s})) e^{\alpha} \tau^{(1)\beta}
+\frac{1}{2} \sinh(s_2-\widetilde{s}) \tau^{(1)\gamma} \nabla_{\gamma} \delta g_{\alpha\beta}(x(\widetilde{s})) e^{\alpha} e^{\beta} \big) \notag \\
&-\frac{\cosh(s_2-s_1)}{\sinh(s_2-s_1)} C^{\tau^{(1)}}(s_1)
+\frac{1}{\sinh(s_2-s_1)} C^{\tau^{(1)}}(s_2) \notag \\
C_{(n^{(1)})}^{e}(s_1)=&
\frac{1}{\sinh(s_2-s_1)}
\int_{s_1}^{s_2} d\widetilde{s}
\big( -\cosh(s_2-\widetilde{s}) \delta g_{\alpha\beta}(x(\widetilde{s})) e^{\alpha} n^{(1)\beta}
+\frac{1}{2}\sinh(s_2-\widetilde{s}) n^{(1)\gamma} \nabla_{\gamma} \delta g_{\alpha\beta}(x(\widetilde{s})) e^{\alpha} e^{\beta} \big) \notag \\
&+\frac{\cosh(s_2-s_1)}{\sinh(s_2-s_1)} C^{n^{(1)}}(s_1)
-\frac{1}{\sinh(s_2-s_1)} C^{n^{(1)}}(s_2) \notag \\
C_{(n^{(1)})}^{\tau^{(1)}}(s_1)=&C_{(\tau^{(1)})}^{n^{(1)}}(s_1)
+\delta g_{\alpha\beta}(x(s_1)) \tau^{(1)\alpha} n^{(1)\beta} \notag \\
C_{(\tau^{(1)})}^{\tau^{(1)}}(s_1)=& \frac{1}{2} \delta g_{\alpha\beta}(x(s_1)) \tau^{(1)\alpha} \tau^{(1)\beta} \notag \\
C_{(n^{(1)})}^{n^{(1)}}(s_1)=&
-\frac{1}{2} \delta g_{\alpha\beta}(x(s_1))n^{(1)\alpha}n^{(1)\beta},
\end{align}
and
\begin{align}\label{Cb2}
C_{(\tau^{(2)})}^{e}(s_2)=&
-\frac{1}{\sinh(s_2-s_1)} \int_{s_1}^{s_2}
d\widetilde{s} \big(
\cosh(\widetilde{s}-s_1) \delta g_{\alpha\beta}(x(\widetilde{s})) e^{\alpha} \tau^{(2)\beta}
+\frac{1}{2} \sinh(\widetilde{s}-s_1) \tau^{(2)\gamma} \nabla_{\gamma} \delta g_{\alpha\beta}(x(\widetilde{s})) e^{\alpha}e^{\beta} \big) \notag \\
&-\frac{1}{\sinh(s_2-s_1)} C^{\tau^{(2)}}(s_1)+\frac{\cosh(s_2-s_1)}{\sinh(s_2-s_1)} C^{\tau^{(2)}}(s_2) \notag \\
C_{(n^{(2)})}^{e}(s_2)=&
-\frac{1}{\sinh(s_2-s_1)} \int_{s_1}^{s_2} d\widetilde{s}
\big( \cosh(\widetilde{s}-s_1) \delta g_{\alpha\beta}(x(\widetilde{s})) e^{\alpha} n^{(2)\beta}
+\frac{1}{2} \sinh(\widetilde{s}-s_1) n^{(2)\gamma} \nabla_{\gamma} \delta g_{\alpha\beta}(x(\widetilde{s})) e^{\alpha} e^{\beta} \big) \notag \\
&+\frac{1}{\sinh(s_2-s_1)} C^{n^{(2)}}(s_1)-\frac{\cosh(s_2-s_1)}{\sinh(s_2-s_1)} C^{n^{(2)}}(s_2) \notag \\
C_{(n^{(2)})}^{\tau^{(2)}}(s_2)=&
C_{(\tau^{(2)})}^{n^{(2)}}(s_2)+\delta g_{\alpha\beta}(x(s_2))\tau^{(2)\alpha} n^{(2)\beta} \notag \\
C_{(\tau^{(2)})}^{\tau^{(2)}}(s_2)=&
\frac{1}{2} \delta g_{\alpha\beta}(x(s_2)) \tau^{(2)\alpha} \tau^{(2)\beta} \notag \\
C_{(n^{(2)})}^{n^{(2)}}(s_2)=&
-\frac{1}{2} \delta g_{\alpha\beta}(x(s_2)) n^{(2)\alpha} n^{(2)\beta}.
\end{align}

We now apply the relations for the boundary values of the $C$s (\ref{Cb1}), (\ref{Cb2}) to (\ref{Cftau10}), (\ref{Cfn10}), (\ref{Cftau20}), (\ref{Cfn20}), from which we get the final expressions of the variation of the normal frames as
\begin{align}\label{Cftau1}
C_{(\tau^{(1)})}^{e}(s)=&
-\frac{\cosh(s_2-s)}{\sinh(s_2-s_1)}
\int_{s_1}^{s} d\widetilde{s} \big( \cosh(\widetilde{s}-s_1) \delta g_{\alpha\beta}(x(\widetilde{s})) e^{\alpha} \tau^{(1)\beta}
+\frac{1}{2} \sinh(\widetilde{s}-s_1) \tau^{(1)\gamma} \nabla_{\gamma} \delta g_{\alpha\beta} (x(\widetilde{s})) e^{\alpha} e^{\beta} \big) \notag \\
&+\frac{\cosh(s-s_1)}{\sinh(s_2-s_1)}
\int_{s}^{s_2} d\widetilde{s}
\big(-\cosh(s_2-\widetilde{s}) \delta g_{\alpha\beta}(x(\widetilde{s})) e^{\alpha} \tau^{(1)\beta}
+\frac{1}{2} \sinh(s_2-\widetilde{s}) \tau^{(1)\gamma} \nabla_{\gamma} \delta g_{\alpha\beta} (x(\widetilde{s})) e^{\alpha} e^{\beta} \big)
\notag \\
&-\frac{\cosh(s_2-s)}{\sinh(s_2-s_1)} C^{\tau^{(1)}}(s_1)
+\frac{\cosh(s-s_1)}{\sinh(s_2-s_1)} C^{\tau^{(1)}}(s_2) \notag \\
C_{(\tau^{(1)})}^{\tau^{(1)}}(s)=& \frac{1}{2} \delta g_{\alpha\beta}(x(s)) \tau^{(1)\alpha} \tau^{(1)\beta} \notag \\
C_{(\tau^{(1)})}^{n^{(1)}}(s)=&
\int_{s_1}^{s} d\widetilde{s} \big(-\frac{1}{2} \tau^{(1)\gamma} \nabla_{\gamma} \delta g_{\alpha\beta}(x(\widetilde{s})) e^{\alpha} n^{(1)\beta}
+\frac{1}{2} n^{(1)\gamma} \nabla_{\gamma} \delta g_{\alpha\beta}(x(\widetilde{s})) e^{\alpha} \tau^{(1)\beta} \big) \notag \\
&-\frac{1}{2} \delta g_{\alpha\beta}(x(s)) \tau^{(1)\alpha} n^{(1)\beta}
+\frac{1}{2} \delta g_{\alpha\beta}(x(s_1)) \tau^{(1)\alpha} n^{(1)\beta}
+C_{(\tau^{(1)})}^{n^{(1)}}(s_1),
\end{align}
for $\delta^{(c)} \tau^{(1)\mu}$,
\begin{align}\label{Cfn1}
C_{(n^{(1)})}^{e}(s)=&
-\frac{\cosh(s_2-s)}{\sinh(s_2-s_1)}
\int_{s_1}^{s} d\widetilde{s}
\big(\cosh(\widetilde{s}-s_1) \delta g_{\alpha\beta}(x(\widetilde{s})) e^{\alpha} n^{(1)\beta}
+\frac{1}{2} \sinh(\widetilde{s}-s_1) n^{(1)\gamma} \nabla_{\gamma} \delta g_{\alpha\beta} (x(\widetilde{s})) e^{\alpha} e^{\beta} \big) \notag \\
&+\frac{\cosh(s-s_1)}{\sinh(s_2-s_1)} \int_{s}^{s_2} d\widetilde{s}
\big(-\cosh(s_2-\widetilde{s}) \delta g_{\alpha\beta}(x(\widetilde{s})) e^{\alpha} n^{(1)\beta}
+\frac{1}{2} \sinh(s_2-\widetilde{s}) n^{(1)\gamma} \nabla_{\gamma} \delta g_{\alpha\beta}(x(\widetilde{s})) e^{\alpha} e^{\beta} \big) \notag \\
&+\frac{\cosh(s_2-s)}{\sinh(s_2-s_1)} C^{n^{(1)}}(s_1)
-\frac{\cosh(s-s_1)}{\sinh(s_2-s_1)} C^{n^{(1)}}(s_2) \notag \\
C_{(n^{(1)})}^{\tau^{(1)}}(s)=&
\int_{s_1}^{s} d\widetilde{s}
\big(-\frac{1}{2} \tau^{(1)\gamma} \nabla_{\gamma} \delta g_{\alpha\beta}(x(\widetilde{s})) e^{\alpha} n^{(1)\beta}
+\frac{1}{2} n^{(1)\gamma} \nabla_{\gamma} \delta g_{\alpha\beta}(x(\widetilde{s})) e^{\alpha} \tau^{(1)\beta} \big) \notag \\
&+\frac{1}{2} \delta g_{\alpha\beta}(x(s)) \tau^{(1)\alpha} n^{(1)\beta}
+\frac{1}{2} \delta g_{\alpha\beta}(x(s_1))\tau^{(1)\alpha} n^{(1)\beta}+C_{(\tau^{(1)})}^{n^{(1)}}(s_1) \notag \\
C_{(n^{(1)})}^{n^{(1)}}(s)=& -\frac{1}{2} \delta g_{\alpha\beta}(x(s)) n^{(1)\alpha} n^{(1)\beta},
\end{align}
for $\delta^{(c)} n^{(1)\mu}$,
\begin{align}\label{Cftau2}
C_{(\tau^{(2)})}^{e}(s)=&
-\frac{\cosh(s_2-s)}{\sinh(s_2-s_1)} \int_{s_1}^{s} d\widetilde{s}
\big( \cosh(\widetilde{s}-s_1) \delta g_{\alpha\beta}(x(\widetilde{s})) e^{\alpha} \tau^{(2)\beta}
+\frac{1}{2} \sinh(\widetilde{s}-s_1) \tau^{(2)\gamma} \nabla_{\gamma} \delta g_{\alpha\beta} (x(\widetilde{s})) e^{\alpha} e^{\beta} \big) \notag \\
&+\frac{\cosh(s-s_1)}{\sinh(s_2-s_1)} \int_{s}^{s_2} d\widetilde{s}
\big( -\cosh(s_2-\widetilde{s}) \delta g_{\alpha\beta}(x(\widetilde{s})) e^{\alpha} \tau^{(2)\beta}
+\frac{1}{2} \sinh(s_2-\widetilde{s})
\tau^{(2)\gamma} \nabla_{\gamma} \delta g_{\alpha\beta} (x(\widetilde{s})) e^{\alpha} e^{\beta} \big) \notag \\
&-\frac{\cosh(s_2-s)}{\sinh(s_2-s_1)} C^{\tau^{(2)}}(s_1)
+\frac{\cosh(s-s_1)}{\sinh(s_2-s_1)} C^{\tau^{(2)}}(s_2) \notag \\
C_{(\tau^{(2)})}^{\tau^{(2)}}(s)=&
\frac{1}{2} \delta g_{\alpha\beta}(x(s)) \tau^{(2)\alpha} \tau^{(2)\beta} \notag \\
C_{(\tau^{(2)})}^{n^{(2)}}(s)=&
\int_{s}^{s_2} d\widetilde{s}
\big( \frac{1}{2} \tau^{(2)\gamma} \nabla_{\gamma} \delta g_{\alpha\beta}(x(\widetilde{s})) e^{\alpha} n^{(2)\beta}
-\frac{1}{2} n^{(2)\gamma} \nabla_{\gamma} \delta g_{\alpha\beta}(x(\widetilde{s})) e^{\alpha} \tau^{(2)\beta} \big) \notag \\
&-\frac{1}{2} \delta g_{\alpha\beta}(x(s))\tau^{(2)\alpha} n^{(2)\beta}
+\frac{1}{2} \delta g_{\alpha\beta}(x(s_2))\tau^{(2)\alpha} n^{(2)\beta} +C_{(\tau^{(2)})}^{n^{(2)}}(s_2),
\end{align}
for $\delta^{(c)}\tau^{(2)\mu}$, and
\begin{align}\label{Cfn2}
C_{(n^{(2)})}^{e}(s)=&
-\frac{\cosh(s_2-s)}{\sinh(s_2-s_1)} \int_{s_1}^{s} d\widetilde{s}
\big( \cosh(\widetilde{s}-s_1) \delta g_{\alpha\beta}(x(\widetilde{s})) e^{\alpha} n^{(2)\beta}
+\frac{1}{2} \sinh(\widetilde{s}-s_1) n^{(2)\gamma} \nabla_{\gamma} \delta g_{\alpha\beta}( x(\widetilde{s})) e^{\alpha} e^{\beta} \big) \notag \\
&+\frac{\cosh(s-s_1)}{\sinh(s_2-s_1)} \int_{s}^{s_2} d\widetilde{s}
\big( -\cosh(s_2-\widetilde{s}) \delta g_{\alpha\beta} (x(\widetilde{s})) e^{\alpha} n^{(2)\beta}
+\frac{1}{2} \sinh(s_2-\widetilde{s}) n^{(2)\gamma} \nabla_{\gamma} \delta g_{\alpha\beta}(x(\widetilde{s})) e^{\alpha} e^{\beta} \big) \notag \\
&+\frac{\cosh(s_2-s)}{\sinh(s_2-s_1)}C^{n^{(2)}}(s_1)
-\frac{\cosh(s-s_1)}{\sinh(s_2-s_1)} C^{n^{(2)}}(s_2) \notag \\
C_{(n^{(2)})}^{\tau^{(2)}}(s)=& \int_{s}^{s_2} d\widetilde{s}
\big( \frac{1}{2} \tau^{(2)\gamma} \nabla_{\gamma} \delta g_{\alpha\beta}(x(\widetilde{s})) e^{\alpha} n^{(2)\beta}
-\frac{1}{2} n^{(2)\gamma} \nabla_{\gamma} \delta g_{\alpha\beta}(x(\widetilde{s})) e^{\alpha} \tau^{(2)\beta} \big) \notag \\
&+\frac{1}{2} \delta g_{\alpha\beta}(x(s)) \tau^{(2)\alpha} n^{(2)\beta}
+\frac{1}{2} \delta g_{\alpha\beta}(x(s_2)) \tau^{(2)\alpha} n^{(2)\beta} +C_{(\tau^{(2)})}^{n^{(2)}}(s_2) \notag \\
C_{(n^{(2)})}^{n^{(2)}}(s)=&-\frac{1}{2} \delta g_{\alpha\beta}(x(s))n^{(2)\alpha} n^{(2)\beta},
\end{align}
for $\delta^{(c)} n^{(2)\mu}$.\footnote{Actually, most of the expressions in (\ref{Cftau1}), (\ref{Cfn1}), (\ref{Cftau2}), (\ref{Cfn2}) can be directly read out from the variational version of the orthonormal conditions (\ref{varorthonormal}). Here, we instead view this shortcut as a crosscheck of our systematical computation.}
For the application in subsection \ref{cdeltazeta}, we also provide the explicit expressions for $C_{(\tau^{(1)})}^{n^{(1)}}(s_1)$, $C_{(\tau^{(2)})}^{n^{(2)}}(s_2)$ as
\begin{align}\label{C1s1C2s2}
&C_{(\tau^{(1)})}^{n^{(1)}}(s_1)=
\delta^{(c)} \tau^{(1)\mu}(s_1) n^{(1)}_{\phantom{(1)}\mu}(s_1) \notag \\
&C_{(\tau^{(2)})}^{n^{(2)}}(s_2)=
\delta^{(c)} \tau^{(2)\mu}(s_2) n^{(2)}_{\phantom{(2)}\mu}(s_2).
\end{align}

\section{The preserving of the constraints under the system's evolution generated by the twist}\label{evoconstraints}

In this appendix, we verify that the system's evolution (\ref{evotwist}), which is generated by the twist, indeed preserves the constraints (\ref{constraintsequ}).
Precisely speaking, our goal is the following. We start from an arbitrary choice of the set of initial data $(\sigma_{ab},K_{ab})$, that satisfies the constraints (\ref{constraintsequ}).
We then take an evolution on the set of initial data in terms of (\ref{evotwist}), and act this evolution on the constraints (\ref{constraints}). We would like to verify that this evolution on the constraints indeed vanishes.

For convenience of the computations, we make a special choice for the quantities $\rho$, $e^a$, $n^a$ in a finite region around the geodesic $\gamma$.
Here, we still use the definition of $\rho$ below equation (\ref{deltaA}); we choose $n^a$ to be
\be\label{drho} n_a=D_a \rho, \ee
which works not only on the geodesic $\gamma$ but also on the finite region;
we then determine $e^a$ from $n^a$ through the orthonormal conditions
\begin{align}\label{orthonormalApp}
&\sigma_{ab}e^ae^b=\sigma_{ab}n^an^b=1 \notag \\
&\sigma_{ab}e^an^b=0. \end{align}
Under such choice of $e^a$, $n^a$, we have more relations for their derivative. Here, we only list two of them for the following discussion, including
\be\label{prerel2}
D_a e^a=0,
\ee
which works for the finite region around the geodesic $\gamma$, and
\be\label{prerel3}
D_ae_b|_{\gamma}=D_an_b|_{\gamma}=0, \ee
which only works on the geodesic $\gamma$.
We can directly check the relations (\ref{prerel2}), (\ref{prerel3}) in the Gaussian normal coordinates, where the induced metric are represented as
\be \sigma_{ab}dx^adx^b=d\rho^2+l(\rho,s)^2ds^2, \ee
with\footnote{In (\ref{lnearzero}), the first equation can be realized by a proper reparametrization of the coordinate $s$, and the second equation is because the curve $\rho=0$ is a geodesic.}
\begin{align}\label{lnearzero}
&l(\rho=0,s)=1 \notag \\
&\partial_{\rho}l(\rho=0,s)=0, \end{align}
and the vectors $e^a$, $n^a$ are represented as
\begin{align}
&e^a=(0,\frac{1}{l(\rho,s)}) \notag \\
&n^a=(1,0). \end{align}

We are now ready to compute the action of the evolution (\ref{evotwist}) on the constraints (\ref{constraints}).

First, we compute the evolution on ${\cal{H}}$ as
\be\label{DeltaH}
\Delta {\cal{H}}=\mbox{I}+\mbox{II}+\mbox{III}+o(\lambda),
\ee
where
\begin{align}\label{termsinH}
\mbox{I}=&\frac{1}{32\pi G} \sqrt{\sigma} (K^{ab}K_{ab}-K^2-\widetilde{R}-2) \sigma^{mn} \Delta \sigma_{mn} \notag \\
\mbox{II}=&\frac{1}{16\pi G} \sqrt{\sigma} \big( (-2K^{am} K^{b}_{\phantom{b}m}+2KK^{ab}+\widetilde{R}^{ab}) \Delta \sigma_{ab}
+2(K^{ab}-K\sigma^{ab}) \Delta K_{ab} \big) \notag \\
\mbox{III}=& \frac{1}{16\pi G} \sqrt{\sigma}
\big(-D^a D^b \Delta \sigma_{ab}+D^2(\sigma^{ab}\Delta \sigma_{ab}) \big),
\end{align}
and the $\Delta \sigma_{ab}$, $\Delta K_{ab}$ in (\ref{termsinH}) should be interpreted as being applied with (\ref{evotwist}).
With some computations, we get
\begin{align}\label{termsinHf} \mbox{I}=&0 \notag \\
\mbox{II}=&o(\lambda) \notag \\
\mbox{III}=&o(\lambda), \end{align}
where we have used (\ref{constraintsequ}), (\ref{constraints}), (\ref{drho}), (\ref{orthonormalApp}), (\ref{prerel2}), (\ref{prerel3}), the last equation of (\ref{relationongeo}), and
\be \widetilde{R}_{ab}=\frac{1}{2} \sigma_{ab} \widetilde{R} \ee
which can be derived by noting that in two dimension the Riemann tensor only has one independent component as
\be \widetilde{R}_{abcd}=\frac{1}{2} (\sigma_{ac}\sigma_{bd}-\sigma_{ad}\sigma_{bc}) \widetilde{R}.
\ee
By applying (\ref{termsinHf}) to (\ref{DeltaH}), we get
\be\label{dHf} \Delta{\cal{H}}=o(\lambda). \ee

Second, we compute the evolution on ${\cal{H}}_a$ as
\be\label{DeltaHa} \Delta {\cal{H}}_a=\mbox{I}+\mbox{II}+\mbox{III}+\mbox{IV}+\mbox{V}+o(\lambda), \ee
where
\begin{align}\label{termsinHa}
\mbox{I}=&-\frac{1}{16\pi G} \sqrt{\sigma} D^b (K_{ab}-K\sigma_{ab}) \sigma^{mn} \Delta \sigma_{mn} \notag \\
\mbox{II}=& \frac{1}{8\pi G} \sqrt{\sigma} \Delta \sigma_{bc} D^c (K_a^{\phantom{a}b}-K\delta_{a}^{\phantom{a}b}) \notag \\
\mbox{III}=&
\frac{1}{8\pi G} \sqrt{\sigma}  D^{b}
(K\Delta \sigma_{ab}-K^{mn}\Delta \sigma_{mn} \sigma_{ab}-\Delta K_{ab}+\sigma^{mn} \Delta K_{mn} \sigma_{ab}) \notag \\
\mbox{IV}=& \frac{1}{8\pi G}\sqrt{\sigma} \sigma^{bc} \Delta \widetilde{\Gamma}_{ca}^m (K_{mb}-K\sigma_{mb}) \notag \\
\mbox{V}=& \frac{1}{8\pi G}\sqrt{\sigma} \sigma^{bc} \Delta \widetilde{\Gamma}_{cb}^m (K_{am}-K\sigma_{am}),
\end{align}
the $\Delta \widetilde{\Gamma}_{ab}^c$ in (\ref{termsinHa}) is determined from $\Delta \sigma_{ab}$ as
\be\label{deltaconnection} \Delta\widetilde{\Gamma}_{ab}^c=\frac{1}{2} \sigma^{cd}(D_{a} \Delta \sigma_{bd}
+D_{b} \Delta \sigma_{ad} -D_{d} \Delta \sigma_{ab})
+o(\lambda), \ee
and the $\Delta \sigma_{ab}$, $\Delta K_{ab}$ in (\ref{termsinHa}) and (\ref{deltaconnection}) are interpreted as being applied with (\ref{evotwist}).
With some computations, we get
\begin{align}\label{termsinHaf}
\mbox{I}=&0 \notag \\
\mbox{II}=&\sqrt{\sigma} \lambda \big(-e^p D_p(K_{mn}e^m n^n)+n^pD_p(K_{mn}n^mn^n)\big) \delta(\rho) e_a \notag \\
&-\sqrt{\sigma} \lambda n^p D_p(K_{mn}e^m n^n) \delta(\rho)n_a+o(\lambda) \notag \\
\mbox{III}=&
-\sqrt{\sigma} \lambda K_{mn}n^mn^n \delta'(\rho) e_a+2\sqrt{\sigma} \lambda K_{mn}e^m n^n \delta'(\rho)n_a \notag \\
&+\sqrt{\sigma} \lambda \big( e^p D_p(K_{mn}e^mn^n)-n^pD_p(K_{mn}n^mn^n) \big) \delta(\rho) e_a \notag \\
&+\sqrt{\sigma} \lambda \big(2n^p D_p(K_{mn}e^mn^n)-e^pD_p(K_{mn}n^mn^n) \big) \delta(\rho) n_a+o(\lambda) \notag \\
\mbox{IV}=&-\sqrt{\sigma} \lambda  K_{mn} e^m n^n \delta'(\rho) n_a+o(\lambda) \notag \\
\mbox{V}=&\sqrt{\sigma} \lambda  K_{mn}n^mn^n \delta'(\rho) e_a
-\sqrt{\sigma} \lambda K_{mn}e^mn^n \delta'(\rho) n_a+o(\lambda),
\end{align}
where we have used (\ref{constraintsequ}), (\ref{constraints}), (\ref{drho}), (\ref{orthonormalApp}), (\ref{prerel3}), the last equation of (\ref{relationongeo}), and
\be\label{evoconnection} \Delta \widetilde{\Gamma}_{ab}^c
=-8\pi G \lambda n_an_b e^c \delta'(\rho)+o(\lambda)
\ee
which is computed from (\ref{deltaconnection}) together with (\ref{drho}), (\ref{prerel3}). We also point out the following relation
\be\label{constraintsongamma} \big( n^pD_p(K_{mn}e^mn^n)-e^pD_p(K_{mn}n^mn^n) \big) \big|_{\gamma}=0, \ee
which can be derived from contracting the second equation of (\ref{constraintsequ}) with $e^a$, restricting to the geodesic $\gamma$, and taking use of (\ref{constraints}), (\ref{orthonormalApp}), (\ref{prerel3}). By applying (\ref{termsinHaf}) to (\ref{DeltaHa}) and taking use of (\ref{constraintsongamma}), we get
\be\label{dHaf} \Delta {\cal{H}}_a=o(\lambda). \ee

From the equations (\ref{dHf}) and (\ref{dHaf}), we verify that the evolution generated by the twist represented in (\ref{evotwist}) indeed preserves the constraints (\ref{constraintsequ}).

\section{More on the example appearing in subsection \ref{twistill}}\label{Cauchycurvature}

In this appendix, we study more on the example appearing in subsection \ref{twistill}. Here, we revisit the equivalence check of the system's evolution on a general Cauchy surface that contains the $y$-axis.
From the computation, we also manifest how the evolution of the extrinsic curvature in (\ref{evotwist}) arises from the relative shift.

We first clarify some conventions. We denote the coordinates of the three dimensional spacetime as $x^{\mu}=(t,x,y)$, and the coordinates of the Cauchy surface as $\bar{x}^a=(\rho,s)$. Here, $\rho$ takes the definition below equation (\ref{deltaA}), and $s$ is another coordinate of the Cauchy surface. We also denote the embedding of the Cauchy surface as $x^{\mu}(\bar{x})$.

With these conventions, we now represent the set of initial data for the original system (\ref{3dgo}).
We first introduce the future-pointing normal vector $\tau^{(o)\mu}(\bar{x})$ of the Cauchy surface for the original metric (\ref{3dgo}), which satisfies the following conditions
\begin{align}\label{exnormal1} &\eta_{\mu\nu} \tau^{(o)\mu}(\bar{x}) \frac{\partial x^{\nu}}{\partial \bar{x}^a}=0 \notag \\
&\eta_{\mu\nu} \tau^{(o)\mu}(\bar{x}) \tau^{(o)\nu}(\bar{x})=-1. \end{align}
With the embedding of the Cauchy surface $x^{\mu}(\bar{x})$ and the normal vector $\tau^{(o)\mu}(\bar{x})$, we can represent the set of initial data for the original system (\ref{3dgo}) as
\begin{align}\label{genino}
&\sigma_{ab}^{(o)}=\eta_{\mu\nu} \frac{\partial x^{\mu}}{\partial \bar{x}^a} \frac{\partial x^{\nu}}{\partial \bar{x}^b} \notag \\
&K_{ab}^{(o)}=\eta_{\mu\nu} \frac{\partial \tau^{(o)\mu}}{\partial \bar{x}^a} \frac{\partial x^{\nu}}{\partial \bar{x}^b}.
\end{align}
For the following application, we also provide the following relations on the $y$-axis for the normal vector $\tau^{(o)\mu}$ as
\be\label{tauy} \tau^{(o)\mu}(\bar{x}) \delta_{\mu}^{\phantom{\mu}y} \big|_{\rho=0}=0, \ee
and for $e_a$, $n_a$ as
\begin{align}\label{eagen}
e_a \big|_{\rho=0}=&\frac{\partial x^{\mu}}{\partial\bar{x}^a} \delta_{\mu}^{\phantom{\mu}y} \bigg|_{\rho=0} \notag \\
n_a \big|_{\rho=0}=&\delta_a^{\phantom{a}\rho}.
\end{align}
Here, (\ref{tauy}) is derived as
\begin{align}
0=& \eta_{\mu\nu} \tau^{(o)\mu} \frac{\partial x^{\nu}}{\partial \bar{x}^a} e^a \bigg|_{\rho=0} \notag \\
=& \eta_{\mu\nu} \tau^{(o)\mu} \delta^{\nu}_{\phantom{\nu}y} \big|_{\rho=0} \notag \\
=& \tau^{(o)\mu} \delta_{\mu}^{\phantom{\mu}y} \big|_{\rho=0},
\end{align}
where we have used the first equation of (\ref{exnormal1}), and the fact that on the $y$-axis the pushforward of $e^a$ to the three dimensional spacetime is $\frac{\partial}{\partial y}$ as
\be\label{epushforward} e^a\frac{\partial x^{\mu}}{\partial \bar{x}^a} \bigg|_{\rho=0}=\delta^{\mu}_{\phantom{\mu}y}. \ee
The first equation of (\ref{eagen}) is derived as
\begin{align}
e_a \big|_{\rho=0}=&\sigma_{ab} e^{b} \big|_{\rho=0} \notag \\
=&\eta_{\mu\nu} \frac{\partial x^{\mu}}{\partial \bar{x}^a} \frac{\partial x^{\nu}}{\partial \bar{x}^b} e^{b} \bigg|_{\rho=0} \notag \\
=&\eta_{\mu\nu} \frac{\partial x^{\mu}}{\partial \bar{x}^a} \delta^{\nu}_{\phantom{\nu}y} \bigg|_{\rho=0} \notag \\
=& \frac{\partial x^{\mu}}{\partial \bar{x}^a} \delta_{\mu}^{\phantom{\mu}y} \bigg|_{\rho=0},
\end{align}
where we have used (\ref{genino}) and (\ref{epushforward}).
The second equation of (\ref{eagen}) is from (\ref{nadrho}).

We now switch to the study of the set of initial data for the evolved system (\ref{3dge}).
Before doing the computation, we would like to first reformulate the question to a more convenient version.
The current question is to compute the set of initial data on the Cauchy surface $x^{\mu}(\bar{x})$ for the evolved system (\ref{3dge}), where the evolved system (\ref{3dge}) is the pullback of the original system (\ref{3dgo}) by the diffeomorphism (\ref{flatshift}).
While, our reformulation is to view the Cauchy surface from a different perspective: instead of taking a pullback of the three dimensional metric, we can take a pushforward of the Cauchy surface by the same diffeomorphism (\ref{flatshift}).
Precisely speaking, we view the Cauchy surface as the following embedding
\be\label{Cauchy2}
x^{(e)\mu}(\bar{x})=x^{\mu}(\bar{x})-8\pi G \lambda \delta^{\mu}_{\phantom{\mu}y} \theta(\rho), \ee
in the original system (\ref{3dgo}).
And the reformulated question is to compute the set of initial data on this Cauchy surface.
For the following discussion, we also provide the derivative of (\ref{Cauchy2}) as
\be\label{dCauchy2}
\frac{\partial x^{(e)\mu}}{\partial \bar{x}^a}
=\frac{\partial x^{\mu}}{\partial \bar{x}^a}-8\pi G \lambda \delta^{\mu}_{\phantom{\mu}y} \delta_a^{\phantom{a}\rho} \delta(\rho).
\ee

We now study our reformulated question, namely to compute the set of initial data on the Cauchy surface (\ref{Cauchy2}) for the original system (\ref{3dgo}).
We first compute the normal vector $\tau^{(e)\mu}(\bar{x})$ from the following conditions
\begin{align}\label{exnormal2} &\eta_{\mu\nu} \tau^{(e)\mu}(\bar{x}) \frac{\partial x^{(e)\nu}}{\partial \bar{x}^a}=0 \notag \\
&\eta_{\mu\nu} \tau^{(e)\mu}(\bar{x}) \tau^{(e)\nu}(\bar{x})=-1. \end{align}
By applying (\ref{dCauchy2}) to (\ref{exnormal2}), taking use of (\ref{tauy}), and comparing with (\ref{exnormal1}), we realize that $\tau^{(e)\mu}(\bar{x})$ has the same expression as $\tau^{(o)\mu}(\bar{x})$
\be\label{tau2} \tau^{(e)\mu}(\bar{x})=\tau^{(o)\mu}(\bar{x}). \ee
Since we are now in the flat spacetime (\ref{3dgo}), we can still use the following expressions to compute the set of initial data
\begin{align}\label{genine}
&\sigma_{ab}^{(e)}=\eta_{\mu\nu} \frac{\partial x^{(e)\mu}}{\partial \bar{x}^a} \frac{\partial x^{(e)\nu}}{\partial \bar{x}^b} \notag \\
&K_{ab}^{(e)}=\eta_{\mu\nu} \frac{\partial \tau^{(e)\mu}}{\partial \bar{x}^a} \frac{\partial x^{(e)\nu}}{\partial \bar{x}^b}.
\end{align}
By applying (\ref{dCauchy2}), (\ref{tau2}) to (\ref{genine}), and taking use of (\ref{genino}), (\ref{eagen}), the last equation of (\ref{relationongeo}), we get the set of initial data for the evolved system
\begin{align}
&\sigma^{(e)}_{ab}=\sigma^{(o)}_{ab}-8\pi G \lambda \delta(\rho)(e_an_b+n_ae_b)+o(\lambda) \notag \\
&K^{(e)}_{ab}=K^{(o)}_{ab}-8\pi G \lambda \delta(\rho)K_{mn}^{(o)}e^mn^n n_an_b,
\end{align}
which is precisely the system's evolution represented in (\ref{evotwist}).

\section{Deriving the brackets of the boundary stress tensor}\label{deTTbracket}

In this appendix, we derive the brackets of the boundary stress tensor. Here, the derivation is based on the asymptotic symmetries and their charges.

\subsection{Asymptotic symmetries and their charges}

Before deriving the brackets of the boundary stress tensor, we first review the asymptotic symmetries and their charges.

Under the Fefferman-Graham gauge, the asymptotic symmetry can be represented as an infinitesimal diffeomorphism
\be\label{asydiff} \Delta_{\xi} g_{\mu\nu}={\cal{L}}_{\xi}g_{\mu\nu}, \ee
whose diffeomorphism parameter has the following expression
\begin{align}\label{diffeo}
&\xi^Z=\frac{1}{2}Z \big(\xi_0^{U}{}'(U)+\xi_0^{V}{}'(V)\big) \notag \\
&\xi^U=\xi_0^{U}(U)+\frac{\frac{1}{2}Z^2}{1-\frac{36}{c^2}Z^4T_{UU}(U)T_{VV}(V)} \xi_0^{V}{}''(V)
-\frac{\frac{3}{c}Z^4T_{VV}(V)}{1-\frac{36}{c^2}Z^4T_{UU}(U)T_{VV}(V)} \xi_0^{U}{}''(U) \notag \\
&\xi^V=\xi_0^{V}(V)+\frac{\frac{1}{2}Z^2}{1-\frac{36}{c^2}Z^4T_{UU}(U)T_{VV}(V)}\xi_0^{U}{}''(U)
-\frac{\frac{3}{c}Z^4T_{UU}(U)}{1-\frac{36}{c^2}Z^4T_{UU}(U)T_{VV}(V)}\xi_0^{V}{}''(V).
\end{align}
Here, the $T_{UU}$, $T_{VV}$ are the boundary stress tensor of the general solution (\ref{generalsol}) on which we act the diffeomorphism.
And the asymptotic symmetry has the following action on the boundary stress tensor
\begin{align}\label{dxiT}
&\Delta_\xi T_{UU}(U)=\xi_0^U(U)T_{UU}^{\phantom{UU}\prime}(U)+2\xi_0^{U\prime}(U)T_{UU}(U)+\frac{c}{12}\xi_0^{U\prime\prime\prime}(U) \notag \\
&\Delta_\xi T_{VV}(V)=\xi_0^V(V)T_{VV}^{\phantom{VV}\prime}(V)+2\xi_0^{V\prime}(V)T_{VV}(V)+\frac{c}{12}\xi_0^{V\prime\prime\prime}(V).
\end{align}

In the remaining part of this subsection, we will compute the charge of the asymptotic symmetry with the covariant phase space formalism.

We first recast the pure AdS$_3$ into the covariant phase space formalism.
Specifically, we need to compute the symplectic form.
(Here, we use the convention in \cite{Harlow:2019yfa}.)
Following the standard procedure of the covariant phase space formalism, we rewrite the Lagrangian density in the differential form as
\be
\mathbf{L}=\frac{1}{16\pi G}(R+2)\frac{1}{3!}\epsilon_{\mu_0\mu_1\mu_2}dx^{\mu_0}\wedge dx^{\mu_1}\wedge dx^{\mu_2}.
\ee
We take a variation of the Lagrangian density and represent it as
\be\label{Lden}
\delta \mathbf{L}=\mathbf{E}^{\mu\nu}\delta g_{\mu\nu}+d\mathbf{\Theta},
\ee
where
\begin{align}\label{ETheta}
&\mathbf{E}^{\mu\nu}=\frac{1}{16\pi G} \big(-R^{\mu\nu}+\frac{1}{2}Rg^{\mu\nu}+g^{\mu\nu} \big)
\frac{1}{3!}\epsilon_{\mu_0\mu_1\mu_2}dx^{\mu_0}\wedge dx^{\mu_1}\wedge dx^{\mu_2} \notag \\
&\mathbf{\Theta}=\frac{1}{32\pi G} \big( g^{\mu\rho}\nabla^\nu \delta g_{\mu\nu}-\nabla^\rho(g^{\mu\nu}\delta g_{\mu\nu}) \big) \epsilon_{\rho\mu_1\mu_2}dx^{\mu_1}\wedge dx^{\mu_2}.
\end{align}
And we finally define the symplectic form as
\be\label{Omega}
\Omega=\int_\Sigma \delta\mathbf{\Theta}.
\ee
Here, $\Sigma$ can be any Cauchy surface with a proper asymptotic behavior.

Having defined the symplectic form, we now compute the charge of the asymptotic symmetry. Specifically, we need to compute the $Q_{\xi}$ from the following equation
\be\label{Hamiltonianequ} -X_{\xi}\cdot \mathbf{\Omega}=\delta Q_{\xi}, \ee
where
\be\label{Xxi}
X_\xi=\int d^3x\mathcal{L}_\xi g_{\mu\nu}(x)\frac{\delta}{\delta g_{\mu\nu}(x)},
\ee
with $\xi^{\mu}$ given in (\ref{diffeo}).
We first take a reformulation for the left hand side of (\ref{Hamiltonianequ}) as
\begin{align}\label{reformulation}
-X_\xi \cdot \mathbf{\Omega}=&\int_\Sigma -X_\xi \cdot \delta\mathbf{\Theta} \notag \\
=&\int_{\Sigma} \Big( -{\cal{L}}_{X_{\xi}}\mathbf{\Theta}+\delta(X_{\xi} \cdot \mathbf{\Theta}) \Big) \notag \\
=&\int_{\Sigma} \Big( -{\cal{L}}_{\xi} \mathbf{\Theta}-X_{\delta \xi} \cdot \mathbf{\Theta} +\delta(X_{\xi} \cdot \mathbf{\Theta}) \Big) \notag \\
=&\int_{\Sigma} \Big( -\xi \cdot d \mathbf{\Theta}-d(\xi \cdot \mathbf{\Theta})
-X_{\delta \xi} \cdot \mathbf{\Theta} +\delta(X_{\xi} \cdot \mathbf{\Theta}) \Big) \notag \\
=&\int_{\Sigma} \Big( -\xi \cdot \delta \mathbf{L} -d(\xi \cdot \mathbf{\Theta})
-X_{\delta \xi} \cdot \mathbf{\Theta} +\delta(X_{\xi} \cdot \mathbf{\Theta}) \Big) \notag \\
=&\int_{\Sigma} \delta (X_{\xi} \cdot \mathbf{\Theta}-\xi\cdot \mathbf{L})
-\int_{\Sigma} (X_{\delta \xi} \cdot \mathbf{\Theta}-\delta \xi \cdot \mathbf{L})
-\int_{\partial\Sigma} \xi \cdot \mathbf{\Theta} .
\end{align}
Here, we have used the Cartan's magic formula, the covariance of the symplectic form, (\ref{Lden}), and the equations of motion.
We now compute the three terms in the last expression of (\ref{reformulation}) respectively as
\begin{align}\label{XO1}
\int_\Sigma \delta(X_\xi \cdot \mathbf{\Theta}-\xi \cdot \mathbf{L})&=\delta \int_{\partial \Sigma}\frac{1}{32\pi G}(\nabla^\mu \xi^\rho-\nabla^\rho \xi^\mu)\epsilon_{\rho\mu\mu_2}dx^{\mu_2} \notag \\
&=-\frac{1}{2\pi}\int dU\delta T_{UU}(U)\xi_0^U(U)-\frac{1}{2\pi}\int dV\delta T_{VV}(V)\xi_0^V(V),
\end{align}
and
\be\label{XO2}
\int_\Sigma (-1)(X_{\delta \xi} \cdot \mathbf{\Theta}-\delta \xi \cdot \mathbf{L})=\int_{\partial \Sigma}(-1)\frac{1}{32\pi G}(\nabla^\mu \delta\xi^\rho-\nabla^\rho \delta\xi^\mu)\epsilon_{\rho\mu\mu_2}dx^{\mu_2}=0,
\ee
and
\be\label{XO3}
\int_{\partial \Sigma}(-1)\xi \cdot \mathbf{\Theta}=\int_{\partial \Sigma}(-1)\frac{1}{16\pi G}(g^{\mu\rho}\nabla^\nu\delta g_{\mu\nu}-\nabla^\rho(g^{\mu\nu}\delta g_{\mu\nu}))\xi^\sigma \epsilon_{\rho\sigma\mu_2}dx^{\mu_2}=0,
\ee
where, in deriving these equations, we have used the equations of motion, the general solution (\ref{generalsol}), and the diffeomorphism (\ref{diffeo}). Applying (\ref{XO1}), (\ref{XO2}), (\ref{XO3}) to (\ref{reformulation}), we get
\be\label{XOMEGA}
-X_\xi \cdot \Omega=-\frac{1}{2\pi}\int dU\delta T_{UU}(U)\xi_0^U(U)-\frac{1}{2\pi}\int dV\delta T_{VV}(V)\xi_0^V(V).
\ee
Moreover, by applying (\ref{XOMEGA}) to (\ref{Hamiltonianequ}), we get the charge of the asymptotic symmetry as
\be\label{concharge}
Q_\xi=-\frac{1}{2\pi}\int dU T_{UU}(U)\xi_0^U(U)-\frac{1}{2\pi}\int dV T_{VV}(V)\xi_0^V(V).
\ee

\subsection{The brackets of the boundary stress tensor}

Having reviewed the asymptotic symmetries and their corresponding charges, we now derive the brackets of the boundary stress tensor.
Since $Q_{\xi}$ is the charge of the the asymptotic symmetry, we have the following relation between the bracket with $Q_{\xi}$ and the action of the asymptotic symmetry as
\be \{g_{\mu\nu},Q_{\xi}\}=\Delta_{\xi} g_{\mu\nu}, \ee
and
\begin{align}\label{bracketQT}
&\{T_{UU}(U),Q_\xi \}=\Delta_{\xi} T_{UU}(U) \notag \\
&\{T_{VV}(V),Q_\xi \}=\Delta_{\xi} T_{VV}(V).
\end{align}
By applying (\ref{dxiT}), (\ref{concharge}) to (\ref{bracketQT}), we can directly read out the brackets of the boundary stress tensor as
\begin{align}
&\{T_{UU}(U),T_{UU}(\widetilde{U})\}=
-2\pi\Big[\frac{c}{12}\delta'''(U-\widetilde{U})+2T_{UU}(U)\delta'(U-\tilde{U})
+T_{UU}'(U)\delta(U-\tilde{U})\Big] \notag \\
&\{T_{VV}(V),T_{VV}(\widetilde{V})\}
=-2\pi\Big[\frac{c}{12}\delta'''(V-\tilde{V})+2T_{VV}(V)\delta'(V-\tilde{V})
+T_{VV}'(V)\delta(V-\tilde{V})\Big].
\end{align}

\section{Killing fields}
In this appendix, we review some relevant properties of the Killing fields of the vacuum solution
\be ds^2=\frac{dz^2-dudv}{z^2}. \ee

We list the independent Killing fields as
\begin{align}\label{xiu}
&\hat{\xi}_{(u,-1)}=\frac{\partial}{\partial u} \notag \\
&\hat{\xi}_{(u,0)}=\frac{1}{2}z\frac{\partial}{\partial z}+u\frac{\partial}{\partial u} \notag \\
&\hat{\xi}_{(u,1)}=uz\frac{\partial}{\partial z}+u^2\frac{\partial}{\partial u}+z^2\frac{\partial}{\partial v},
\end{align}
and
\begin{align}\label{xiv}
&\hat{\xi}_{(v,-1)}=\frac{\partial}{\partial v} \notag \\
&\hat{\xi}_{(v,0)}=\frac{1}{2}z\frac{\partial}{\partial z}+v\frac{\partial}{\partial v} \notag \\
&\hat{\xi}_{(v,1)}=vz\frac{\partial}{\partial z}+z^2\frac{\partial}{\partial u}+v^2\frac{\partial}{\partial v},
\end{align}
which correspond to the following global conformal Killing fields as
\begin{align}\label{etauv}
\hat{\eta}_{(u,i)}=u^{i+1}\frac{\partial}{\partial u} \notag \\
\hat{\eta}_{(v,i)}=v^{i+1}\frac{\partial}{\partial v}.
\end{align}

We provide the following property of the Killing fields on their derivative
\begin{align}\label{derivativexi}
&\nabla_{\mu}\xi_{(u,i)\nu}=-\epsilon_{\mu\nu\rho}\xi_{(u,i)}^{\phantom{(u,i)}\rho} \notag \\
&\nabla_{\mu}\xi_{(v,i)\nu}=\epsilon_{\mu\nu\rho}\xi_{(v,i)}^{\phantom{(v,i)}\rho},
\end{align}
where the $\epsilon_{\mu\nu\rho}$ is the volume form whose orientation is specified by its non-zero components as
\begin{align}
&\epsilon_{zuv}=\epsilon_{uvz}=\epsilon_{vzu}=\frac{1}{2z^3} \notag \\
&\epsilon_{zvu}=\epsilon_{vuz}=\epsilon_{uzv}=-\frac{1}{2z^3}.
\end{align}

We also use the Killing fields to represent the tangent vector of a geodesic.
Here, we consider a spacelike geodesic $\gamma$ connecting two points $(u_1,v_1)$, $(u_2,v_2)$ on the asymptotic boundary with
\begin{align}
&u_1>u_2 \notag  \\
&v_1<v_2.
\end{align}
The geodesic $\gamma$ can be represented as
\begin{align}
&z(s)=\frac{1}{2}\sqrt{(u_1-u_2)(v_2-v_1)} \frac{1}{\cosh s} \notag \\
&u(s)=\frac{u_2-u_1}{2} \frac{\sinh s}{\cosh s} + \frac{u_1+u_2}{2} \notag \\
&v(s)=\frac{v_2-v_1}{2} \frac{\sinh s}{\cosh s} + \frac{v_1+v_2}{2},
\end{align}
with $s$ being the proper length up to a shift.
And the tangent vector of geodesic $\gamma$ is
\be
e^{\mu}\equiv \frac{d x^{\mu}}{ds},
\ee
where $x^{\mu}=\{z,u,v\}$.
We directly provide two Killing fields to represent the tangent vector of the geodesic $\gamma$ as
\begin{align}\label{xiuve}
&\hat{\xi}_{(u,e)}=\frac{2}{u_1-u_2}\hat{\xi}_{(u,1)}-\frac{2(u_1+u_2)}{u_1-u_2}\hat{\xi}_{(u,0)}
+\frac{2u_1 u_2}{u_1-u_2} \hat{\xi}_{(u,-1)} \notag \\
&\hat{\xi}_{(v,e)}=-\frac{2}{v_2-v_1}\hat{\xi}_{(v,1)}+\frac{2(v_1+v_2)}{v_2-v_1}\hat{\xi}_{(v,0)}
-\frac{2v_1 v_2}{v_2-v_1} \hat{\xi}_{(v,-1)},
\end{align}
which can be equivalently represented by their corresponding global conformal Killing fields as
\begin{align}\label{etae}
&\hat{\eta}_{(u,e)}=\frac{2(u-u_1)(u-u_2)}{u_1-u_2} \frac{\partial}{\partial u} \notag \\
&\hat{\eta}_{(v,e)}=-\frac{2(v-v_1)(v-v_2)}{v_2-v_1} \frac{\partial}{\partial v}.
\end{align}
We can directly verify that the Killing fields $\hat{\xi}_{(u,e)}$, $\hat{\xi}_{(v,e)}$, when restricted on the geodesic $\gamma$, indeed coincide with the tangent vector of the geodesic $\gamma$ as
\be\label{Killingequ}
\hat{\xi}_{(u,e)} \big|_{\gamma}=\hat{\xi}_{(v,e)} \big|_{\gamma}=\hat{e}|_{\gamma}.
\ee

\section{The incompatibility between the system's evolutions (\ref{evolutionA}), (\ref{evotwist}) and the asymptotic boundary conditions (\ref{asy0})}\label{incompatibility}

In this appendix, we use a concrete computation to show the incompatibility between the system's evolutions (\ref{evolutionA}), (\ref{evotwist}) and the asymptotic boundary conditions (\ref{asy0}).
In particular, we act the system's evolutions (\ref{evolutionA}), (\ref{evotwist}) on the vacuum solution, and we find that the evolved system is not compatible with the asymptotic boundary conditions (\ref{asy0}).

We now introduce the setup. We introduce a coordinate system $(Z,T,X)$.
In this coordinate system, we choose the Cauchy surface $\Sigma$ at $T=0$.
We choose the end points of the geodesic $\gamma$ on the asymptotic boundary at $(T=0,X=-a)$ and $(T=0,X=a)$.
And we choose the unevolved system to be the vacuum solution represented as
\be\label{vacuumZTX} ds^2=\frac{dZ^2-dT^2+dX^2}{Z^2}. \ee

For the following discussion, we also provide some structures for the vacuum solution (\ref{vacuumZTX}).
First, we point out that the geodesic $\gamma$ is on the Cauchy surface $\Sigma$ and is located at
\be Z=\sqrt{a^2-X^2}, \ee
for $-a<X<a$.
Second, we express $e_a$, $n_a$ on the geodesic as
\begin{align}\label{en}
&e_a|_{\gamma}=\Big(-\frac{X}{aZ},\frac{1}{a}\Big) \notag \\
&n_a |_{\gamma}=\Big(-\frac{1}{a},-\frac{X}{aZ} \Big). \end{align}
Third, we choose $\rho$ to be
\be\label{rho}
\rho=\frac{1}{Z}\big(a-\sqrt{Z^2+X^2}\big), \ee
and we compute $\delta(\rho)$ as
\begin{align}\label{deltarho11}
\delta(\rho)&=\frac{aZ}{\sqrt{a^2-Z^2}}
\Big( \delta(X+\sqrt{a^2-Z^2})+\delta(X-\sqrt{a^2-Z^2}) \Big),
\end{align}
for $Z<a$.

Given these structures, we now compute the system's evolutions (\ref{evolutionA}) and (\ref{evotwist}) on the set of initial data.
In particular, we only focus on their actions on special components, from which we show the incompatibility between the system's evolutions (\ref{evolutionA}), (\ref{evotwist}) and the asymptotic boundary conditions (\ref{asy0}).

First, we consider the system's evolution (\ref{evotwist}) and we focus on its action on $\sigma_{ZX}$.
By applying (\ref{en}), (\ref{deltarho11}) to (\ref{evotwist}) and by taking a small $Z$ expansion, we get the evolution on $\sigma_{ZX}$ as
\be\label{sigmacom1}
\Delta g_{ZX}=\Delta\sigma_{ZX}=\lambda\bigg[-8\pi G \frac{1}{Z} \Big(\delta(X+a)+\delta(X-a) \Big)+{\cal{O}}(Z)\bigg]+o(\lambda).
\ee
The expression (\ref{sigmacom1}) explicitly shows the incompatibility between the system's evolution (\ref{evotwist}) and the asymptotic boundary conditions (\ref{asy0}).

Second, we consider the system's evolution (\ref{evolutionA}) and we focus on its action on $K_{XX}$.
By applying (\ref{en}), (\ref{deltarho11}) to (\ref{evolutionA}) and by taking a small $Z$ expansion, we get the evolution on $K_{XX}$ as
\be\label{DeltaKXX} \Delta K_{XX}=\lambda \Big[ 8\pi G \frac{1}{Z} \big( \delta(X+a)+\delta(X-a) \big)+{\cal{O}}(Z) \Big]+o(\lambda). \ee
And by comparing (\ref{DeltaKXX}) with (\ref{Kab}), we then get
\be\label{DeltagTZ} \Delta g_{TZ}=\Delta \beta_{Z}=
\lambda \Big[ 8\pi G \frac{1}{Z} \big( \delta(X+a)+\delta(X-a) \big) +o\big(\frac{1}{Z}) \Big] +o(\lambda). \ee
Here, we have also used the following expressions of the connection of the induced metric for the unevolved system
\be
\left\{
\begin{array}{l}
\widetilde{\Gamma}_{ZZ}^Z=-\frac{1}{Z} \\ \widetilde{\Gamma}_{ZZ}^X=0 \end{array} \right.~~~
\left\{\begin{array}{l}
\widetilde{\Gamma}_{ZX}^Z=0 \\ \widetilde{\Gamma}_{ZX}^X=-\frac{1}{Z} \end{array} \right.~~~
\left\{ \begin{array}{l}
\widetilde{\Gamma}_{XX}^Z=\frac{1}{Z} \\ \widetilde{\Gamma}_{XX}^X=0 \end{array} \right. . \ee
And we have assumed
\be \Delta g_{\mu\nu}=\lambda \cdot o \Big( \frac{1}{Z^2} \Big)+o(\lambda). \ee
The expression (\ref{DeltagTZ}) shows the incompatibility between the system's evolution (\ref{evolutionA}) and the asymptotic boundary conditions (\ref{asy0}).

\end{document}